\documentclass[prx, 11pt, reprint, amsmath,amssymb,showpacs,floatfix,longbibliography, aps, twocolumn, superscriptaddress, nofootinbib]{revtex4-2}
\usepackage[utf8]{inputenc}
\usepackage{CJKutf8}

\usepackage{natbib}

\usepackage[left=1in,right=1in,top=1in,bottom=1in]{geometry}

\usepackage{color}
\usepackage[dvipsnames]{xcolor}
\usepackage{hyperref}
\usepackage{enumitem}
\usepackage{cancel}
\hypersetup{colorlinks=true,citecolor=blue,linkcolor=blue, urlcolor=blue}
\hypersetup{linktocpage}
\usepackage{graphicx}
\usepackage{bm} 
\usepackage[dvipsnames]{xcolor}
\usepackage{mathtools} 

\usepackage{environ} 
\usepackage[caption=false]{subfig}
\NewEnviron{eqs}{%
\begin{equation}\begin{split}
    \BODY
\end{split}\end{equation}
}

\usepackage[most]{tcolorbox}

\definecolor{examplecolor}{RGB}{0,0,157}
\newcounter{tcbexample}
\newenvironment{tcbexample}[1][]{%
  \refstepcounter{tcbexample}%
  \begin{tcolorbox}[
    enhanced,
    breakable,
    pad at break*=4mm,
    attach boxed title to top left={yshift=-3mm, yshifttext=-1mm},
    colback=examplecolor!2!white,
    colframe=examplecolor!95!black,
    colbacktitle=examplecolor!95!black,
    fonttitle=\bfseries,
    boxed title style={size=small,colframe=examplecolor!65!black},
    parbox=false,
    title=Example~\thetcbexample\if\relax\detokenize{#1}\relax\else: #1\fi
  ]%
}{%
  \end{tcolorbox}%
}

\makeatletter
\newcommand{\pagefootnotemark}{%
  \refstepcounter{footnote}%
  \protected@xdef\@thefnmark{\thefootnote}%
  \@footnotemark
}
\newcommand{\pagefootnotetext}[1]{%
  \protected@xdef\@thefnmark{\thefootnote}%
  \@footnotetext{#1}%
}
\makeatother

\usepackage{tabularray} 
\UseTblrLibrary{booktabs} 

\usepackage{physics}
\usepackage{dsfont}

\usepackage{centernot}

\usepackage[normalem]{ulem}

\usepackage{subcaption}

\usepackage{caption}
\usepackage{ragged2e} 
\makeatletter 
\def\l@subsubsection#1#2{}
\makeatother

\newcommand{\LL}{\mathsf{L}}

\newcommand{\Z}{{\mathbb Z}}

\newcommand{\knit}{\mathbin{\ltimes\mkern-13.9mu\rtimes}}

\definecolor{tyler}{rgb}{1,.2,0}

\definecolor{bluem}{rgb}{0.0, 0.5, 0.69}

\begin{document}

\title{Universal quantum computation with group surface codes}

\author{Naren Manjunath}
\thanks{These authors contributed equally.}
\affiliation{Perimeter Institute for Theoretical Physics, Waterloo, Ontario N2L 2Y5, Canada}
\author{Vieri Mattei}
\thanks{These authors contributed equally.}
\affiliation{Department of Physics and Astronomy, Purdue University, West Lafayette, IN, 47907}
\author{Apoorv Tiwari}
\affiliation{Center for Quantum Mathematics \& 
Danish Institute for Advanced Study (Danish IAS), Southern Denmark University, Campusvej 55, 5230 Odense, Denmark}
\author{{Tyler D.~Ellison}}
\email{tdelliso@purdue.edu}
\affiliation{Perimeter Institute for Theoretical Physics, Waterloo, Ontario N2L 2Y5, Canada}
\affiliation{Department of Physics and Astronomy, Purdue University, West Lafayette, IN, 47907}
\affiliation{Purdue Quantum Science and Engineering Institute, Purdue University, West Lafayette, IN, 47907}

\begin{abstract}
    We introduce group surface codes, which are a natural generalization of the $\mathbb{Z}_2$ surface code, and equivalent to quantum double models of finite groups with specific boundary conditions. We show that group surface codes can be leveraged to perform non-Clifford gates in $\mathbb{Z}_2$ surface codes, thus enabling universal computation along with established means of performing logical Clifford gates. Moreover, for suitably chosen groups, we demonstrate that arbitrary reversible classical gates can be implemented transversally in the group surface code. We present the logical operations in terms of a set of elementary logical operations, which include transversal logical gates, a means of transferring encoded information into and out of group surface codes, and preparation and readout. By composing these elementary operations, we implement a wide variety of logical gates and provide a unified perspective on recent constructions in the literature for sliding group surface codes and preparing magic states. We furthermore use tensor networks inspired by ZX-calculus to construct spacetime implementations of the elementary operations. This spacetime perspective also allows us to establish explicit correspondences with topological gauge theories. Our work extends recent efforts in performing universal quantum computation in topological orders without the braiding of anyons, and shows how certain group surface codes allow us to bypass the restrictions set by the Bravyi-K{\"o}nig theorem, which limits the computational power of topological \textit{Pauli} stabilizer models.
\end{abstract}

\maketitle

\setcounter{tocdepth}{1}
\tableofcontents

\section{Introduction}

Quantum computers have the potential for transformative applications across science and technology. Realizing this potential, however, requires performing long and reliable quantum computations in the presence of noise and imperfect operations. This necessitates redundantly encoding the information using a quantum error-correcting code. 

Among the many quantum error-correcting codes that have been proposed, the surface code, referred to here as the $\Z_2$ surface code, has emerged as one of the most promising candidates for intermediate-term quantum hardware. The surface code exhibits high thresholds against local stochastic errors and admits an implementation using geometrically local interactions in a planar layout. These favorable properties have led to substantial experimental progress across multiple physical platforms, including superconducting qubits, trapped ions, and neutral atoms~\cite{Wallraff2020surfacedetection, Marques2021surfacedetecting, Wallraff2022distance3, Zhao2022surfacecode, GoogleAI2023scaling, GoogleAI2025belowthreshold, Quantinuum2024adaptive, Quantinuum2025z3, Bluvstein2023reconfigurable, Bluvstein2025architecture}. In particular, recent experiments have demonstrated logical error suppression consistent with operation below the $\Z_2$ surface code thresholds~\cite{GoogleAI2025belowthreshold, Bluvstein2025architecture}.

The same features that make the $\Z_2$ surface code robust against errors also make the logical information harder to manipulate in computations. Logical degrees of freedom are encoded nonlocally, and consequently, logical operations must be implemented through carefully designed fault-tolerant protocols. A variety of such methods are now well established, including transversal operations, braiding of defects, and lattice surgery~\cite{Steane1996CSS, Moussa2016folded, Fowler2012practical, Bombin2010twist, Yoder2017twist, Horsman2012latticesurgery, Litinski2019gameof}. Collectively, these techniques enable the implementation of logical Clifford gates and Pauli measurements in a way that is fault-tolerant. However, it is a fundamental limitation of the $\Z_2$ surface code that these native operations are restricted to the Clifford operations~\cite{Bravyi2013bravyikonig, Beverland2016protected, Webster2018locality, Webstr2020defects, Webster2022stabilizer}.

To be able to perform universal quantum computations, it is necessary that at least one non-Clifford operation can be implemented fault tolerantly~\cite{boykin1999CliffodplusT, nebe2000invariantscliffordgroups, Bravyi2005magic}. Several approaches have been developed to supplement the $\Z_2$ surface code with such an operation. The most widely studied method is magic-state distillation~\cite{Bravyi2005magic}, in which noisy ancillary states are distilled into high-fidelity magic states using Clifford operations. Magic-state distillation is, however, notoriously resource intensive, often dominating the space-time overhead of fault-tolerant quantum algorithms~\cite{Campbell2017overheads}. 

An alternative strategy is to perform dimensional jumping, wherein the encoded information is temporarily transferred into a higher-dimensional quantum error-correcting code that admits a transversal non-Clifford gate~\cite{Bombin2015GCC, Bombin2016dimensionaljumping, Brown2016GCC, Beverland2021comparison, Brown20202DnonClifford}. This approach, however, requires hardware capable of accommodating a higher-dimensional connectivity.

A third approach, which is the focus of this work, is to leverage non-Abelian topological orders.~It has long been appreciated that non-Abelian anyons can be used to store and process quantum information through braiding and fusion~\cite{Kitaev2003computationbyanyons, Nayak2008nonAbelian}--in some cases, allowing for a universal set of quantum gates~\cite{freedman2000modularfunctoruniversalquantum, Bonesteel2005braid, Cui2015weakly, Cui2015metaplectic, Kaufmann2025doublebraid}. Here, we consider a different paradigm: rather than processing information entirely within the nonlocal internal space of the non-Abelian anyons, we code switch into a quantum error-correcting code based on a non-Abelian topological order only to enact a non-Clifford operation. Afterwards, the quantum information is transferred back into $\Z_2$ surface codes, where it continues to be protected and manipulated using standard $\Z_2$ surface-code techniques.

This approach was, to the best of our knowledge, pioneered in Ref.~\cite{Wootton2019S3}, through transferring information back and forth between non-Abelian anyons and holes in a $\Z_2$ surface code. More recently, there has been an effort to leverage non-Abelian topological orders without the need of manipulating anyons or using space-consuming defects of the $\Z_2$ surface code~\cite{Brown20202DnonClifford, huang2025D4, davydova2025D4, bauer2025planarfaulttolerantcircuitsnonclifford, sajith2025noncliffordS3}. These offer practical benefits over the scheme presented in Ref.~\cite{Wootton2019S3} and provide promising alternatives for performing non-Clifford operations in $\Z_2$ surface codes.

Our work extends these recent efforts and develops a unifying framework based on elementary group theory, avoiding the abstract language of group cohomology and category theory. In particular, we formally introduce group surface codes (GSCs), which generalize the $\Z_2$ surface code to arbitrary finite groups. GSCs based on non-Abelian groups exhibit non-Abelian topological order, which we exploit to perform non-Clifford gates. 
We describe how GSCs can be interfaced with $\Z_2$ surface codes to allow for implementing these non-Clifford operations on the $\Z_2$ surface codes. In this way, GSCs provide a mechanism for completing a universal gate set in $\Z_2$ surface codes, without resorting to magic-state distillation or higher-dimensional quantum error-correcting codes. 

Importantly, the formalism of GSCs provides flexibility in the specific non-Clifford operations that can be implemented within the $\Z_2$ surface codes. This is to say that the group can be engineered for a desired non-Clifford operation. We demonstrate that an arbitrary set of reversible classical circuits can be performed transversally using a GSC, for a suitably engineered group. 

While any non-Clifford gate would be sufficient for universal quantum computation, the ability to engineer the non-Clifford operations offers the potential for algorithm-specific advantages. In particular, selecting the group to directly implement a gate may reduce circuit depth or ancillary overhead. Thus, beyond merely achieving universality, group engineering opens a pathway toward more resource-efficient fault-tolerant implementations within $\Z_2$ surface-code architectures.

\begin{table*}[t]
    \centering
    \begin{tabular}{|p{0.2\textwidth}|p{0.4\textwidth}|p{0.2\textwidth}|}
    \hline

    \multicolumn{3}{|c|}{\textbf{Transversal logical gates} (L: Left multiplication, Aut: Automorphism)}  \\ \hline 
        {\bf Group} & {\bf Logical action} & {\bf Reference} \\ \hline
        $\Z_2\times\Z_2$ & L: $X_1, \,X_2$; Aut: $\mathrm{CX}_{12},\, \mathrm{SWAP}_{12}$ & Section~\ref{sec:GSC-TransversalGates} \\ \hline
       $D_4$ & L: $X_1, X_2, X_3\mathrm{CX}_{12}$ & Section~\ref{sec:GSC-TransversalGates} \\
       & Aut: $\text{non-Clifford, e.g., } \mathrm{SWAP}_{13}\mathrm{CCX}_{132}
        $ & Section~\ref{sec:GSC-TransversalGates}\\ 
        \hline
       $D_{2^n}$ & L: $n$th level of Clifford hierarchy   &  Appendix~\ref{d2n appendix} \\ 
       &Aut: $(n+1)$th level & Appendix~\ref{d2n appendix}  \\
       \hline
       $G_{\mathrm{CCX}}$ (CSC) & L: $\mathrm{CCX}$ & Section~\ref{sec:Egs-CCX} \\ \hline
       $G_{\mathrm{C}^n\mathrm{X}}$ (CSC) & L: $C^n X$ &Section~\ref{sec:Egs-CCX} \\ \hline
       $G_\Pi$ (CSC) & L: Arbitrary reversible classical gate & Section~\ref{sec:group_engg}  \\
       \hline 
    \end{tabular}
    
    \vspace*{0.6 cm}
    
    \begin{tabular}{|p{0.2\textwidth}|p{0.4\textwidth}|p{0.2\textwidth}|}
    \hline
    \multicolumn{3}{|c|}{\textbf{Sliding}} \\ \hline
        {\bf Group} &  {\bf Logical action} & {\bf Reference} \\ \hline
        $D_4$ & $\mathrm{CCX}$ & Section~\ref{sec: extension and splitting} \\ \hline
       $D_{2^n}$ & $(n+1)$-th level of Clifford hierarchy &  Appendix~\ref{d2n appendix} \\ \hline
       $H \rtimes K$ &Controlled conjugation: $\ket{h} \otimes \ket{k} \rightarrow \ket{\bar{k} h k} \otimes \ket{k}$ & Section~\ref{sec:Egs-sliding} \\ \hline
    \end{tabular}
    
    \vspace*{0.6 cm}
    
    \begin{tabular}{|p{0.2\textwidth}|p{0.4\textwidth}|p{0.2\textwidth}|}
    \hline
    \multicolumn{3}{|c|}{\textbf{Magic state preparation}} \\ \hline
        {\bf Group} &  {\bf Final state} & {\bf Reference} \\ \hline 

        $D_4$ &$\mathrm{CX}$ state: $\left( 2\ket{00}+\ket{10}+\ket{11} \right)/{\sqrt{6}}$ & Section~\ref{sec:Egs-magicstate} \\ \hline

        $D_4$ &$T$ state: $\ket{0} + e^{i\pi/4} \ket{1}$ & Section~\ref{sec:Egs-magicstate} \\ \hline
    \end{tabular}
    \caption{Example logical operations, described in more detail in Section~\ref{sec: examples}. CSC denotes the coset surface code, introduced in Section~\ref{Sec:Egs-CSC}. }
    \label{tab:Summary_logical_gates}
\end{table*}

\begin{table*}[t]
\centering

\begin{tblr}{
  colspec = {|Q[c,m,wd=0.22\textwidth]|
              Q[c,m,wd=0.22\textwidth]|
              Q[c,m,wd=0.22\textwidth]|
              Q[c,m,wd=0.22\textwidth]|},
  rowsep = 6pt,
  hlines, vlines,
  cell{2-Z}{1-4} = {valign=m}, 
}

Identity & Left multiplication &Right multiplication & Automorphism \\
\includegraphics[scale=.2]{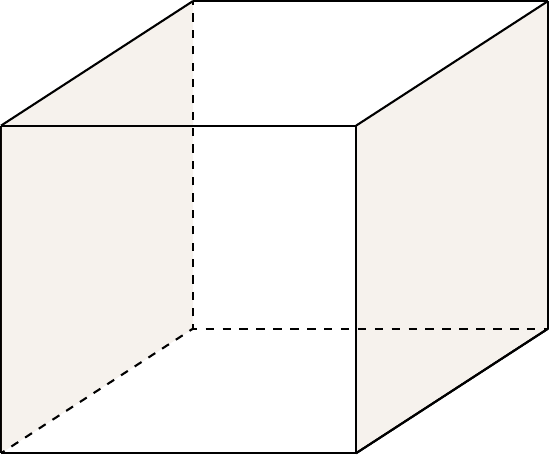}\vspace{-.0cm} & \includegraphics[scale=.2]{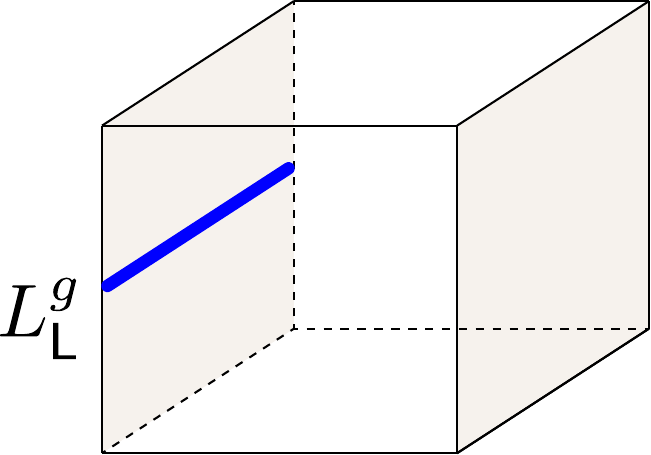}\vspace{-.0cm} & \includegraphics[scale=.2]{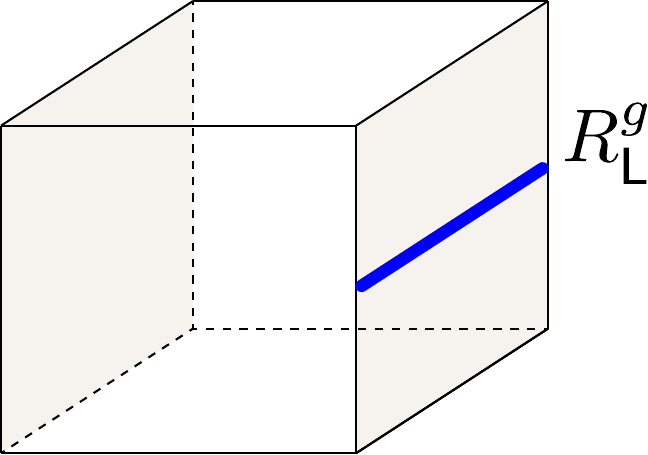}\vspace{-.0cm} & \vspace{.0cm}\includegraphics[scale=.2]{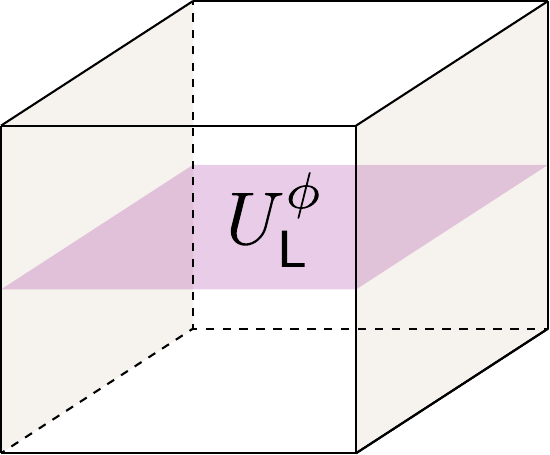}\vspace{-.0cm} \\

Extension & Splitting & Preparation & Readout \\
\includegraphics[scale=.15]{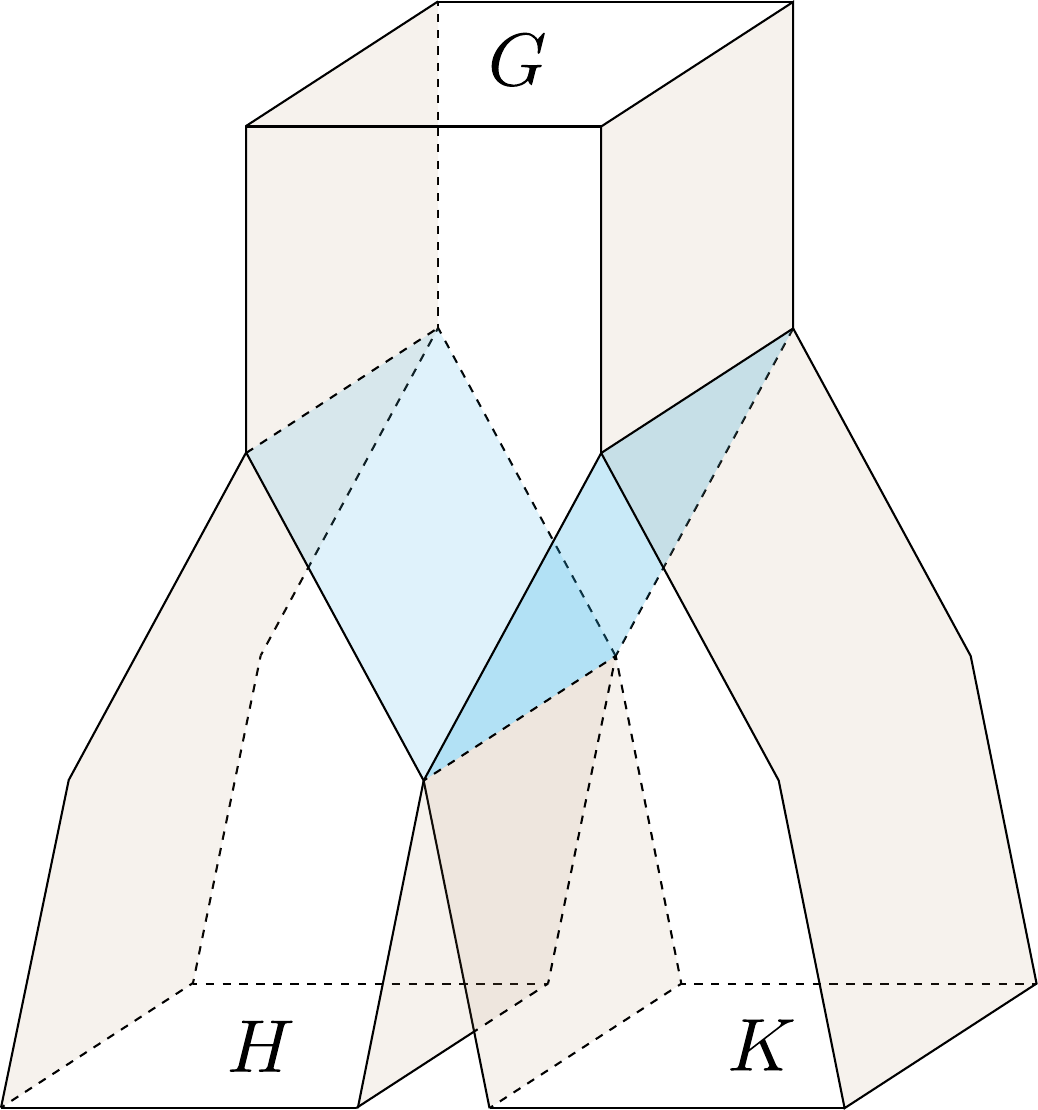} & \includegraphics[scale=.15]{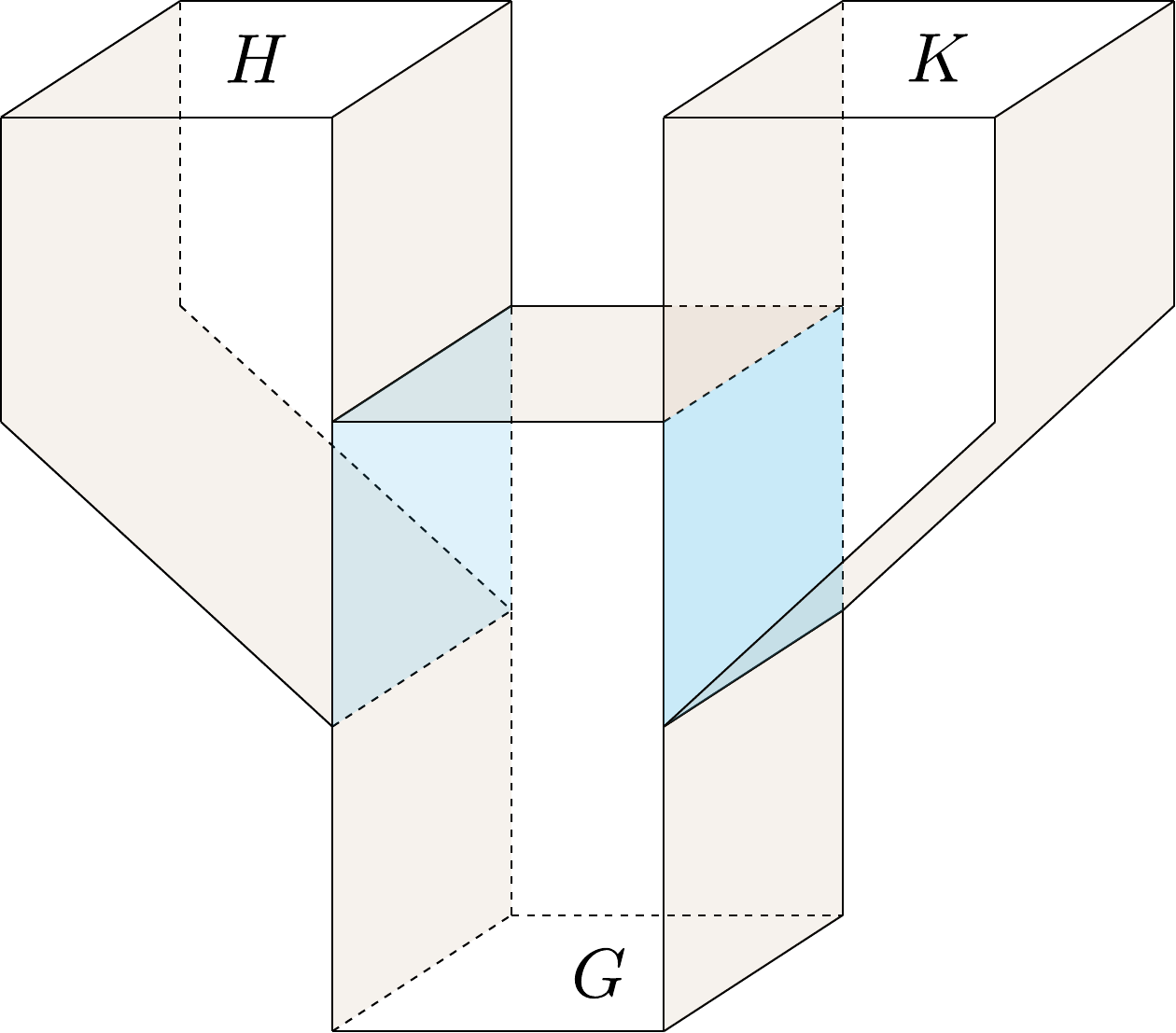} & \includegraphics[scale=.15]{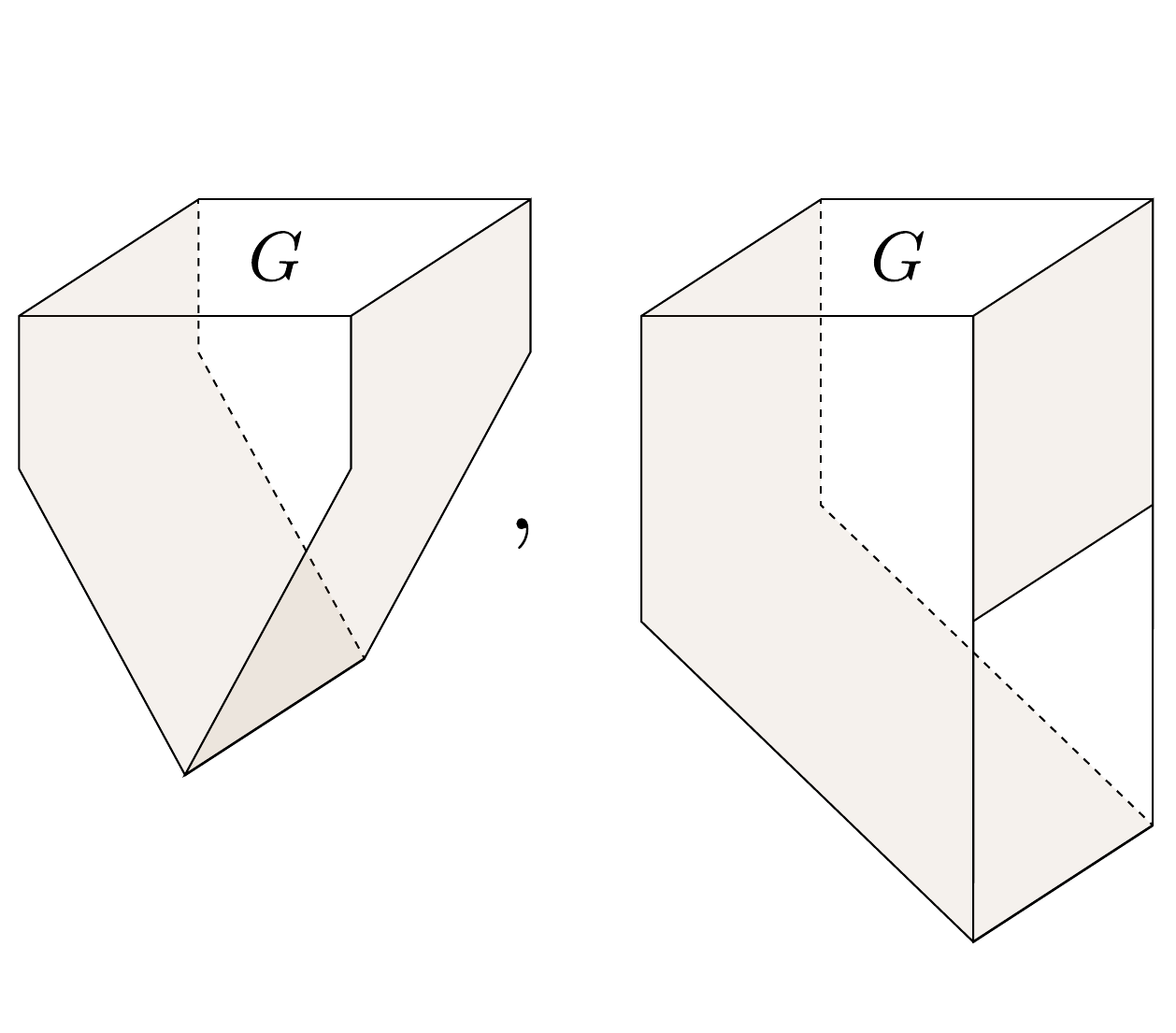} & \includegraphics[scale=.15]{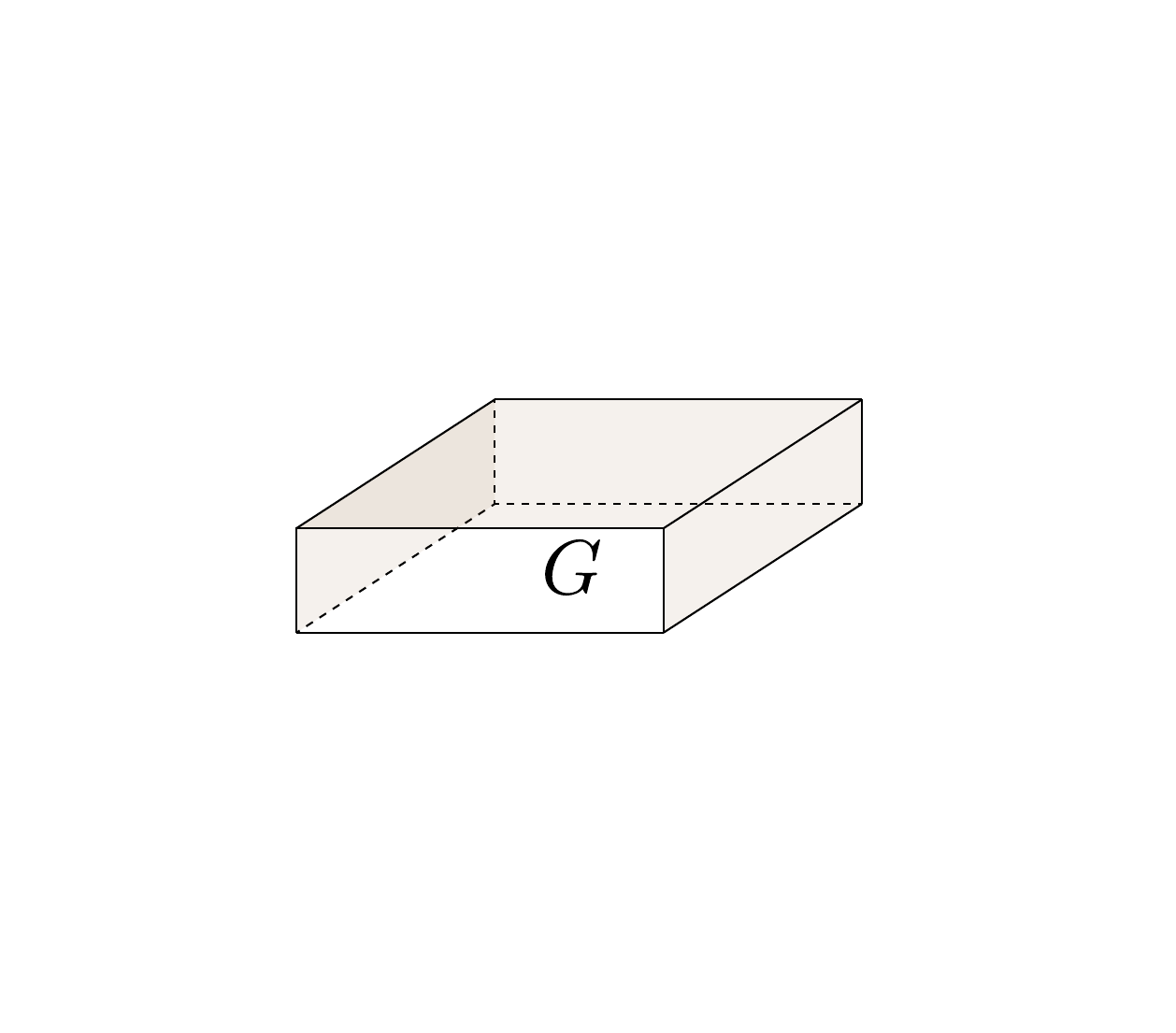}

\end{tblr}

\caption{Summary of elementary logical operations and spacetime logical blocks.}
\label{tab:logical_blocks_intro_summary}
\end{table*}

We organize our presentation of GSCs around a set of elementary logical operations. These include transversal logical gates, a method for switching between GSCs based on different groups, and a protocol for preparing and reading out logical information. The elementary logical operations can then be composed to perform a variety of non-Clifford operations on $\Z_2$ surface codes---including controlled gates through sliding GSCs, and the preparation of magic states. These constructions generalize the magic-state preparation in Refs.~\cite{huang2025D4, davydova2025D4} and the protocols for sliding in Refs.~\cite{davydova2025D4, sajith2025noncliffordS3}. We give representative examples in Table~\ref{tab:Summary_logical_gates}. 

We further develop a tensor-network formalism---similar to ZX-calculus---to provide circuit implementations of the elementary logical operations. This also enables spacetime representations of the logical operations, which we refer to as spacetime logical blocks~\cite{Bombin2023logicalblocks}, depicted in Table~\ref{tab:logical_blocks_intro_summary}. Although we do not rigorously prove the fault tolerance of the spacetime logical blocks, we anticipate that the tensor network formalism is a valuable step in this direction. 

Moreover, the tensor-network formalism makes for an explicit connection to topological quantum field theories (TQFTs), as first described in Refs.~\cite{Bauer2024topologicalerror, Bauer2025lowoverheadnon, Bauer2025xplusy, davydova2025D4, bauer2025planarfaulttolerantcircuitsnonclifford}. We show that the syndrome-extraction circuit for a $G$-GSC is precisely the spacetime partition function of a $G$ topological gauge theory. This allows us to directly port the rich understanding of TQFTs into the design of logical operations for GSCs.

Finally, we remark that while this manuscript was in preparation, we became aware of several related works with complementary results. Most notably, Ref.~\cite{huang2026hybridlatticesurgerynonclifford} describes GSCs in the context of hybrid lattice surgery (see also a partial treatment of lattice surgery with GSCs in Ref.~\cite{cowtan2025homologyhopfalgebrasquantum}). In this text, we focus on operations on the GSCs that preserve the logical space. 

In Refs.~\cite{kobayashi2025cliffordhierarchystabilizercodes, hsin2025automorphismgaugetheorieshigher, warman2026transversalcliffordhierarchygatesnonabelian}, transversal and constant-depth logical gates were introduced for codes that are closely related to GSCs. Similar to Refs.~\cite{kobayashi2025cliffordhierarchystabilizercodes, hsin2025automorphismgaugetheorieshigher}, we consider logical operations arising from group automorphisms. In contrast, however, GSCs correspond to untwisted quantum double models, as opposed to the twisted versions that appear in their work, and which rely on group cohomology. Ref.~\cite{warman2026transversalcliffordhierarchygatesnonabelian} considers untwisted quantum double models but implements logical gates associated to group cohomology. Similar to these works, we identify transversal logical gates in arbitrary levels of the Clifford hierarchy.

The rest of the manuscript is organized as follows. In Section~\ref{Sec:GroupSurfaceCodes}, we introduce GSCs and prove that, for a finite group $G$, the code space has dimension $|G|$. In Section~\ref{sec: elementary logical operations}, we define a set of elementary logical operations on GSCs, including transversal logical gates, extension and splitting of GSCs, and preparation and readout of logical states. In Section~\ref{Sec:Lattice to continuum}, we transition from a spatial description of GSCs to a spacetime description through Euclidean time evolution. In Section~\ref{sec: spacetime logical blocks}, we then define spacetime logical blocks, which are the spacetime analog of the elementary logical operations. Finally, in Section~\ref{sec: examples}, we give explicit examples of how the spacetime logical blocks can be stitched together to perform non-Clifford operations in Abelian surface codes. 

The appendices contain technical details and numerous results that may be of independent interest. 
In Appendix~\ref{app: representation on qubits}, we give an explicit representation of the $G=D_4$ GSC on qubits. We show that the GSC can be mapped explicitly to the $\Z_2^3$ twisted quantum double model of Ref.~\cite{kobayashi2025cliffordhierarchystabilizercodes}. In Appendix~\ref{app: charge and flux}, we specify the detectors of the GSC. Physically, these correspond to measurements of fluxes and charges. We further demonstrate how the fluxes and charges can, in some cases, be moved deterministically. In Appendix~\ref{app: anyon measurements}, we argue that the flux and charge detectors are sufficient for resolving the local syndromes. In Appendix~\ref{d2n appendix}, we provide a minimal family of examples that realize transversal non-Clifford gates at arbitrary levels of the Clifford hierarchy. Lastly, in Appendix~\ref{app: ResInd}, we clarify the charges that can be condensed at the interfaces between GSCs that appear in extension and splitting.

\section{Group surface codes}
\label{Sec:GroupSurfaceCodes}

Let us get started by introducing GSCs. We begin by describing the underlying physical Hilbert space and specifying the code space. We prove that, for a finite group $G$, the GSC encodes a Hilbert space of dimension $|G|$. In the process, we introduce certain gauge-fixed states, which are convenient for describing the elementary logical operations in Section~\ref{sec: elementary logical operations}.

Before getting into the details of GSCs, we would like to emphasize that GSCs are nothing more than quantum double models~\cite{Kitaev2003computationbyanyons} with specific boundary conditions. Throughout the text, however, we do not assume any familiarity with quantum double models.\footnote{For further background on quantum double models, we recommend the original work of Ref.~\cite{Kitaev2003computationbyanyons} and the pedagogical treatments in Refs.~\cite{cui2018topological,lo2025universalquantumcomputations3}. For background on the boundaries of quantum double models, we refer to Refs.~\cite{beigi2011boundaries, cong2016boundaries, Cong2017boundarydefects}.} While the mathematical structure of the quantum double is useful for describing the excitations of the underlying Hamiltonian, the key properties of GSCs can be understood entirely in terms of group multiplication. For this reason, we refer to them as GSCs.

\subsection{Physical Hilbert space}

\begin{figure}
    \centering
    \includegraphics[width=.7\linewidth]{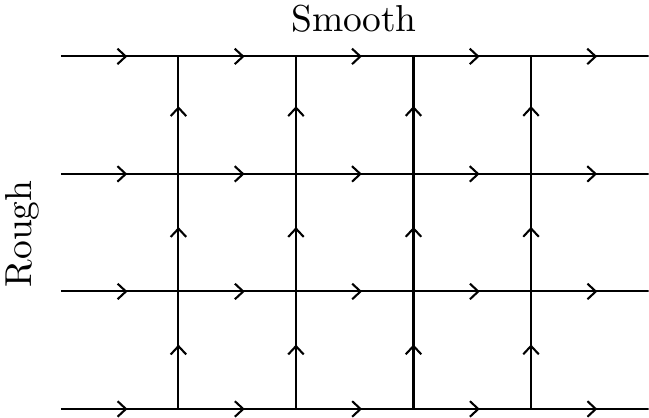}
    \caption{Lattice of the GSC. The GSC is defined on a square lattice with rough boundaries on the left and right and smooth boundaries on the top and bottom. The edges are oriented either to the right or upwards.}
    \label{fig:Lattice Reference}
\end{figure}

GSCs are defined on a square lattice with rough and smooth boundaries, as shown in Fig.~\ref{fig:Lattice Reference}. We take as a convention that the rough boundaries are on the left and right, while the smooth boundaries are on the top and bottom. We also assign an orientation to each edge so that the horizontal edges are oriented to the right, and the vertical edges are oriented upward.

For a GSC based on the finite group $G$, we place a $|G|$-dimensional Hilbert space on every edge $e$. The basis states of the edge Hilbert space can be labeled by the elements of $G$, e.g., $|g\rangle$ for $g \in G$. The full Hilbert space is given by the tensor product of the edge Hilbert spaces and admits a basis labeled by configurations of group elements. For example, letting $\bm{g_e}$ denote a choice of $g_e \in G$ for each edge $e$, a basis state for the full Hilbert space may be written as $|\bm{g_e}\rangle$. We refer to this basis as the group basis.\footnote{More formally, each edge hosts the Hilbert space $\mathbb C[G]$, i.e., the group algebra of the finite group $G$. The full Hilbert space is $\mathbb C[G]^{\otimes N_E}$, where $N_E$ is the number of edges.}

Lastly, we define two operators $L_e^g$ and $R_e^g$, which act at an edge $e$ as left or right group multiplication, respectively. More precisely, for $g,h \in G$ and a state $|h\rangle$ at $e$, we have:
\begin{eqs}
\label{eq:left right mult ops}
    L_e^g |h \rangle = |gh\rangle, \quad R_e^g |h\rangle = |h\bar{g}\rangle,
\end{eqs}
where $\bar{g}$ denotes the inverse of $g$. Here, $R^g$ acts with $\bar{g}$ on the right so that it forms a homomorphism of the group, i.e., $R^gR^h = R^{gh}$. 

\subsection{Code space} \label{sec: code space}

To specify the code space of the GSC, we define stabilizers associated to the vertices and plaquettes. 
The vertex stabilizers are a product of left and right action on the edges surrounding a vertex. For any $g \in G$, the vertex stabilizer $A_v^g$ at a bulk vertex $v$ is represented graphically as:
\begin{align}
A_v^g =     \vcenter{\hbox{\includegraphics[scale=.3]{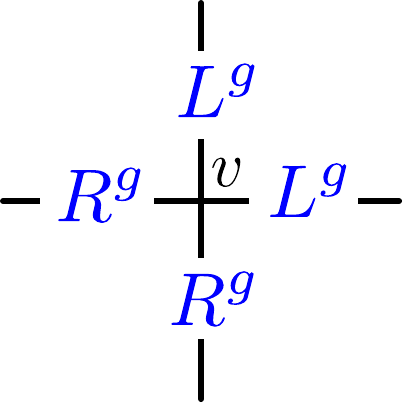}}}.
\end{align}
For the vertices on the top and bottom of the surface code, the vertex stabilizers are truncated to:
\begin{align}
A_v^g = \,  \vcenter{\vspace{.5cm}\hbox{\includegraphics[scale=.3]{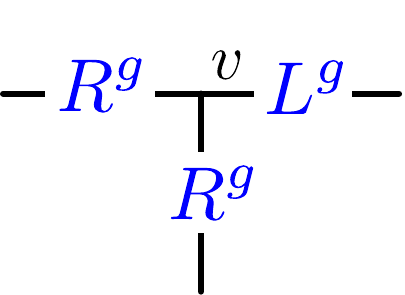}}}\,\,,  \,\,\, \vcenter{\vspace{-.5cm}\hbox{\includegraphics[scale=.3]{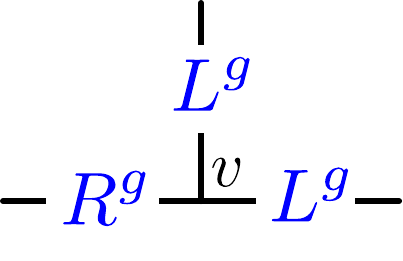}}}.
\end{align}
Here, we have taken the convention that left action is on the edges pointing away from the vertex, while right action is on those pointing toward the vertex. 

Using the language of lattice gauge theory, the $A_v^g$ operators can be interpreted as implementing gauge transformations. The states that are invariant under these operators, which includes the code states, are known as gauge-invariant states.

Next, we define the plaquette stabilizers, which are diagonal in the group basis. For a plaquette $p$ in the bulk, the plaquette stabilizer $B_p$ is given by: 
\begin{align}
    B_p \Bigg|\vcenter{\hbox{\includegraphics[scale=0.2]{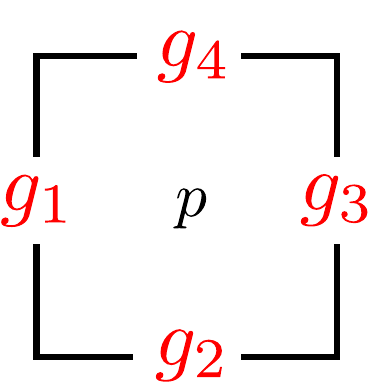}}} \Bigg\rangle 
    = \delta_{\bar{g}_1 g_2 g_3 \bar{g}_4, 1} \Bigg|\vcenter{\hbox{\includegraphics[scale=0.2]{Figures/plaquettegroup.pdf}}} \Bigg\rangle.
\end{align}

While for the plaquettes on the left and the right of the surface code, the stabilizers reduce to:

\begin{align}
    B_p \Bigg|\,\vcenter{\hbox{\includegraphics[scale=0.2]{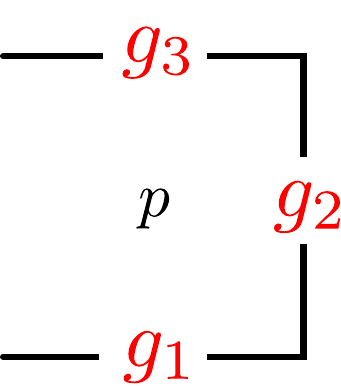}}} \Bigg\rangle 
    &= \delta_{g_1 {g}_2 \bar{g}_3, 1} \Bigg|\, \vcenter{\hbox{\includegraphics[scale=0.2]{Figures/plaquettegroupleft.pdf}}} \Bigg\rangle, \\
    B_p \Bigg| \vcenter{\hbox{\includegraphics[scale=0.2]{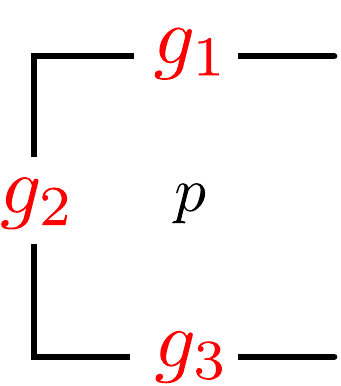}}} \,\Bigg\rangle 
    &= \delta_{\bar{g}_1 \bar{g}_2 {g}_3, 1} \Bigg|\vcenter{\hbox{\includegraphics[scale=0.2]{Figures/plaquettegroupright.pdf}}} \, \Bigg\rangle.
\end{align}
Here, we only show the portion of the group basis state around the plaquette $p$. In words, the $B_p$ stabilize the state if the product of group elements along a counterclockwise path around the plaquette is the identity. Otherwise, they annihilate the state. Note that we multiply by the inverse of the group element if the orientation of the edge is opposite of the counterclockwise path.\footnote{It is also worth noting that the choice of base point, i.e., the initial point of the counterclockwise path does not matter.} This is a path-ordered product of group elements around the plaquette.

In terms of lattice gauge theory, $B_p$ projects onto states with zero flux, in the sense that the product of group elements around a plaquette is $1\in G$. We refer to these states as the flux-free states. The code space then consists of flux-free states that are gauge invariant. We make this more precise below.

Before doing so, we comment on the relations of the stabilizers. Since $L^g$ and $R^g$ are group homomorphisms, $A^g_v$ satisfies:
\begin{align} \label{eq: vertex homomorphism}
    A^g_vA^h_v = A_v^{gh}. 
\end{align}
Note that this implies that, when $G$ is non-Abelian, the generators of gauge transformations $A^g_v$ do not necessarily commute with each other. 
Nonetheless, as shown below, there are common $+1$ eigenstates of the vertex stabilizers. For distinct vertices $v$ and $w$, the vertex stabilizers commute, i.e.,
\begin{align}
    A^g_vA^h_w = A^h_wA^g_v.
\end{align}
This follows from the fact that $L^g$ and $R^h$ commute for any $g,h \in G$.

As for the plaquette stabilizers, they are mutually commuting for any pair of plaquettes $p$ and $q$:
\begin{align}
    B_p B_q = B_q B_p.
\end{align}
This is because the plaquette stabilizers are simultaneously diagonalized in the group basis.

Finally, it can be checked directly from the definitions of $A^g_v$ and $B_p$ that they commute with each other for any $v$ and $p$:
\begin{align}
    A_v^gB_p = B_p A_v^g.
\end{align}
In other words, gauge transformations, implemented by $A^g_v$, do not create any fluxes starting from a flux-free state. 

With this, the code space is defined as the set of states that are stabilized by the vertex and plaquette stabilizers. Explicitly, the code space $\mathcal{H}_C$ is:
\begin{align}
    \mathcal{H}_C=\left \{ |\psi\rangle \,:\, A_v^g|\psi \rangle =B_p|\psi \rangle = |\psi\rangle, \, \forall v,p,g \right \}.
\end{align} 
For later purposes, it is helpful to also introduce a projector $A^G_v$ onto the gauge invariant subspace at the vertex $v$. The projector $A^G_v$ is defined as:
\begin{align}\label{eq: Av}
    A^G_v = \frac{1}{|G|}\sum_{g \in G}A^g_v.
\end{align}

\subsubsection{Dimension of the code space}

Next, we argue that the dimension of $\mathcal{H}_C$ is $|G|$. To see this, we start by defining $\mathcal{F}$ as the set of group basis states that are flux free:
\begin{align}
    \mathcal{F} = \left \{|\bm{g_e}\rangle \,:\, B_p |\bm{g_e}\rangle = |\bm{g_e}\rangle, \, \forall p \right \}.
\end{align}
The states in $\mathcal{F}$ satisfy the plaquette stabilizers, but they do not satisfy the vertex stabilizers. That is, they are not gauge invariant, since the gauge transformations $A_v^g$ permute the group basis states. 

To form code states, we must find gauge invariant superpositions of the states in $\mathcal{F}$. To this end, notice that the gauge transformations partition $\mathcal{F}$ into gauge equivalence classes. In particular, two states $|\bm{g_e}\rangle$ and $|\bm{h_e}\rangle$ in $\mathcal{F}$ are gauge equivalent if there exists a set of gauge transformations that map one to the other. That is, there exists a vertex-dependent product of gauge transformations, such that:
\begin{align}
    \prod_v A_v^{k_v} |\bm{g_e}\rangle = |\bm{h_e}\rangle.
\end{align}
In a slight abuse of the notation, we denote the gauge equivalence class containing the state $\ket{\bm{g_e}}$ as $[\bm{g_e}]$. 

Code states can now be formed by taking an equal-amplitude superposition over all group basis states belonging to a gauge equivalence class:
\begin{align} \label{eq: code state}
    |[\bm{g_e}]\rangle\propto \sum_{\ket{\bm{h_e}} \in [\bm{g_e}]} |\bm{h_e}\rangle.
\end{align}
The states $|[\bm{g_e}]\rangle$, for varying choices of gauge equivalence classes, span the entire code space. This is because, if a code state has a nontrivial overlap with a flux-free state then, to ensure gauge invariance, every state in the gauge equivalence class must appear with an equal amplitude. Furthermore, the states $|[\bm{g_e}]\rangle$ form a basis for the code space, which follows from the fact that $|[\bm{g_e}]\rangle$ and $|[\bm{h_e}]\rangle$ are orthogonal for inequivalent classes $[\bm{g_e}]$ and $[\bm{h_e}]$. This then implies that the dimension of the code space is given by the number of gauge equivalence classes. 

To finish the argument, we show that the number of gauge equivalence classes is equal to $|G|$. We accomplish this by identifying gauge-fixed states that are in one-to-one correspondence with the elements of $G$. We caution that, in this context, we are using gauge fixing in the sense of lattice gauge theory---i.e., a choice of representative state for each gauge equivalence class. This is distinct from gauge fixing in the context of subsystem codes.\footnote{Explicitly, a subsystem code could be defined by the gauge group generated by $A_v^g$ and $B_p$. The operator $A^G_v$ belongs to the center of the gauge group, so any gauge-fixed state (in the sense of subsystem codes) should be a $+1$ eigenstate. That is not the case for the gauge-fixed states in the lattice gauge theory sense. }

\begin{figure*}[t]
        \centering
        \includegraphics[width=0.7\textwidth]{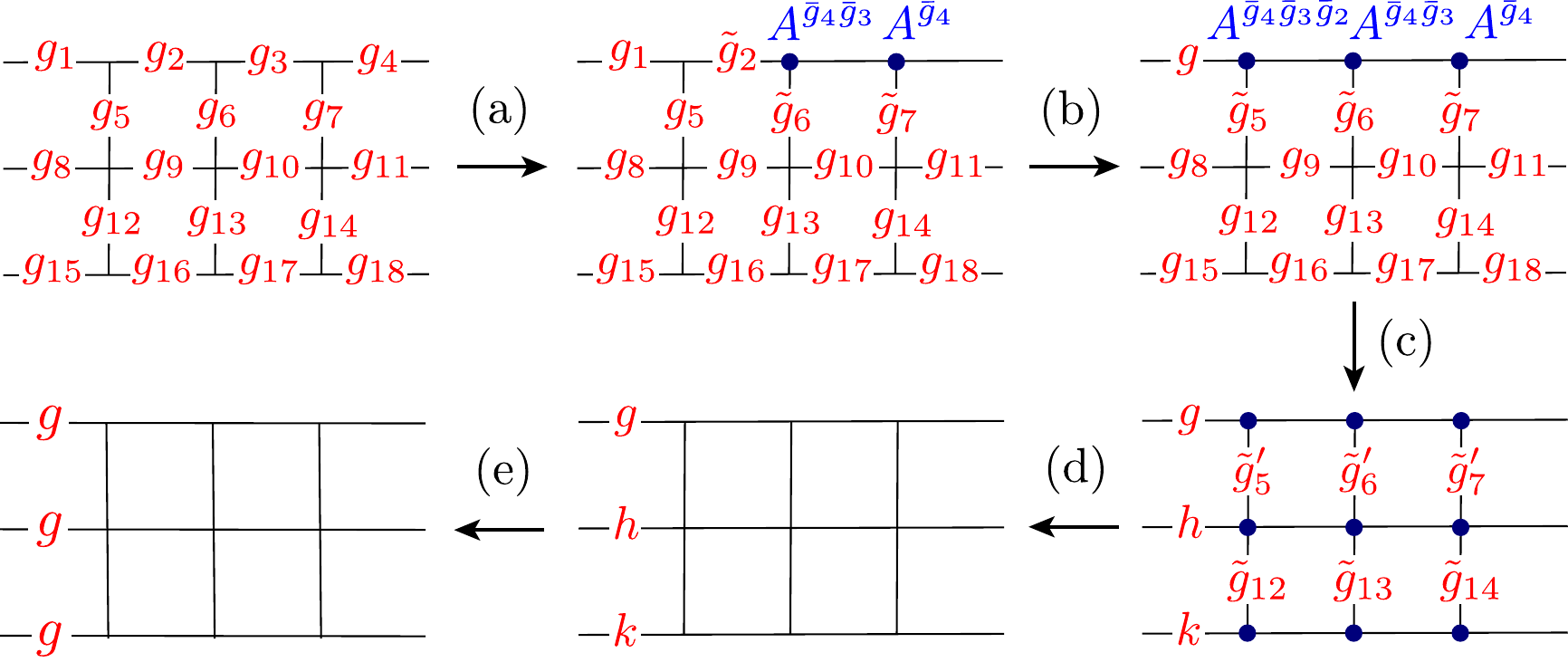}
    \caption{Mapping a flux-free state to the left gauge. (a) Starting with an arbitrary flux-free state, we apply gauge transformations in sequence (from right to left) to map the group labels to the identity. An edge without a label is assumed to be in the identity state $|1\rangle$. (b) Moving across the top row, we are left with some group element $g$ on the left most edge. We also leave behind group elements labeled with a tilde. (c) For each row, we can use gauge transformations to move the group elements on the horizontal edges to the leftmost edges. This gives us some group elements $g,h,k$ on the left. (d) The flux-free condition, and the fact that gauge transformations commute with the plaquette stabilizers, guarantees that the group elements around the plaquette multiply to the identity. This allows us to deduce that the group elements with a tilde are the identity. (e) Again, using the flux-free condition, we see that $g$, $h$, and $k$ must be equal.}
    \label{fig: leftgaugetransformation}
\end{figure*}

Consider an arbitrary group basis state $|\bm{g_e}\rangle$ in $\mathcal{F}$, such as the one pictured in Fig.~\ref{fig: leftgaugetransformation}. By applying gauge transformations in sequence, we can attempt to map the states $|g_e\rangle$ on the horizontal edges to the state $|1\rangle$, where $1$ is the identity element in $G$. Moving from right to left, we can map all of the horizontal edges to $\ket{1}$, except for those on the leftmost edges. This pushes the group elements to the leftmost edges, as depicted in Fig.~\ref{fig: leftgaugetransformation}. 

The flux-free condition then guarantees that \textbf{(i)} the vertical edges are in the $|1\rangle$ state and \textbf{(ii)} the group elements on the leftmost edges are all the same. Thus, by applying gauge transformations, the configuration of group elements $\bm{g_e}$ can be mapped to a configuration with some $g\in G$ on the leftmost edges and $1$ everywhere else. We refer to this choice of representative for each gauge equivalence class as the left gauge.

The gauge equivalence classes can be uniquely labeled by the group element $g$ on the leftmost edges in the left gauge. This is to say that the gauge transformations are unable to change the configuration labeled by $g$ into one labeled by some other element $h$. This can be seen by noting that the product of group elements across the top of the patch is gauge invariant. In lattice gauge theory terms, this product of group elements computes the holonomy across the patch. Therefore, since the holonomy is different for the configuration labeled by $g$ and the one labeled by $h$, they must be gauge inequivalent configurations. 

Consequently, we can label each gauge equivalence class by an element $g \in G$. Furthermore, for every $g \in G$, there is an associated equivalence class, represented by the configuration with $g$ on the leftmost edges and $1$ elsewhere. This proves that the number of gauge equivalence classes is $|G|$, and hence, the dimension of the code space is $|G|$. We label the basis states for the code space by $\{\ket{g}_{\mathsf{L}}\}$, for $g \in G$. Explicitly, the state $\ket{g}_\mathsf{L}$ is the code state in Eq.~\eqref{eq: code state} for which the product of group elements across the top boundary multiply to $g$. Intuitively, this is the logical state that has $g$ flux in the large (fictitious) plaquette formed by the edges on the top boundary.

\section{Elementary logical operations} \label{sec: elementary logical operations}

We now introduce a set of elementary logical operations that can be performed with GSCs. These are summarized in Table~\ref{tab:logical_operations_summary} and consist of:
\textbf{(i)} transversal logical gates, \textbf{(ii)} extension, \textbf{(iii)} splitting, and \textbf{(iv)} preparation and readout. 
Throughout, we illustrate the action of the logical operations by making use of the left gauge, described in the previous section. We give examples of how these elementary operations can be composed to perform non-Clifford operations on $\Z_2$ surface codes.

\subsection{Transversal logical gates}\label{sec:GSC-TransversalGates}
The first set of elementary logical operations that we discuss are transversal logical gates. That is, they are logical gates that can be implemented as a tensor product of single-site unitaries.  We describe two types of transversal logical gates for GSCs. The first act as left and right group multiplication, while the second perform group automorphisms on the logical basis states.

\subsubsection{Left and right group multiplication}

We define the logical operators $L^g_{\mathsf{L}}$ and $R^g_{\mathsf{L}}$ as:
\begin{align}
    L^g_{\mathsf{L}} = \prod_{e \in E_\mathrm{left}} L^g_e, \qquad R^g_{\mathsf{L}} = \prod_{e \in E_\mathrm{right}} R_e^g.
\end{align}
Here, $E_\mathrm{left}$ and $E_\mathrm{right}$ are the sets of the leftmost and rightmost horizontal edges, respectively. Therefore, the operators are fully supported on one of the boundaries and are composed of either left or right multiplication operators. 

These operators commute with the vertex stabilizers, thanks to the commutation of $L^g_e$ and $R^g_e$. It can also be checked that they commute with all of the plaquette stabilizers. This implies that they preserve the code space.

As the notation suggests, $L^g_{\mathsf{L}}$ and $R^g_{\mathsf{L}}$ act on the logical basis state $\ket{h}_{\mathsf{L}}$ as:
\begin{align} \label{eq: left and right logicals}
   \boxed{{L^g_\mathsf{L}} \ket{h}_\mathsf{L} = {\ket{gh}_\mathsf{L}}, \quad R^g_{\mathsf{L}} \ket{h}_{\mathsf{L}} = \ket{h\bar{g}}_\mathsf{L}.}
\end{align}
This can be seen explicitly in the left gauge. Acting on a representative gauge-fixed state, we find:
\begin{align}
    L^g_{\mathsf{L}}: \vcenter{\hbox{\includegraphics[scale=.3]{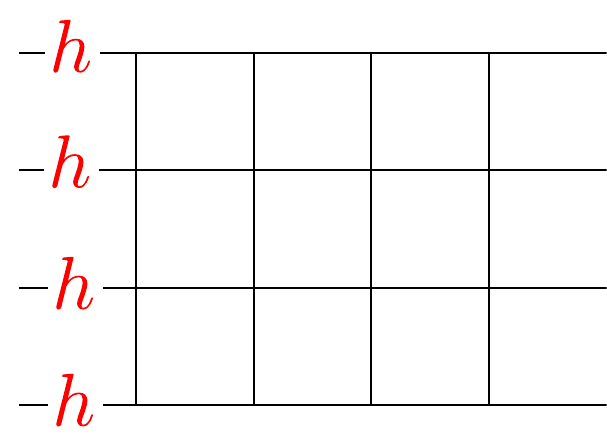}}} \mapsto \vcenter{\hbox{\includegraphics[scale=.3]{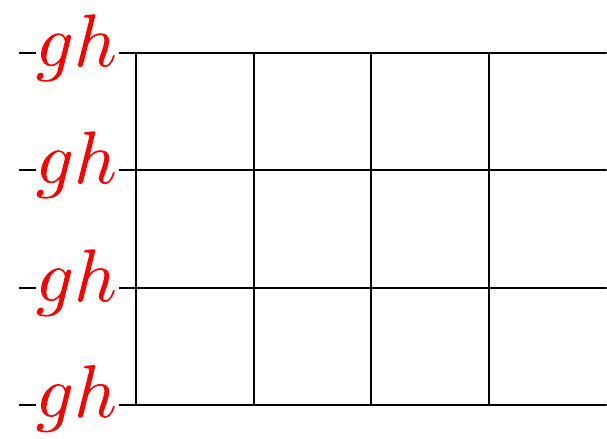}}}.
\end{align}
For right action, we have:
\begin{align}
    R^g_{\mathsf{L}}: \vcenter{\hbox{\includegraphics[scale=.3]{Figures/hleftgauge.pdf}}} \mapsto \vcenter{\hbox{\includegraphics[scale=.3]{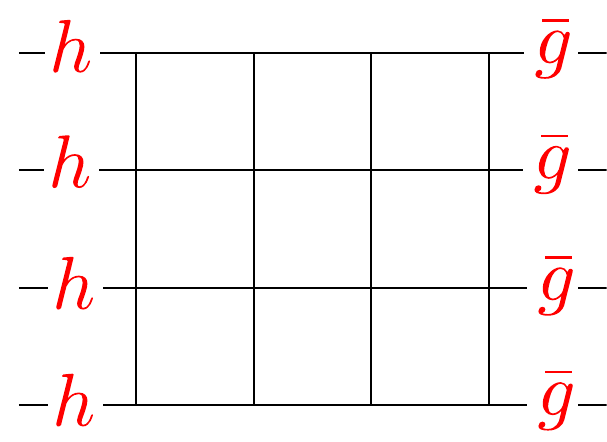}}},
\end{align}
for which the right-hand side is gauge equivalent to:
\begin{align}
    \vcenter{\hbox{\includegraphics[scale=.3]{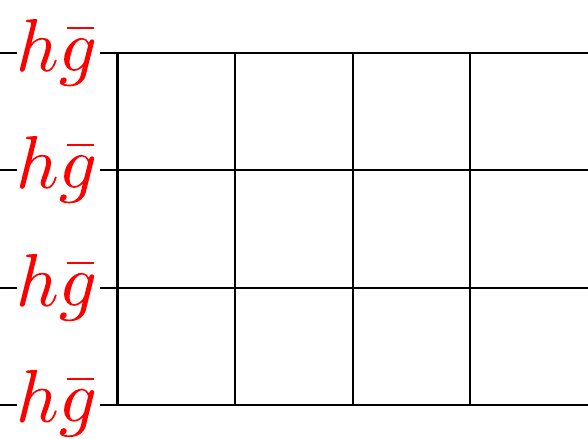}}}.
\end{align}
Thus, in the left gauge, we reproduce the action in Eq.~\eqref{eq: left and right logicals}. Note that the action on the gauge-fixed states is well defined, since the logical operators commute with all of the stabilizers.

At an intuitive level, $L^g_\mathsf{L}$ and $R^g_\mathsf{L}$ tunnel flux across the patch to change the holonomy. It is worth noting that the operators that tunnel flux across the patch only take on this simple representation near the boundary. If the same operators were translated into the bulk, they would fail to commute with the bulk vertex stabilizers.\footnote{This is because fluxes labeled by group elements (as opposed to conjugacy classes) are not gauge invariant in the bulk. The gauge transformations change the value of the flux up to conjugation. Near the rough boundaries, on the other hand, the fluxes are well defined, since there are no corresponding vertex stabilizers that conjugate the value of the flux.}  

\subsubsection{Group automorphisms}
The second set of logical operators that we consider enacts group automorphisms of $G$ on the logical basis states. Let $\phi$ be an automorphism of $G$. We then define $U_e^\phi$ as the unitary operator at the edge $e$ that acts on the state $|g_e\rangle$ as:
\begin{align}
    U_e^\phi\ket{g_e}=\ket{\phi(g_e)}.
\end{align}
We further define $U^\phi_\mathsf{L}$ as the transversal operator:
\begin{align}    U^\phi_\mathsf{L}=\prod_e U^\phi_e.
\end{align}
The operator $U^\phi_\mathsf{L}$ does not commute with individual vertex stabilizers, in general. However, it preserves the set of vertex stabilizers, since it maps vertex stabilizers to vertex stabilizers. 
In particular, conjugating $A_v^g$ by $U^\phi_\mathsf{L}$ gives:
\begin{align}
    U^\phi_\mathsf{L} A_v^g (U^\phi_\mathsf{L})^\dagger = A_v^{\phi(g)}.
\end{align}
The plaquette stabilizers commute with $U^\phi_\mathsf{L}$, given that $\phi$ acts invariantly on flux-free states, as guaranteed by the fact that $\phi$ is a homomorphism.
The action of $U^\phi_\mathsf{L}$ on a logical state $\ket{h}_\mathsf{L}$ is:
\begin{align}
    \boxed{U^\phi_\mathsf{L} \ket{h}_\mathsf{L} = \ket{\phi(h)}_\mathsf{L}.}
\end{align}
This can be determined from the action of $U^\phi_\mathsf{L}$ on a state in the left gauge. Explicitly, $U^\phi_\mathsf{L}$ acts as:
\begin{align} \label{eq: automorphism logical}
    U^\phi_\mathsf{L}: \vcenter{\hbox{\includegraphics[scale=.3]{Figures/hleftgauge.pdf}}} \mapsto \vcenter{\hbox{\includegraphics[scale=.3]{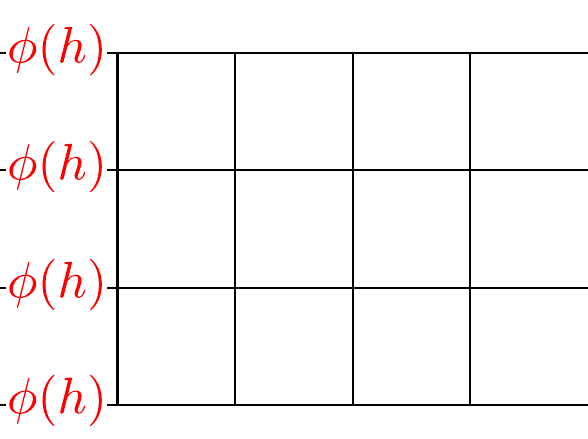}}}.
\end{align}
This means that $U^\phi_\mathsf{L}$ implements the automorphism $\phi$ on the logical basis states as in Eq.~\eqref{eq: automorphism logical}. 

We note that group automorphisms can be divided into inner automorphisms and outer automorphisms. Inner automorphisms are those that can be enacted by conjugation with an element in $G$. For an inner automorphism $\phi_g$ enacted by conjugation with $g$, we can write $U^{\phi_g}_\mathsf{L}$ explicitly as:
\begin{align}
    U^{\phi_g}_\mathsf{L} = \prod_e L^g_eR^g_e,
\end{align}
which is equivalent to:
\begin{align} \label{eq: inner automorphism}
    U^{\phi_g}_\mathsf{L}= L_\mathsf{L}^gR_\mathsf{L}^g \prod_v A^g_v.
\end{align}
Therefore, within the code space, the inner automorphism $\phi_g$ can be implemented by $L^g_\mathsf{L}R^g_\mathsf{L}$.

\begin{tcbexample}[$D_4$] \label{ex: transversal gates D4}

\vspace{.2cm}

Here, we consider the transversal gates of the GSC for the group $G=D_4$. We recall that $D_4$ admits the following presentation:
\begin{align}
    D_4 = \langle a,b,c \,|\, a^2=b^2=c^2=1, cac=ab \rangle.
\end{align}
This implies that $D_4$ can be written as $(\Z_2^a \times \Z_2^b) \rtimes \Z_2^c$. Here the factor of $\Z_2^a \times \Z_2^b$ is generated by $a$ and $b$, while the additional factor of $\Z_2^c$ is generated by $c$. The semidirect product $\rtimes$ captures the fact that $c$ implements a group automorphism on $\Z_2^a\times \Z_2^b$. This is reflected in the relation $cac=ab$.

In Appendix~\ref{app: representation on qubits}, we give explicit stabilizers for the $D_4$ GSC defined on qubits. Here, for simplicity, we focus on the code space. The code space has dimension $|D_4|=2^3$, and the logical basis states can be labeled by the elements of $D_4$. A generic element of $D_4$ can be written as $a^{\alpha}b^{\beta}c^\gamma$, with $\alpha$, $\beta$, and $\gamma$ valued in $\{0,1\}$. Therefore, the logical basis states take the form $\ket{a^{\alpha}b^{\beta}c^\gamma}_\mathsf{L}$. 

It is convenient to keep only the exponents, so that the logical basis states are labeled as $\ket{{\alpha},{\beta},\gamma}_\mathsf{L}$. This makes it explicit that the $D_4$ GSC encodes three qubits.

Let us next consider the action of the logical operators $L^g_\mathsf{L}$ and $R^g_\mathsf{L}$ on the logical state $\ket{{\alpha},{\beta},\gamma}_\mathsf{L}$. We note that $L^a_\mathsf{L}$ and $L^b_\mathsf{L}$ act as Pauli $X$ operators on the first two qubits:
\begin{eqs}
    L^a_\mathsf{L}\ket{{\alpha},{\beta},\gamma}_\mathsf{L}&=\ket{{\alpha}+1,{\beta},\gamma}_\mathsf{L}, \\
    L^b_\mathsf{L}\ket{{\alpha},{\beta},\gamma}_\mathsf{L}&=\ket{{\alpha},{\beta}+1,\gamma}_\mathsf{L},
\end{eqs}
where addition is taken modulo 2.
The action by $L^c_\mathsf{L}$ is more interesting, due to the nontrivial relation $cac = ab$:
\begin{align}
    L^c_\mathsf{L}\ket{{\alpha},{\beta},\gamma}_\mathsf{L}=\ket{{\alpha},{\beta}+\alpha,\gamma+1}_\mathsf{L}.
\end{align}
This implies that $L^c_\mathsf{L}$ performs a $\mathrm{CX}$ gate on the first two qubits and a Pauli $X$ on the third qubit. 

Notice that, to enact a $\mathrm{CX}$ on the first two qubits, without affecting the third qubit, we can combine left action and right action:
\begin{align} \label{eq: inner c}
    L^c_\mathsf{L}R^c_\mathsf{L}\ket{{\alpha},{\beta},\gamma}_\mathsf{L}=\ket{{\alpha},{\beta}+\alpha,\gamma}_\mathsf{L}.
\end{align}
We see that the GSC based on $D_4$ admits a transversal $\mathrm{CX}$ gate. 

The action of $L^c_\mathsf{L}R^c_\mathsf{L}$ can alternatively be viewed as an inner automorphism of $D_4$. To make this explicit, we define the automorphism $\phi_c$ as $\phi_c(g) = c g c$. Then, in agreement with Eq.~\eqref{eq: inner automorphism}, we have: 
\begin{equation}
    L^c_{\mathsf{L}} R^c_{\mathsf{L}} = U_{\mathsf{L}}^{\phi_c}\prod_v A_v^c.
\end{equation}
Therefore, the action of $L^c_{\mathsf{L}} R^c_{\mathsf{L}}$ within the code space is equivalent to $U_{\mathsf{L}}^{\phi_c}$.
We note this automorphism is also an outer automorphism of $\Z_2\times \Z_2$. As an outer automorphism, it implements the transversal $\mathrm{CX}$ of the $\Z_2\times \Z_2$ surface code~\cite{Steane1996CSS, Sahay2025tCNOT, cain2025correlateddecodinglogicalalgorithms}.

The full automorphism group of $D_4$ is isomorphic to $D_4$ itself. There is a $\Z_2 \times \Z_2$ subgroup generated by inner automorphisms -- specifically, conjugation by $a$ or $c$. One can check that the inner automorphisms of $D_4$ result in transversal Clifford gates. 

The full automorphism group of $D_4$ is obtained by extending the $\Z_2 \times \Z_2$ inner automorphisms by a single order 2 outer automorphism. 
We may pick the outer automorphism, which we denote as  $\phi$, to permute the group elements as follows:
\begin{equation}
    \phi(a)=c, \quad \phi(b)=b, \quad \phi(c)=a.
\end{equation}
The operator $U^\phi_\mathsf{L}$ acts on the logical states as:
\begin{align}
  \hspace{-7pt}  U^\phi_\mathsf{L}|a^{\alpha} b^{\beta} c^{\gamma}\rangle = |c^{\alpha} b^{\beta} a^{\gamma}\rangle 
    =  |a^{\gamma} b^{\beta + \alpha \gamma}c^{\alpha}\rangle,
\end{align}
where in the second equality, we used the relation $c a c = ab$. This shows that $U^\phi_\mathsf{L}$ implements the gate $\mathrm{SWAP}_{13}\mathrm{CCX}_{132}$ on the logical qubits. We thus see that the outer automorphism $U_{\mathsf{L}}^{\phi}$ implements a transversal non-Clifford gate (see also the recent results of Ref.~\cite{hsin2025automorphismgaugetheorieshigher}).
This shouldn't be surprising as $\phi$ acts non-trivially on the group of inner automorphisms generated by $\phi_a$ and $\phi_c$, which implement Clifford gates on the codespace. 

In fact, for the GSC for $G = D_{2^n}$, where $D_{2^n}$ is the dihedral group of $2^{n+1}$ elements, we find that the automorphisms of $D_{2^n}$ admit transversal logical gates in the $n$th level of the Clifford hierarchy.
This has been detailed in Appendix~\ref{d2n appendix}.
For example, the group $D_8$ admits a gate in the third level of the Clifford hierarchy. 
This is related to the above result for $D_4$ through the fact that outer automorphisms of $D_4$ are inner automorphisms of $D_8$.
This also demonstrates that GSCs admit transversal gates in arbitrarily high levels of the Clifford hierarchy by appropriately extending the group. 
\end{tcbexample}

\subsection{Extension and splitting} \label{sec: extension and splitting}

The next elementary operations that we introduce are extension and splitting. These operations allow us to switch between GSCs with different underlying groups. This is essential for transferring information between Abelian GSCs and non-Abelian GSCs.
The extension and splitting operations may be summarized as follows:
\begin{itemize}
    \item Extension takes an $H$-GSC and a $K$-GSC and produces a single GSC for a group $G=H\knit K$, where the symbol $\knit$ denotes a knit product, as defined below. 
    \item Splitting takes a GSC for a group $H\knit K$ and splits it into an $H$-GSC and a $K$-GSC.
\end{itemize}

We note that similar operations were introduced in Ref.~\cite{sajith2025noncliffordS3} for a pair of Abelian groups $H$ and $K$ and a group $G$ that is a semidirect product $H\rtimes K$. Here, we allow $H$ and $K$ to be non-Abelian and consider cases in which $G$ is not a semidirect product.

Before describing the operations, we introduce the knit product.\footnote{This is more commonly referred to as the Zappa-Sz{\'e}p product, but we opt for the more descriptive name knit product.} A group $G$ is a knit product of $H$ and $K$, denoted as $G=H \knit K$, if it has the following three properties:
\begin{enumerate}
    \item $H$ and $K$ are subgroups of $G$.
    \item The only common element of $H$ and $K$ in $G$ is the identity, i.e., $H \cap K = 1$,
    \item Every $g \in G$ can be uniquely decomposed as $g=hk$, for $h \in H$ and $k \in K$.
\end{enumerate}
We provide examples of knit products below to help build intuition.

\begin{tcbexample}[Knit products] \label{ex: knit products}

\vspace{.2cm}

The group $D_4$ is a knit product of $\Z_2\times \Z_2$ and $\Z_2$. First, both $\Z_2\times \Z_2$ and $\Z_2$ are subgroups of $D_4$, as demonstrated in Example~\ref{ex: transversal gates D4}. Using the notation from Example~\ref{ex: transversal gates D4}, we can take $\Z_2 \times \Z_2$ to be generated by $a$ and $b$, and $\Z_2$ to be generated by $c$. These subgroups then only intersect with the identity. Finally, 
every element of $D_4$ can be uniquely written as $a^\alpha b^\beta c^\gamma$, for $\alpha, \beta, \gamma \in \{0,1\}$. Thus, $D_4$ satisfies the three conditions of a knit product.

The group $D_4$ is also a knit product of $\Z_4$ and $\Z_2$. Considering $D_4$ as the symmetries of a square, we can take $\Z_4$ to be generated by the order four rotation $r$ and $\Z_2$ to be generated by the reflection $s$. The only common element of these transformations is the identity, and every element of $D_4$ can be written as $r^\alpha s^\beta$, with $\alpha \in \{0,\ldots,3\}$ and $\beta \in \{0,1\}$.

More generally, any group $H \rtimes K$ that is a semidirect product of $H$ and $K$ is also a knit product of $H$ and $K$. In fact, the semidirect product can be defined as a knit product in which one of the subgroups is normal.

An example of a knit product that is not a semidirect product is the alternating group $A_5$. It is the knit product of $\Z_5$ and $A_4$, where $\Z_5$ is generated by the order five permutation $(1\,2\,3\,4\,5)$, and $A_4$ is the subgroup of permutations that fix the object 1. These are subgroups of $A_5$, and their only common element is the identity. Furthermore, since the order of $A_5$ satisfies $|A_5|=|\Z_5|\cdot |A_4|$, every element of $A_5$ can be uniquely decomposed as a product of elements in $\Z_5$ and $A_4$. 

Therefore, $A_5$ satisfies all three conditions and can be written as $A_5 = \Z_5\knit A_4$. It is not a semidirect product, because it is simple.  

It is also useful to consider a non-example. The group $\Z_4$ is not a knit product of two copies of $\Z_2$. The first condition holds, since $\Z_2$ is a subgroup. However, the second condition fails, since there is only one $\Z_2$ subgroup of $\Z_4$. In general, central extensions, such as $\Z_4$ are not knit products. 

\end{tcbexample}

\subsubsection{Extension} \label{sec: extension}

To perform an extension, we start with an $H$-GSC and a $K$-GSC with overlapping rough boundaries:
\begin{eqs}\label{eq:extension overlap}
\vcenter{\hbox{\includegraphics[scale=0.3]{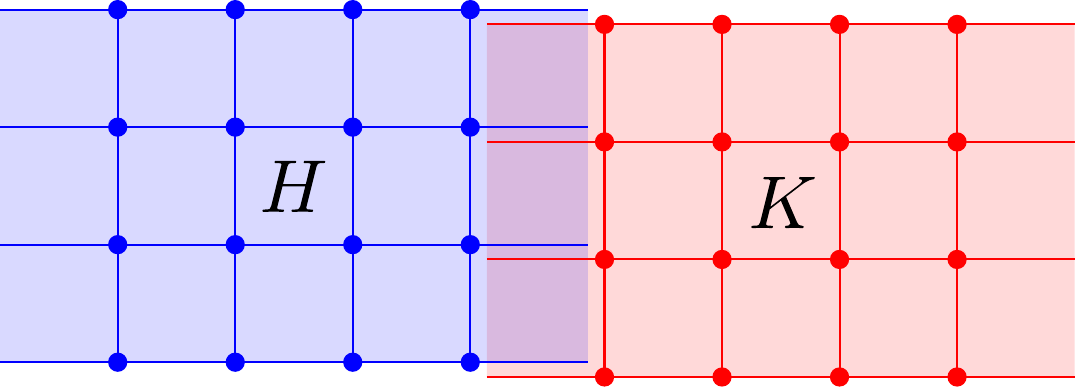}}}.
\end{eqs}
Here, we have used blue and red lines to denote the $H$ and $K$ qudits, the dots mark the satisfied vertex stabilizers, and the shaded plaquettes imply that the plaquette stabilizers are satisfied. Blue corresponds to the $H$-GSC and red corresponds to the $K$-GSC.

The next step is to sew together the GSCs through measurements of the $G=H\knit K$ stabilizers along the overlapping boundary. More specifically, we consider measuring the projectors $A^G_v$, as in Eq.~\eqref{eq: Av}, which project onto the $G$ gauge-invariant states. We describe more refined vertex measurements in Section~\ref{Sec:Lattice to continuum}.

To be able to measure the projectors $A^G_v$, we first need $G$ qudits on each edge of the shared boundary. These can be formed from $H$ and $K$ qudits using the identification in the group basis:
\begin{align} \label{eq: G and HK hilbert space}
    |h\rangle \otimes |k\rangle \leftrightarrow |hk\rangle.
\end{align}
Note that we have explicitly chosen to represent the state as $hk \in G$, where $h \in H$ is on the left and $k \in K$ is on the right. \\

\noindent \textbf{Remark:} If the group $G$ is not a direct product, then the operators $L^g$ and $R^g$ do not factorize on the tensor product of $H$ and $K$ qudits. Due to the convention in Eq.~\eqref{eq: G and HK hilbert space}, however, if $h\in H$ and $k \in K$, then the operators $L^h$ and $R^k$ factorize as:
\begin{align}
    L^h = L^h \otimes I,\quad R^k=I \otimes R^k,
\end{align}
where the operators on the left-hand side enact the multiplication by elements in $G$. For this reason, it is important that the $H$-GSC is on the left side of the $K$-GSC. This ensures that the vertex stabilizer $A_v^h$ is satisfied when $h$ is viewed as an element of $H$ or as an element of $G$; similarly for the $K$ vertex stabilizers.  \\

Therefore, we can add $H$ and $K$ qudits initialized in the $|1\rangle$ state to form $G$ qudits along the seam. We then measure the projectors $A_v^G$ on the vertices of the shared boundary. This glues together the GSCs as:
\begin{align*}
    \vcenter{\hbox{\includegraphics[scale=0.3]{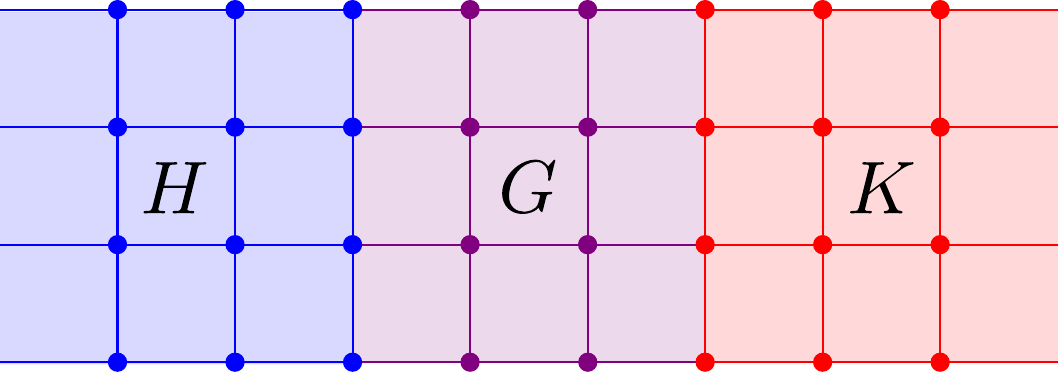}}}.
\end{align*} 
Here, the violet edges are the effective $G$ qudits, the violet dots are the satisfied $G$ vertex stabilizers, and the violet shaded plaquettes are satisfied plaquette stabilizers. The plaquette stabilizers are satisfied by virtue of the fact that we initialized the ancillary qudits in the $|1\rangle$ state.

To simplify the discussion, here we have assumed that the measurements of the projectors $A^G_v$ have $+1$ measurement outcomes, meaning that the state has been successfully projected into the gauge invariant subspace. 
 
In Section~\ref{sec: spacetime logical blocks}, we describe how the procedure needs to be adapted according to different measurement outcomes. 

Next, we shrink the $H$- and $K$-GSCs from the left and right, respectively. 
They are shrunk by measuring a column of qudits on the boundary in the group basis.
Depending on the measurement outcomes, we apply gauge transformations to map the measured qudits into the $|1\rangle$ state.\footnote{Note that this can be done in software. The gauge transformations are purely to ensure that any nontrivial holonomy on the $H$ and $K$ patches is pushed into the $G$-GSC.} In the presence of errors, we may need to make corrections to remove fluxes before applying gauge transformations. This results in:
\begin{align*}
    \vcenter{\hbox{\includegraphics[scale=0.3]{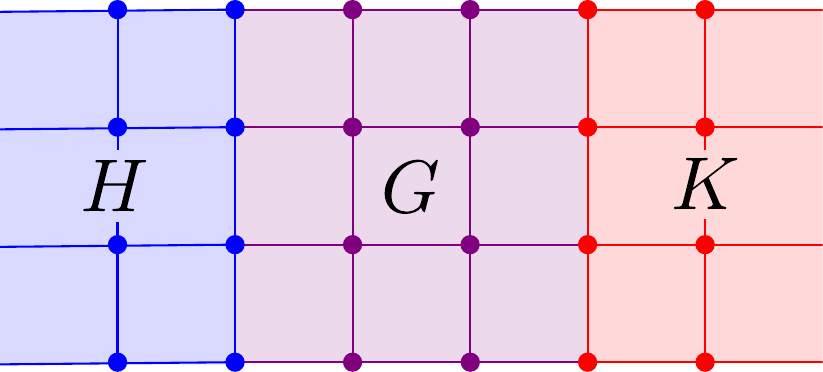}}}.
\end{align*}

By repeating the steps of growing the $G$-GSC by measuring $A_v^G$ and shrinking the $H$ and $K$ boundaries, we eventually arrive at a single patch of GSC for $G=H\knit K$: 
\begin{align*}    
\vcenter{\hbox{\includegraphics[scale=0.3]{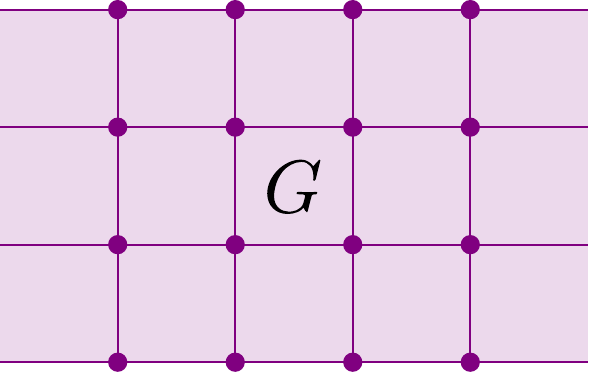}}}.
\end{align*}
This concludes the operation of extension. In Section~\ref{sec: spacetime logical blocks}, we give a spacetime figure for the whole process. 

\medskip To determine the effects of extension on the logical states, we find it insightful to track the logical states in the left gauge, as defined in Section~\ref{sec: code space}. Note that the meaning of the left gauge needs to be modified as the set of gauge transformations changes throughout the operation. We take the left gauge to be the one in which the group elements are shuffled as far to the left as possible, using the gauge transformations. 
Initially, a state $|h\rangle_\mathsf{L} \otimes |k\rangle_\mathsf{L}$ is represented in the left gauge as:
\begin{align*}
    \vcenter{\hbox{\includegraphics[scale=.3]{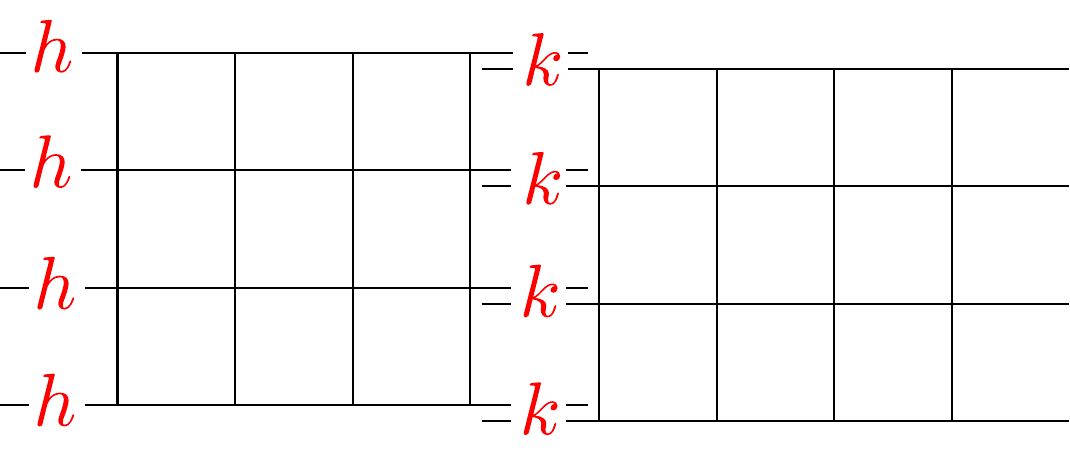}}}.
\end{align*}
Sewing the GSCs together by projecting to the $A^G_v=1$ subspace introduces $G$ gauge transformations at the seam. The state above can then be mapped to:
\begin{align*}
    \vcenter{\hbox{\includegraphics[scale=.3]{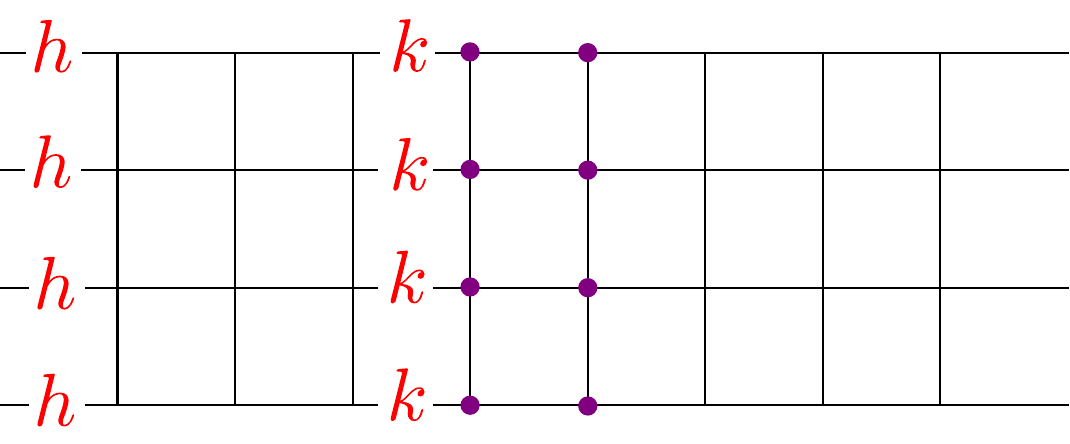}}}.
\end{align*}

Now the system is shrunk from the rough boundaries. Note that as the $H$ and $K$ qudits are measured out, we also lose gauge transformations at the rough boundaries. The code state is mapped to:
\begin{align*}
    \vcenter{\hbox{\includegraphics[scale=.3]{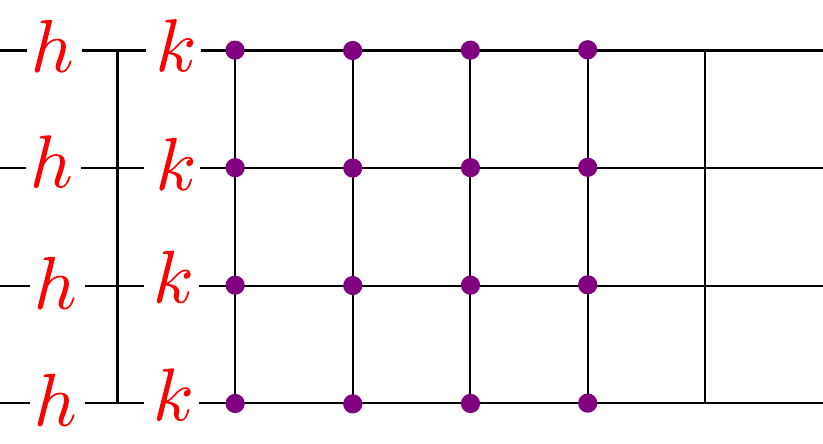}}}.
\end{align*}
We eventually create a single $G$-GSC in the state:
\begin{align*}
    \vcenter{\hbox{\includegraphics[scale=.3]{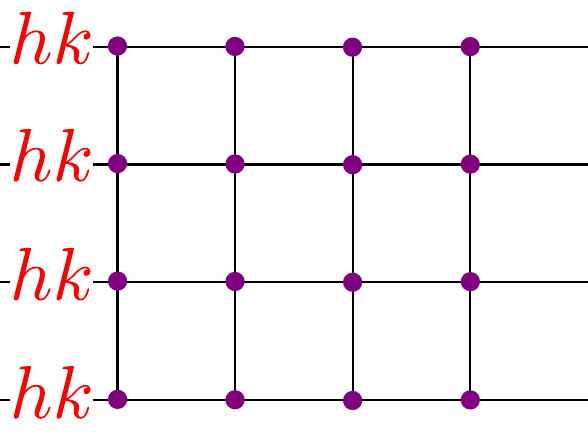}}}.
\end{align*}
Therefore, extending enacts the following transformation on a logical state $|h\rangle_\mathsf{L} \otimes |k\rangle_\mathsf{L}$:
\begin{align}\label{eq: extension HK}
    \boxed{\text{Extend$_{HK}$:} \,\,|h\rangle_\mathsf{L} \otimes |k\rangle_\mathsf{L} \mapsto |hk\rangle_\mathsf{L},}
\end{align}
where the subscript $HK$ is a reminder that the $H$-GSC started to the left of the $K$-GSC.

\subsubsection{Splitting}\label{sec: splitting}

Let us now describe the elementary operation of splitting, which takes a GSC for $G=H\knit K$ and splits it into $H$- and $K$-GSCs. The strategy for splitting is to eject an $H$-GSC from one side and a $K$-GSC from the other. For concreteness, we consider ejecting the $H$-GSC to the left, but emphasize that all of the steps can be performed to eject to the right instead, in contrast to the asymmetry of extension.  

To eject the GSCs, we start by initializing $H$ and $K$ qudits in the $|1\rangle$ state on the left and right side of the patch, respectively. This gives us:
\begin{align*}
    \vcenter{\hbox{\includegraphics[scale=0.3]{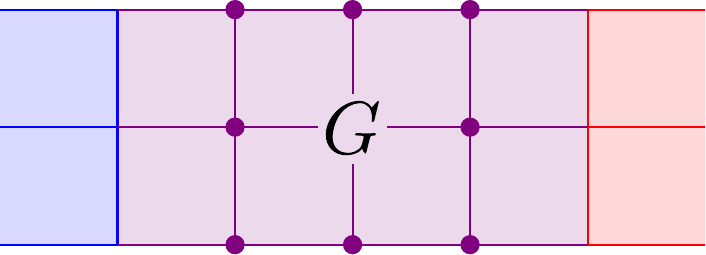}}}.
\end{align*}
We next measure the projectors $A^H_v$ on the left and $A^K_v$ on the right:
\begin{align*}
    \vcenter{\hbox{\includegraphics[scale=0.3]{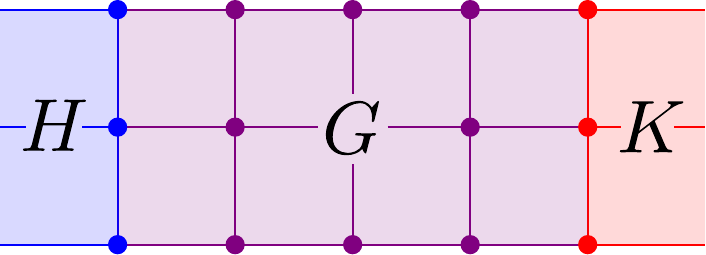}}}.
\end{align*}
The blue (red) dots denote the vertices where $A^H_v$ ($A^K_v$) has been measured. For simplicity, we assume $+1$ measurement outcomes of the projectors and refine this process in Sections~\ref{Sec:Lattice to continuum} and \ref{sec: spacetime logical blocks}. 

We then repeat the process of adding ancilla and measuring the projectors $A^H_v$ and $A^K_v$. This ejects an $H$ and $K$ GSC from the left and right as:
\begin{align*}
    \vcenter{\hbox{\includegraphics[scale=0.3]{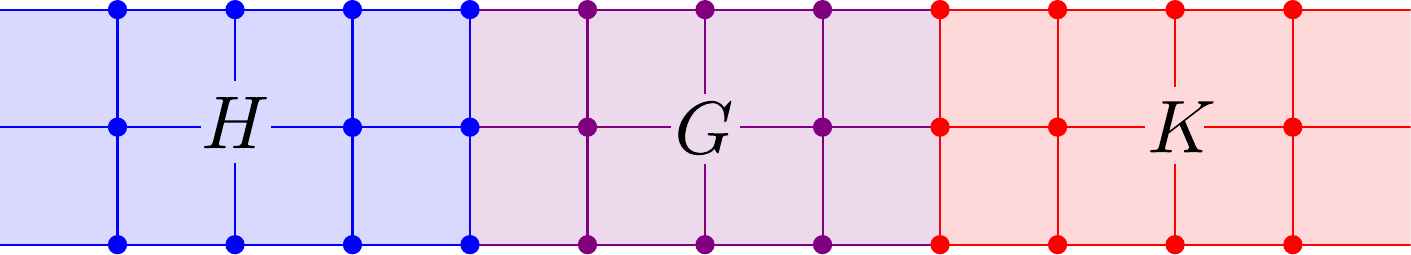}}}.
\end{align*}

To decouple the $H$- and $K$-GSCs, we measure the qudits in the middle in the group basis, leaving us with a fixed configuration of group elements:
\begin{align*}
    \vcenter{\hbox{\includegraphics[scale=0.3]{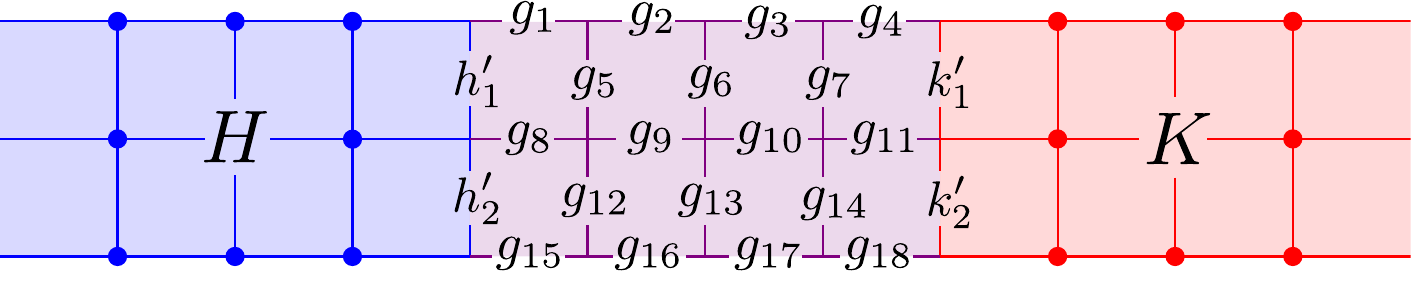}}}.
\end{align*}
After making corrections, we apply $G$ gauge transformations to move $H$ group elements to the left and $K$ group elements to the right. The gauge transformations do not need to be applied in practice. What is important is that we apply right and left multiplication operators to effectively sweep the $H$ and $K$ group elements into their respective patches:
\begin{align*}
    \vcenter{\hbox{\includegraphics[scale=0.3]{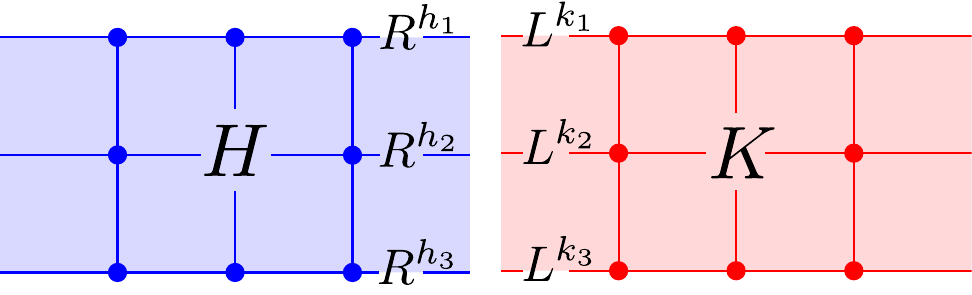}}}.
\end{align*}
This concludes the splitting procedure, with the $G$-GSC having been split into an $H$- and $K$-GSC.

Finally, we can determine the effects of the splitting operation on the logical state by tracking the state in the left gauge. We assume that we start with the $G=H\knit K$ GSC in the logical state $|g\rangle_\mathsf{L}$, represented in the left gauge as:
\begin{align*}
    \vcenter{\hbox{\includegraphics[scale=.3]{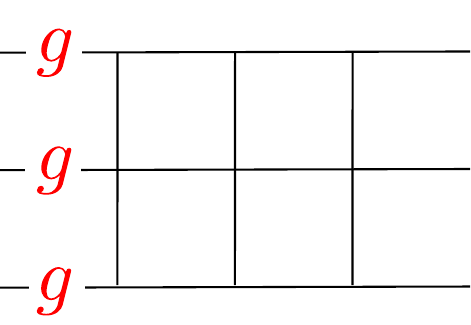}}}.
\end{align*}

Next, we add $H$ and $K$ qudits to the rough boundaries and measure the vertex stabilizers for $H$ and $K$. This allows us to use the $H$ gauge transformations to split $g=hk$ into $h$ and $k$:
\begin{align*}
    \vcenter{\hbox{\includegraphics[scale=.3]{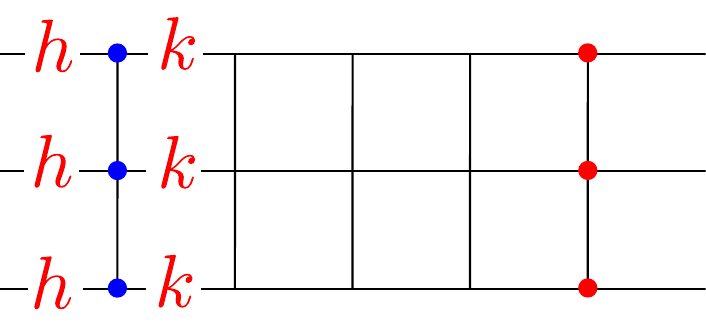}}}.
\end{align*}
Continuing this procedure, we arrive at:
\begin{align*}
    \vcenter{\hbox{\includegraphics[scale=.3]{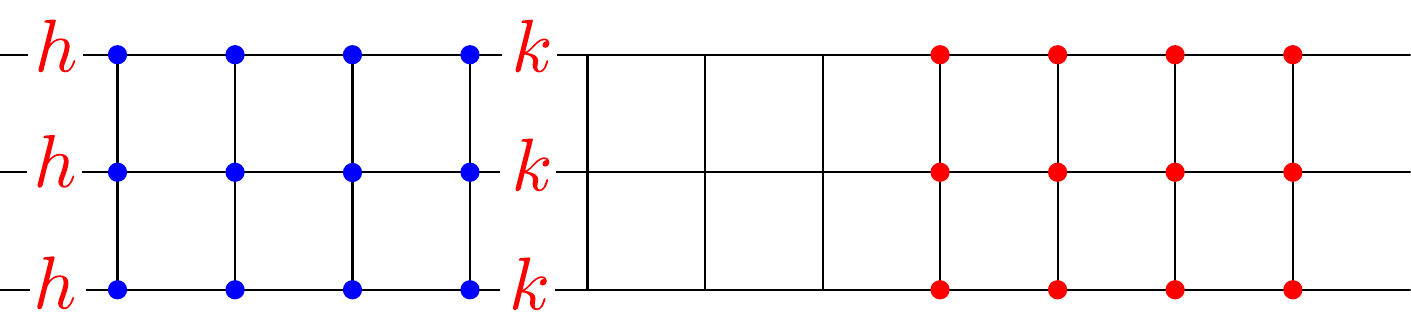}}}.
\end{align*}

Now we measure the qudits in the middle in the group basis and use $G$ gauge transformations to push $H$ elements to the left and $K$ elements to the right:
\begin{align*}
    \vcenter{\hbox{\includegraphics[scale=.3]{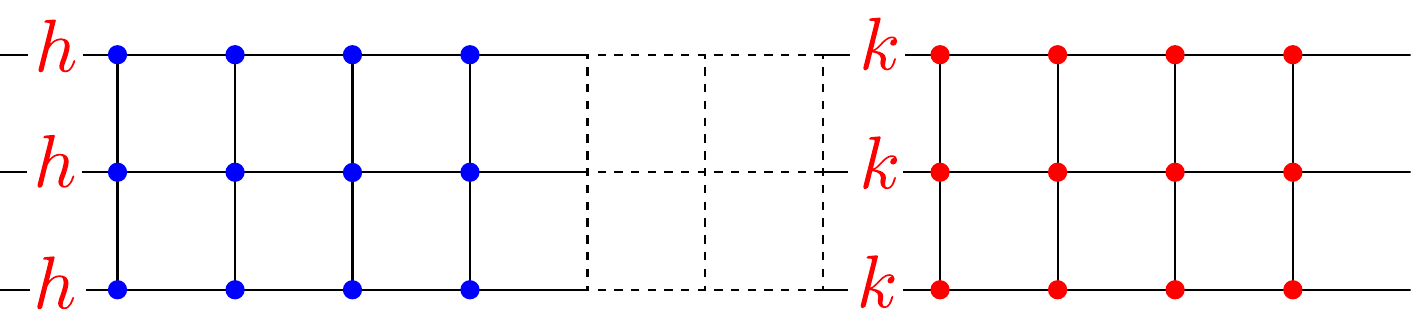}}}.
\end{align*}
Here, we have used dashed lines to denote that the qudits have been removed. When pictured in the left gauge, it is clear that this final step is necessary to initialize the $K$-GSC in the correct logical state.

This implies that the splitting operation maps the state $|g\rangle_\mathsf{L}$ as:
\begin{align}
    \boxed{\text{Split$_{HK}$:} \,\, |g\rangle_\mathsf{L} \mapsto |h\rangle_\mathsf{L} \otimes |k\rangle_\mathsf{L},}
\end{align}
for $g=hk$.
We emphasize that since $G=H\knit K$, there is a unique $h\in H$ and $k\in K$ such that $g=hk$.
We could have also ejected the $H$-GSC to the right. This would result in the operation:
\begin{align}
    \boxed{\text{Split$_{KH}$:} \,\, |g\rangle_\mathsf{L} \mapsto |h'\rangle_\mathsf{L} \otimes |k'\rangle_\mathsf{L},}
\end{align}
where $h'\in H$ and $k'\in K$ are the unique elements satisfying $hk=k'h'$. \\

\noindent \textbf{Remark:} If $H$ is a normal subgroup of $G$, then the knit product is a semidirect product. The operation $\mathrm{Split}_{KH}$ acts, in this case, as:
\begin{align}
    \text{Split$_{KH}$:} \,\, |g\rangle_\mathsf{L} \mapsto |\bar{k} h {k} \rangle_\mathsf{L} \otimes |k\rangle_\mathsf{L}.
\end{align}
Therefore, an inner automorphism is enacted on the element of $H$. We further note that if $H$ is a normal subgroup, then instead of ejecting an $H$-GSC, we can create the $H$-GSC by sequentially measuring out the $K$ qudits of the $G$-GSC. This is analogous to the procedure described in Ref.~\cite{sajith2025noncliffordS3}.\\

\begin{tcbexample}[$D_4$]\label{ex: extending and splitting D4}

\vspace{.2cm}

Let us now demonstrate how the elementary operations of extension and splitting can be composed to perform non-Clifford gates in $\Z_2$ surface codes. In this example, we take $G$ to be $D_4$ and reproduce the sliding protocol in Ref.~\cite{davydova2025D4}. If we instead choose $G$ to be $S_3$, we reproduce the construction in Ref.~\cite{sajith2025noncliffordS3}. 

In Example~\ref{ex: transversal gates D4}, we showed that $D_4$ can be written as $(\Z_2^a \times \Z_2^b)\rtimes \Z_2^c$. This implies that the GSC for $D_4$ can be built starting with a $\Z_2^a\times Z_2^b$ GSC and a $\Z_2^c$ GSC. Performing an extension encodes the logical information of the two Abelian GSCs into the $D_4$ GSC. 

More explicitly, an arbitrary logical basis state for the two Abelian GSCs is:
\begin{align}\label{eq: sliding initial D4}
    |a^\alpha b^\beta\rangle_\mathsf{L} \otimes |c^\gamma\rangle_\mathsf{L},
\end{align}
where $\alpha$, $\beta$, and $\gamma$ are valued in $\{0,1\}$. Letting $H=\Z_2^a\times \Z_2^b$ and $K=\Z_2^c$, the Extend$_{HK}$ operation in Eq.~\eqref{eq: extension HK} produces:
\begin{align} \label{eq: extended D4 state}
    |a^\alpha b^\beta c^\gamma\rangle_\mathsf{L}.
\end{align}

To perform a non-Clifford gate, we split the $D_4$ GSC back into the Abelian GSCs. Importantly, we do so using the operation Split$_{KH}$, such that the $K=\Z_2^c$ GSC is ejected to the left. Intuitively, this results in the $K$-GSC being slid across the $H$-GSC, from right to left. Applying Split$_{KH}$ to Eq.~\eqref{eq: extended D4 state} yields:
\begin{align}
    |{c^\gamma} a^\alpha b^\beta c^\gamma \rangle_\mathsf{L} \otimes |c^\gamma\rangle_\mathsf{L}.
\end{align}
Therefore, using the group relations in Example~\ref{ex: transversal gates D4}, we arrive at the state:
\begin{align}
    |a^\alpha b^{\beta+\alpha \gamma} \rangle_\mathsf{L} \otimes |c^\gamma\rangle_\mathsf{L}.
\end{align}
Comparing to Eq.~\eqref{eq: sliding initial D4}, we have precisely implemented a $\mathrm{CCX}_{132}$ gate, where the $a$ and $c$ logical qubits are the control qubits, and the $b$ qubit is the target.

We refer to this composition of an extension and a splitting operation as sliding. We give a spacetime description of sliding in Section~\ref{sec: examples}.

\end{tcbexample}

\subsection{Preparation and readout} \label{sec: preparation and readout}

The final elementary operations that we need are preparation and readout, allowing us to initialize the logical information and read it out after the computation. Preparation is well established for the $\Z_2$ surface code~\cite{Fowler2012practical}, and information can be read out by measuring arbitrary products of Pauli operators; for example, using lattice surgery with twists~\cite{Litinski2019gameof, Horsman2012latticesurgery}.

Here, we describe how to prepare a $G$-GSC in two different logical states. The first is the logical basis state $|1\rangle_\mathsf{L}$, where $1$ is the identity element of $G$. The second is a logical $|+\rangle_\mathsf{L}$ state, defined as:
\begin{align}
    |+\rangle_\mathsf{L} = \frac{1}{\sqrt{|G|}} \sum_{g \in G}|g\rangle_\mathsf{L}.
\end{align}
We then give a prescription for measuring the logical state in the group basis. We note that, although we expect the operations to be fault tolerant for any choice of finite group, a full proof of fault tolerance falls outside of the scope of this work. 

We conclude this section by providing an example of how preparation and readout can be paired with extension and splitting to generate a magic state in $\Z_2$ GSCs. 

\subsubsection{Preparation}

To prepare a $G$-GSC in the logical state $|1\rangle_\mathsf{L}$, we begin by initializing a column of $G$ qudits in the $|1\rangle$ state, represented as:
\begin{align*}
    \vcenter{\hbox{\includegraphics[scale=0.3]{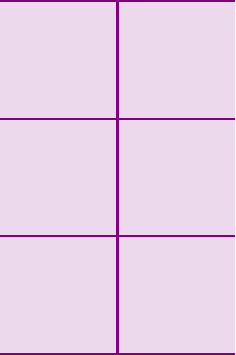}}}.
\end{align*}
The violet lines are $G$ qudits in the $|1\rangle$ state, and the violet shaded plaquettes represent satisfied plaquette stabilizers.
We then create a gauge-invariant state by measuring the projectors $A_v^G$:
\begin{align*}
    \vcenter{\hbox{\includegraphics[scale=0.3]{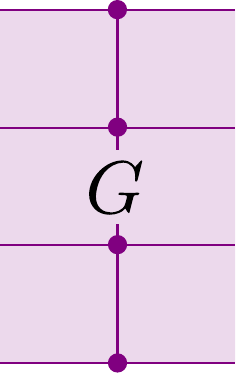}}},
\end{align*}
At this point, we have a thin $G$-GSC.
We continue to grow the $G$-GSC by first initializing qudits in the $|1\rangle$ state to the left and right of the GSC:
\begin{align*}
    \vcenter{\hbox{\includegraphics[scale=0.3]{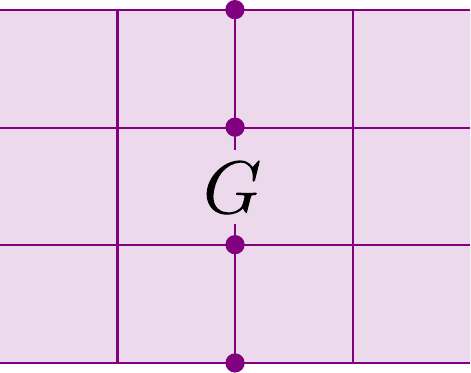}}}.
\end{align*}
The patch is grown by measuring the $A_v^G$ projectors in columns:
\begin{align*}
    \vcenter{\hbox{\includegraphics[scale=0.3]{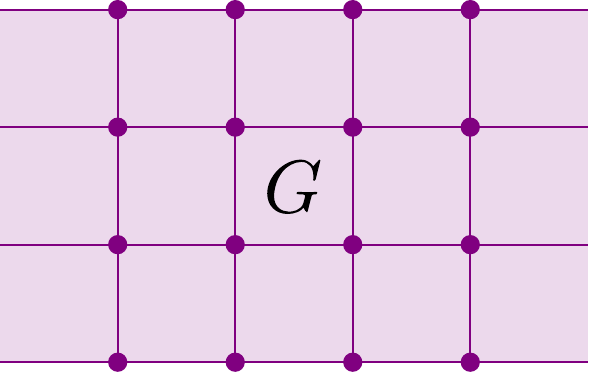}}}.
\end{align*}

This leaves us with a $G$-GSC in the logical state $|1\rangle_\mathsf{L}$. This can be seen by considering the evolution of the logical state in the left gauge---throughout the process, the state in the left gauge is the configuration with $|1\rangle$ on every edge. Here, we have assumed that the measurements successfully project the state to a gauge invariant state. In Section~\ref{sec: spacetime logical blocks}, we give a more careful treatment of the various measurement outcomes.

Next, we consider preparing a $G$-GSC in the $|+\rangle_\mathsf{L}$ state. We begin by initializing a column of $G$ qudits in the $|1\rangle$ state. Notice that, here, we have smooth boundaries on the right:
\begin{align*}
    \vcenter{\hbox{\includegraphics[scale=0.3]{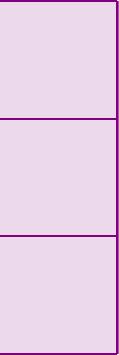}}}.
\end{align*}
The next step is to measure the projectors $A_v^G$ on the rightmost vertices:
\begin{align*}
    \vcenter{\hbox{\includegraphics[scale=0.3]{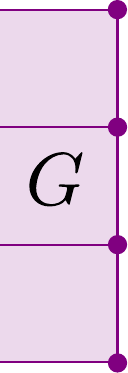}}}.
\end{align*}
We continue to grow the system to the left by repeating. We initialize qudits in the $|1\rangle$ state:
\begin{align*}
    \vcenter{\hbox{\includegraphics[scale=0.3]{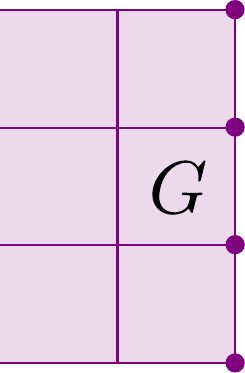}}},
\end{align*}
and subsequently measure $A_v^G$ on a column of vertices:
\begin{align*}
    \vcenter{\hbox{\includegraphics[scale=0.3]{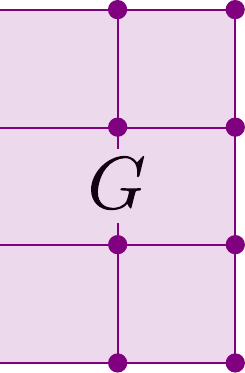}}}.
\end{align*}

Eventually, we obtain a $G$-GSC, with the caveat that there is a smooth boundary on the right side:
\begin{align*} \label{eq: prepplus 3smooth}
    \vcenter{\hbox{\includegraphics[scale=0.3]{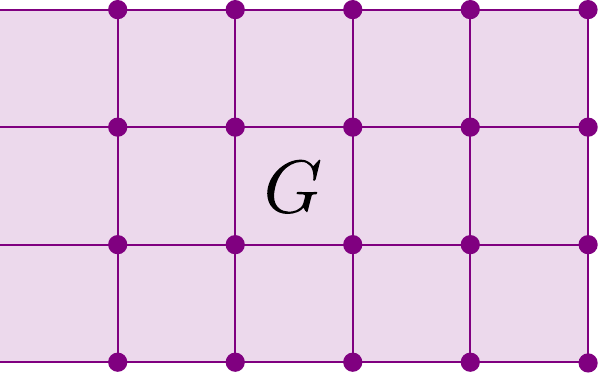}}}.
\end{align*}
At this point, the system does not encode any logical information.
To then turn the smooth boundary into a rough boundary, we measure the rightmost qudits in the group basis:
\begin{align*}
    \vcenter{\hbox{\includegraphics[scale=0.3]{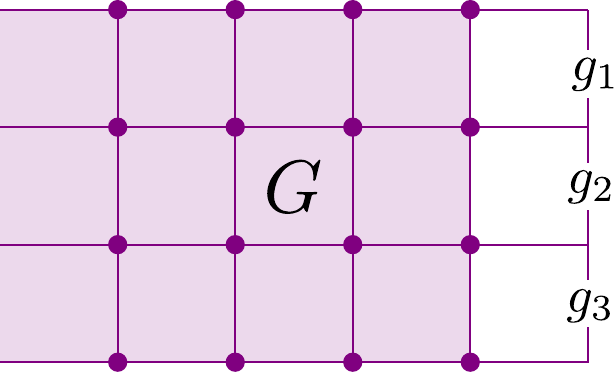}}},
\end{align*}
which yields a configuration of group elements and, in general, violated plaquette stabilizers. We finish the protocol by correcting the plaquette stabilizers, as described in Section~\ref{sec: spacetime logical blocks}:
\begin{align*}
    \vcenter{\hbox{\includegraphics[scale=0.3]{Figures/prepend.pdf}}}.
\end{align*}

To see that this prepares the logical state $|+\rangle_\mathsf{L}$, we consider the operator $U^{\phi_g}$, which acts on each qudit with conjugation by $g\in G$:
\begin{align}
    U^{\phi_g}=\prod_e L^g_e R^g_e.
\end{align}
The $|1\rangle$ state is invariant under conjugation, and the projector $A_v^G$ commutes with $U^{\phi_g}$. Therefore, the each state produced from this protocol is invariant under $U^{\phi_g}$. For the state represented by Eq.~\eqref{eq: prepplus 3smooth}, due to the extra smooth boundary, we can write $U^{\phi_g}$ as:
\begin{align} \label{eq: conjugation Lg identity}
    U^{\phi_g} = L^g_\mathsf{L}\prod_v A_v^g.
\end{align}
Here, we have suggestively used the notation $L^g_\mathsf{L}$, but emphasize that the logical space is one dimensional in Eq.~\eqref{eq: prepplus 3smooth}.

The expression in Eq.~\eqref{eq: conjugation Lg identity} tells us that $U^{\phi_g}=L^g_\mathsf{L}$ in the gauge invariant subspace. Since the state is invariant under $U^{\phi_g}$, it is also invariant under $L^g_\mathsf{L}$. Measuring the rightmost qudits does not affect this, so the final state, after introducing the rough boundary, must also be invariant under $L^g_\mathsf{L}$. The unique logical state that is invariant under $L^g_\mathsf{L}$ is the state $|+\rangle_\mathsf{L}$. Therefore, we have prepared the $|+\rangle_\mathsf{L}$ state.\footnote{We note that a similar approach could be taken to prepare the logical $|1\rangle_\mathsf{L}$ state, i.e., by preparing a patch with rough boundaries on all sides and introducing smooth boundaries in the last step.} \\

\noindent \textbf{Remark:} There are protocols for efficiently preparing certain non-Abelian topological orders through gauging global symmetries~\cite{tantivasadakarn2023hierarchy,Ren2025solvablepreparation}. However, these protocols are not fault tolerant in the presence of measurement errors, in general. 

\subsubsection{Readout}

Next, we briefly describe how the logical state can be read out in the group basis. This is accomplished by measuring each physical qudit in the group basis. This produces a configuration of group elements. As described in Section~\ref{sec: code space}, the logical state can then be determined by computing the holonomy. That is, we multiply the group elements along a path that traverses the patch from left to right. We multiply by $g$ if the orientation of the edge is in the same direction as the path and $\bar{g}$ otherwise. Note that, in the presence of measurement errors, we would first need to perform a round of decoding to modify the measurement outcomes so that there are no fluxes.

\begin{tcbexample}[$D_4$]\label{ex: preparation and readout in D4}

\vspace{.2cm}

In this example, we demonstrate that preparation and readout can be used to prepare a magic state on a $\Z_2 \times \Z_2$ GSC. Using the notation from Example~\ref{ex: transversal gates D4} for the $D_4$ GSC, we start by preparing a $\Z_2^a\times \Z_2^b$ GSC and a $\Z_2^c$ GSC in the state:
\begin{align}
    \frac{1}{{2}}(|a^0\rangle + |a^1\rangle)\otimes|b^0\rangle_\mathsf{L} \otimes (|c^0\rangle_\mathsf{L}+|c^1\rangle_\mathsf{L}).
\end{align}
To simplify the notation, we drop the explicit group elements and write this as:
\begin{align}
    |+0\rangle_\mathsf{L} \otimes |+\rangle_\mathsf{L}.
\end{align}

Next, we apply the extension and splitting in Example~\ref{ex: extending and splitting D4}. This enacts a logical $\mathrm{CCX}_{132}$ to give us
\begin{align}
   \frac{1}{2} (|000\rangle_\mathsf{L} + |100\rangle_\mathsf{L} + |001\rangle_\mathsf{L} + |111\rangle_\mathsf{L})~,
\end{align}
where the effect of the non-Clifford gate is seen in the last summand. We then perform a logical Hadamard on the $c$ qubit and read it out in the group basis. Postselecting on a $+1$ measurement outcome gives us the magic state:
\begin{align}
   |\mathrm{CX}\rangle = \frac{1}{\sqrt{6}}\left( 2\ket{00}_\mathsf{L}+\ket{10}_\mathsf{L}+\ket{11}_\mathsf{L} \right).
\end{align}
To put this in the more familiar form from Ref.~\cite{ibm2024magicstate}, we apply a logical Hadamard to the $b$ qubit. This produces the magic state $|\mathrm{CZ}\rangle$:
\begin{align}
    |\mathrm{CZ}\rangle = \frac{1}{\sqrt{3}}\left( \ket{00}_\mathsf{L}+\ket{10}_\mathsf{L}+\ket{01}_\mathsf{L} \right).
\end{align}

In Section~\ref{sec: examples}, we provide a spacetime picture for this process and illustrate how the preparation and readout can be understood as changing the boundary conditions of the $D_4$ GSC. This also allows us to relate the above construction to the magic state preparations described in Refs.~\cite{huang2025D4,davydova2025D4}.

\end{tcbexample}

\section{From space to spacetime}
\label{Sec:Lattice to continuum}
In this section, we transition from a description of GSCs and their logical operators in space to a spacetime perspective. The benefit of this perspective is threefold: 
\begin{enumerate}[label=(\roman*)]
    \item It gives explicit circuit implementations of the code and the logical operations.
    \item It is more natural to assess the fault tolerance of the operations, wherein a logical error is an undetectable error in spacetime---although, we leave a complete diagnosis of fault tolerance to future studies.
    \item It enables explicit connections between GSCs and spacetime partition functions of topological quantum field theories (TQFTs), namely topological gauge theories.\footnote{These are also known as untwisted Dijkgraaf--Witten theories.} This builds off of the connections developed recently in Refs.~\cite{Bauer2024topologicalerror,Bauer2025lowoverheadnon, Bauer2025xplusy, davydova2025D4, bauer2025planarfaulttolerantcircuitsnonclifford}.
\end{enumerate}
To establish this correspondence, we first describe a convenient set of tensors that generalize those of ZX-calculus~\cite{vandewetering2020zxcalculusworkingquantumcomputer, Bombin2024unifyingflavorsof}. We use these tensors to represent the stabilizers of the GSC and provide explicit circuits for measuring them. We further discuss the potential measurement outcomes and their physical interpretation as fluxes and charges. We derive explicit operators for moving them in spacetime, and argue that the fluxes can be moved deterministically. 
We also remark that generalization of ZX-calculus to groups and more abstract structures can be found in Refs.~\cite{collins2019hopf,duncan2016interacting,majid2021planar,majid2022quantum,lu2026generalized}.  

Through successive measurements of the stabilizers, we obtain a circuit in spacetime, where a timestep represents a round of measurements. We show that the spacetime circuit for a $G$-GSC is equivalent to the spacetime partition function of a $G$ topological gauge theory. This allows us to reformulate the elementary logical operations in spacetime, as described in more detail in Section~\ref{sec: spacetime logical blocks}.

\subsection{Circuit implementation}
\label{sec: circuit implementation}
To give a circuit implementation of the GSC, we begin by introducing two types of tensors, called the copy and multiplication tensors. Each leg of the tensors hosts a $|G|$-dimensional Hilbert space.\footnote{At this point, we do not specify whether the tensor leg is assigned $\mathbb{C}[G]$ or the dual space $\mathbb{C}[G]^*$. This is determined by a direction of time for the circuit. The input to a tensor is taken to be in $\mathbb{C}[G]^*$, while the output is in $\mathbb{C}[G]$. A state in $\mathbb{C}[G]$ can be identified with a state in $\mathbb{C}[G]^*$ by making the following identifications in the group basis: $|g\rangle \leftrightarrow \langle g|$.} 

\medskip \noindent 
{\bf Copy tensor:} In the group basis, the copy tensor evaluates to $+1$ if the group elements on each leg are the same and 0 otherwise.
Graphically, a component of the tensor can be represented as:
\begin{align}
    \vcenter{\hbox{\includegraphics[scale=0.27]{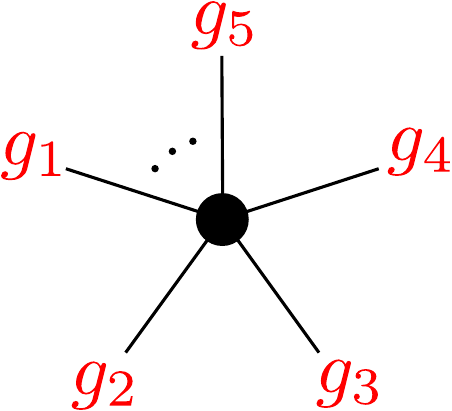}}}=
    \begin{cases}
        1 & \text{if }g_1=g_2=\cdots=g_n, \\
        0 & \text{otherwise}.
    \end{cases}
\end{align}
Here, the ellipses denote the fact that the copy tensor may be defined with arbitrarily many legs. 
The copy tensor satisfies the merge and split property:
\begin{align} \label{eq: merge and split copy}
    \vcenter{\hbox{\includegraphics[scale=0.27]{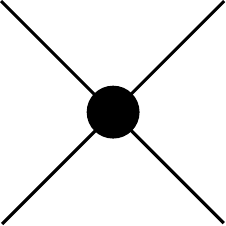}}}\,\,=\,\,\vcenter{\hbox{\includegraphics[scale=0.27]{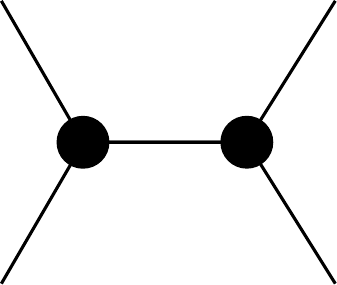}}}.
\end{align}

\medskip \noindent {\bf Multiplication tensor:} The multiplication tensor evaluates to $+1$, if a certain product of the group elements on the legs of the tensor is the identity. 
To make this more precise, we assign an orientation to each leg and define a handedness of the tensor. We multiply by $g$ if the leg points inwards and the inverse $\bar{g}$ otherwise. The handedness then determines whether the group elements are multiplied in a clockwise or counter-clockwise order. For example, we represent the component of a counter-clockwise multiplication tensor as:
\begin{align}
    \vcenter{\hbox{\includegraphics[scale=0.27]{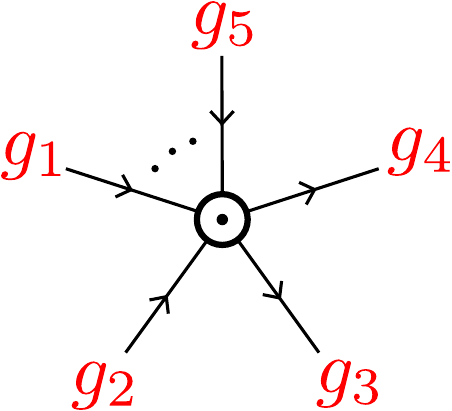}}} = \delta_{g_1g_2\bar{g}_3\bar{g}_4g_5\cdots g_n,1},
\end{align}
while the components of the clockwise multiplication tensor are:
\begin{align}
    \vcenter{\hbox{\includegraphics[scale=0.27]{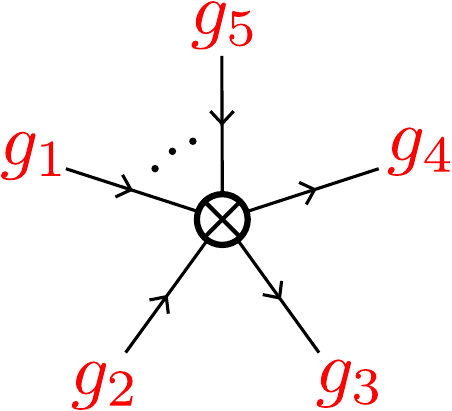}}} = \delta_{g_n \cdots g_5\bar{g}_4\bar{g}_3{g}_2g_1,1}.
\end{align}
Note that, since the product of group elements is set equal to the identity, it is invariant under cyclic permutations. 
Therefore, the tensor is independent of the choice of the first group element in the product.  

We remark that, the handedness of the multiplication tensor is a matter of perspective. That is, viewing a counter clockwise multiplication tensor from behind gives the identity:
\begin{align}
    \vcenter{\hbox{\includegraphics[scale=0.27]{Figures/ccmulti.pdf}}} \,\,\,=\,\,\, \vcenter{\hbox{\includegraphics[scale=0.27]{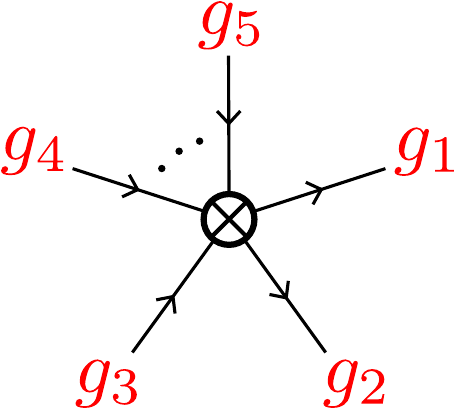}}}.
\end{align}
For multiplication tensors that are part of a tensor network in three dimensions, we fix the handedness by projecting the tensor network to the page; this serves as a reference two-dimensional plane. 

The multiplication tensor also satisfies an analogous merge and split property: 
\begin{align}
    \vcenter{\hbox{\includegraphics[scale=0.27]{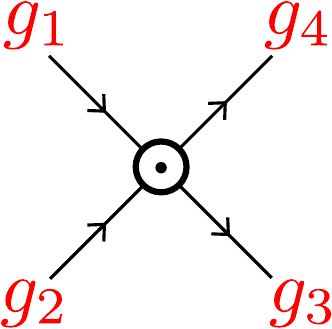}}}\,\,=\,\,\vcenter{\hbox{\includegraphics[scale=0.27]{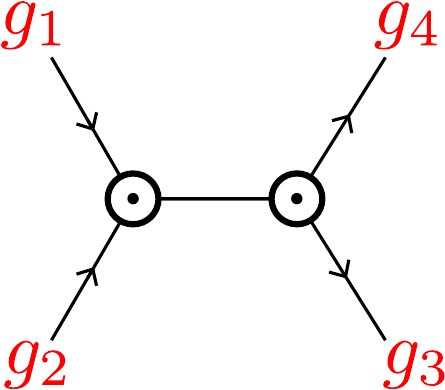}}}.
\end{align}
Here, the orientations of the uncontracted legs have been chosen arbitrarily---with the exception that the orientations on either side of the equation agree with each other. The orientation of the contracted leg is left unspecified, as it can be chosen freely.

The copy and multiplication tensors can be used to represent operators. To do so, we need to specify the input and output legs of the tensor. We associate a dual Hilbert space (bras) with the input legs and a Hilbert space (kets) with the output legs. The tensor components are independent of this assignment, using the identification $|g\rangle \leftrightarrow \langle g|$ of group basis states.

\begin{tcbexample}[Tensor-network operators] \label{ex: controlled gates}

\vspace{.2cm}

    As an example of how copy and multiplication tensors can be contracted to form operators, we define four particularly useful gates. The first is:
    \begin{align}
        \mathrm{CL}_{12}=\vcenter{\hbox{\includegraphics[scale=0.27]{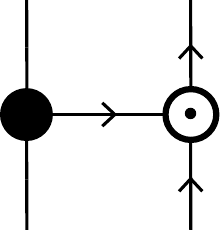}}}.
    \end{align}
    Here, we take the two legs on the bottom to be input and the two legs on the top to be the output. Thus, it defines a two-body operator on a pair of $|G|$-dimensional qudits.

    To determine the effect of $\mathrm{CL}_{12}$, we consider its action on an arbitrary group basis state $|g,h\rangle$. This fixes the indices on the bottom legs to be $g$ and $h$. The relevant components of the tensor are then:
    \begin{align} \label{eq: CL component}
        \vcenter{\hbox{\includegraphics[scale=0.27]{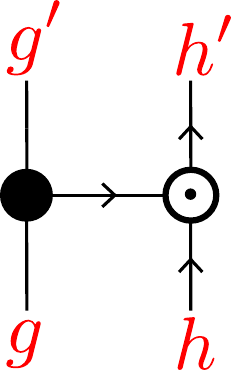}}},
    \end{align}
    for arbitrary $g',h' \in G$. 
    We now consider the constraints imposed on $g'$ and $h'$ by the copy and multiplication tensors. Due to the copy tensor on the left, the component in Eq.~\eqref{eq: CL component} is only nonzero if $g'=g$. The contracted leg is also forced to be in the state $g$. The multiplication tensor then imposes that $h'=gh$. 
    Therefore, the state $|g,h\rangle$ must be mapped as:
    \begin{align}
        \mathrm{CL}_{12}|g,h\rangle = |g,gh\rangle.
    \end{align}
    We see that the tensor performs a left action on the second qudit, controlled on the state of the first qudit. 
    Using similar logic, we define the following operators and give their action on a group basis state $|g,h\rangle$:
    \begin{align}
    \hspace{-8pt}\mathrm{C}\bar{\mathrm{L}}_{12}=\vcenter{\hbox{\includegraphics[scale=0.27]{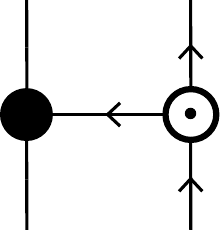}}}, \quad  \mathrm{C}\bar{\mathrm{L}}_{12}|g,h\rangle &=|g,\bar{g}h\rangle,
    \end{align}
    \vspace{-5mm}  
    \begin{align}
\hspace{-8pt}\mathrm{CR}_{12}=\vcenter{\hbox{\includegraphics[scale=0.27]{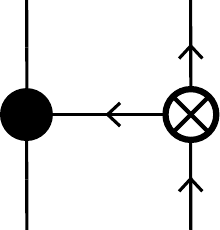}}}, \quad \mathrm{CR}_{12}|g,h\rangle = |g,h\bar{g}\rangle,
    \end{align}
    \vspace{-5mm}
    \begin{align}
   \hspace{-8pt}     \mathrm{C}\bar{\mathrm{R}}_{12}=\vcenter{\hbox{\includegraphics[scale=0.27]{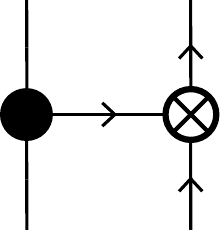}}}, \quad \mathrm{C}\bar{\mathrm{R}}_{12}|g,h\rangle = |g,hg\rangle.
    \end{align}
\end{tcbexample}
\noindent Next we represent the plaquette and vertex stabilizers in terms of copy and multiplication tensors. This naturally leads to a circuit for implementing the stabilizer measurements. We start with the plaquette measurements, since the measurement outcomes are easier to describe.    

\subsubsection{Plaquette measurements}
The plaquette stabilizers can be represented as: 
\begin{align}
    B_p=\,\,\vcenter{\hbox{\includegraphics[scale=0.27]{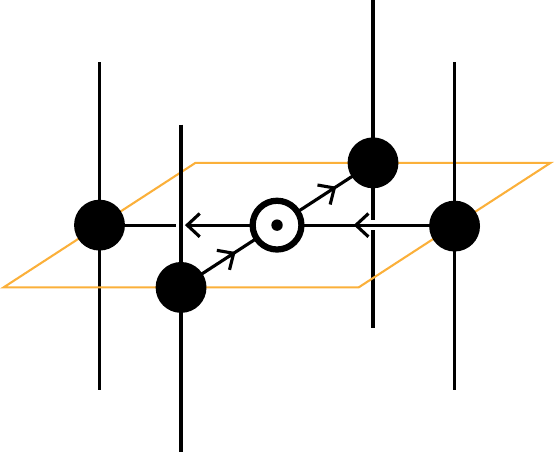}}},
\end{align}
where the bottom legs are the input and the top legs are the output. The yellow edges mark the plaquette of the lattice. This representation of $B_p$ can be checked using a calculation similar to that of Example~\ref{ex: controlled gates}. The copy tensors enforce the operator to be diagonal in the group basis, while the multiplication tensor enforces that there is zero flux.

By judiciously using the merge and split property of the multiplication tensors, the plaquette stabilizer can alternatively be decomposed as: 
\begin{align} \label{eq: decomposition of Bp}
    B_p=\,\,\vcenter{\hbox{\includegraphics[scale=0.27]{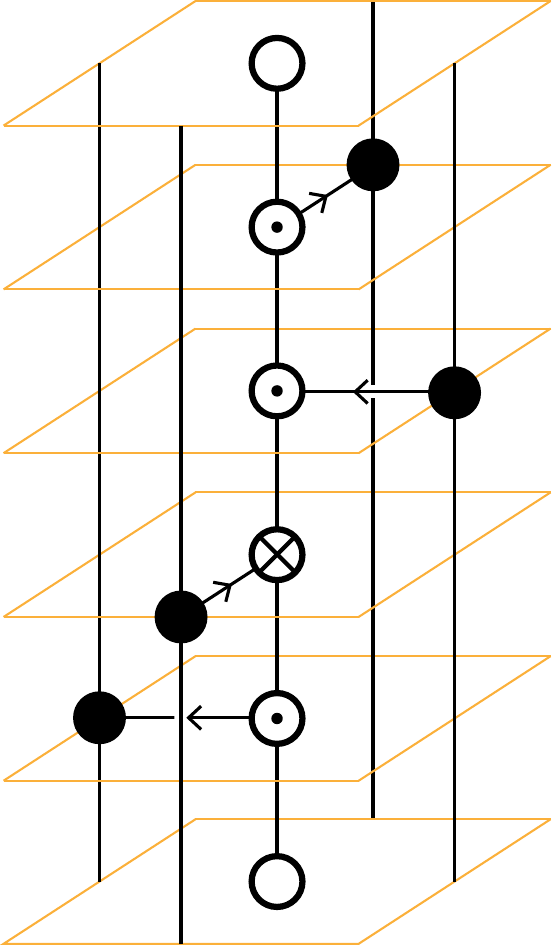}}}\,.
\end{align}
Here, the one-legged tensors are the ket and bra:
\begin{align}
        \vcenter{\hbox{\includegraphics[scale=0.27]{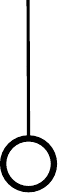}}} =\ket{1}, \quad \vcenter{\hbox{\includegraphics[scale=0.27]{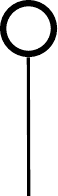}}} = \bra{1}.
\end{align}
Reading from bottom to top, the decomposition in Eq.~\eqref{eq: decomposition of Bp} can be interpreted as follows. First, at the center of the plaquette, an ancillary qudit is initialized in the state $|1\rangle$. Then we apply controlled gates, as defined in Example~\ref{ex: controlled gates}, between the physical qudits and the ancilla. Finally, we project the ancilla onto the $|1\rangle$ state.

A natural way to convert this into a measurement is to replace the projection onto the $|1\rangle$ state with a measurement of the ancilla in the group basis. This can be represented as: 
\begin{align} \label{eq: plaquette measurement}
   \vcenter{\hbox{\includegraphics[scale=0.27]{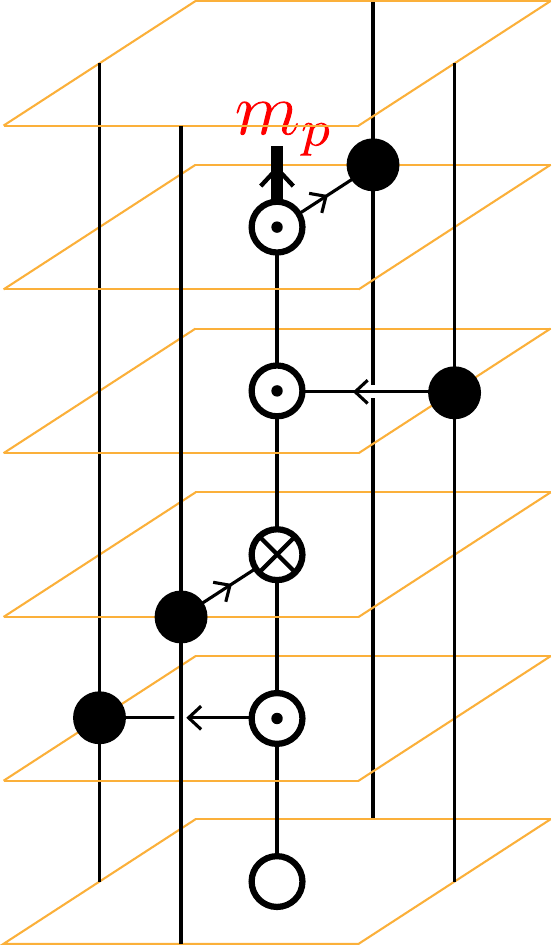}}}\,.
\end{align}
Here, and throughout, we use a bold leg to denote a measurement. When the leg is labeled with the measurement outcome, we assume the index is fixed. In the case shown above, the measurement outcome is $m_p \in G$. 
We refer to the protocol in Eq.~\eqref{eq: plaquette measurement} as a plaquette measurement. A plaquette measurement can similarly be defined at the rough boundary by appropriately truncating the tensors.

The physical interpretation of the outcome $m_p$ is as the group-valued flux at the plaquette $p$. Later in this section, we describe how feedforward allows us to transport the flux to a neighboring plaquette, for example, as part of a decoding algorithm. Note that the plaquette measurement described above conveys more information than a measurement of the projector $B_p$, which only gives a $\{0,1\}$-valued outcome to determine whether there is a flux at the plaquette. The information about the group value is essential for deterministically moving the flux. 

\subsubsection{Vertex measurements}

The projectors $A_v^G$ can be represented with copy and multiplication tensors as:
\begin{align}
   A_v^G = \,\,\vcenter{\hbox{\includegraphics[scale=0.27]{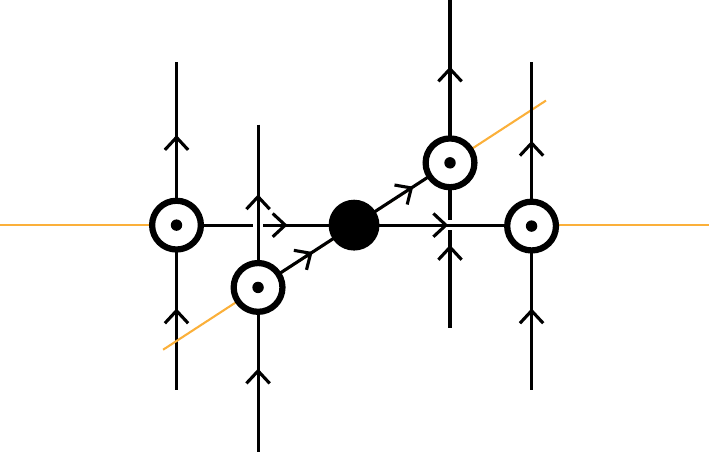}}}.
\end{align}
Intuitively, the copy tensor implements the sum over $g \in G$, while the multiplication tensors implement $L^g_e$ or $R^g_e$ depending on the orientation of the edge. 

Using the merge and split property in Eq.~\eqref{eq: merge and split copy}, the stabilizer $A_v^G$ can be decomposed as:
\begin{align} \label{eq: Avdecomposition}
    A_v^G = \,\,\vcenter{\hbox{\includegraphics[scale=0.27]{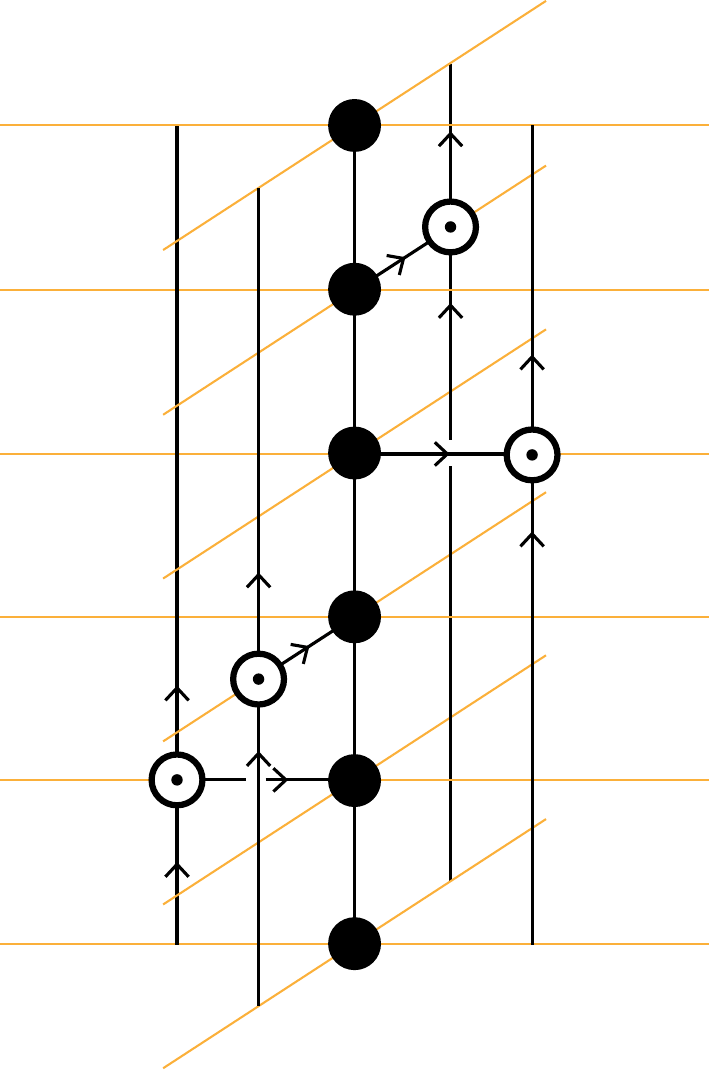}}}\,.
\end{align}
Here, the one-legged tensors are the ket and bra:
\begin{align}
        \vcenter{\hbox{\includegraphics[scale=0.27]{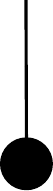}}} =\ket{+}, \quad \vcenter{\hbox{\includegraphics[scale=0.27]{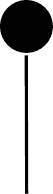}}} = \bra{+},
\end{align}
where the $|+\rangle$ state is defined as:
\begin{align}
    |+\rangle = \frac{1}{\sqrt{|G|}}\sum_{g\in G}|g\rangle.
\end{align}
The diagram in Eq.~\eqref{eq: Avdecomposition} can be interpreted as follows, reading from bottom to top. First, an ancillary $|G|$-dimensional qudit is initialized in the $|+\rangle$ state. Then controlled gates are applied to the ancilla and the surrounding physical qudits. Finally, the ancilla is projected into the $|+\rangle$ state.

To convert the tensor network in Eq.~\eqref{eq: Avdecomposition} into a measurement, we replace the projection onto the $|+\rangle$ state with a measurement. For the vertex stabilizers, it is natural to measure the ancilla in the irreducible-representation basis, or irrep basis. The basis states in the irrep basis are labeled as $\{|R;i,j\rangle\}$. Here, $R$ labels an irrep of the group $G$. It defines a map from $G$ to $d_R \times d_R$ matrices, where $d_R$ is the dimension of the representation.\footnote{\label{foot:GLV}Strictly speaking, a representation $R$ is a map from $G$ to $\mathrm{GL}(V)$, where $\mathrm{GL}(V)$ is the general linear group on a $d_R$-dimensional vector space $V$. Only after choosing a basis for the vector space can we view the representation as a map to $d_R \times d_R$ matrices. To avoid notational clutter, we simply assume that a basis has been chosen for $V$. The choice of basis for the vector space does not play a crucial role in our discussion.} The indices $i,j\in \{1,\ldots,d_R\}$ label a matrix element. Explicitly, the state $|R;i,j\rangle$ is:
\begin{align}
    |R;i,j\rangle = \sqrt{\frac{d_R}{|G|}}\sum_{g\in G}R(g)_{ij}|g\rangle.
\end{align}
The $|+\rangle$ state corresponds to the choice for which $R$ is the trivial irrep. In this case, the indices satisfy $i=j=1$ and $R(g)_{11}=1$, for all $g \in G$.
We depict the resulting measurement as:
\begin{align}
\label{eq: vertex measurement nontrivial}
     \vcenter{\hbox{\includegraphics[scale=0.27]{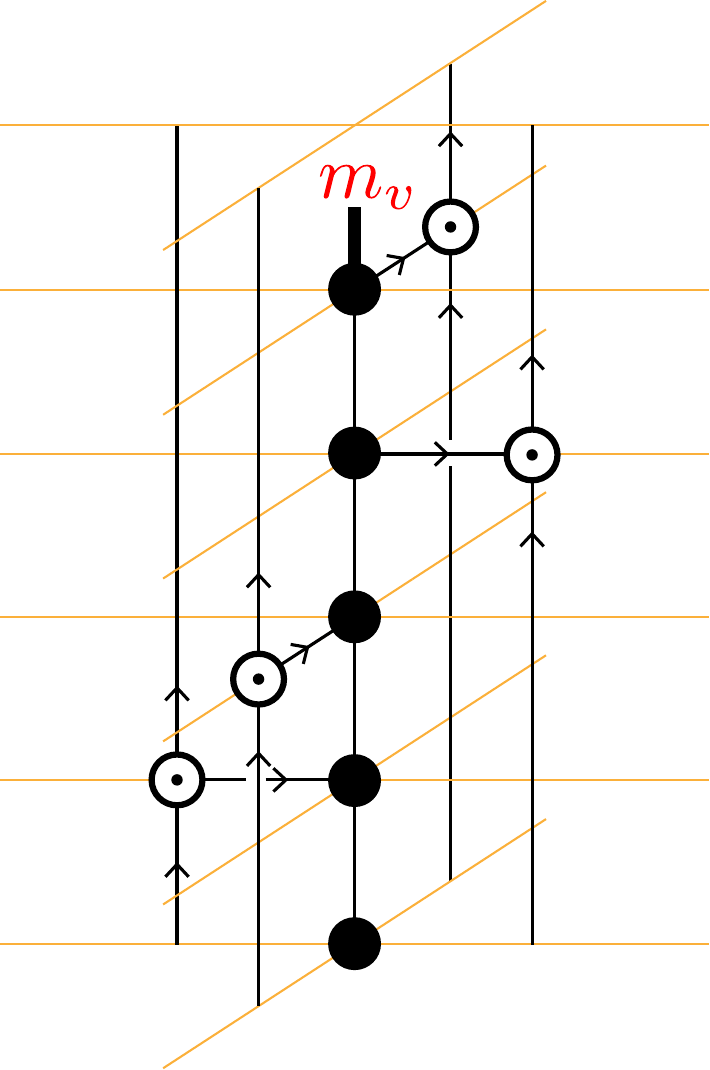}}}\,.
\end{align}
The measurement outcome $m_v$ determines the irrep $R$ and the indices $i,j \in \{1,\ldots, d_R\}$, which specify a matrix element. The tensor leg is depicted in bold to emphasize that the state has been fixed. We refer to this operation as a vertex measurement. Note that a vertex measurement can be defined on the smooth boundaries by appropriately truncating the tensors.

In terms of lattice gauge theory, the measurement outcome $m_v$ corresponds to the presence of a gauge charge. The irrep determines the charge type, while the matrix element specified by $i,j$ determines a transition between internal states of the charge, explained in more detail in Appendix~\ref{app: charge and flux}. 
Note that the internal state after the measurement depends on our choice of basis for the representation $R$. Any basis is fine for our purposes, since the internal state is fully determined after the measurement.\footnote{However, as discussed later, a judicious choice of basis may be useful for increasing the probability of moving a charge with a short movement operator.} 

We describe in Section~\ref{sec: detectors and movement} how charges can be moved to neighboring vertices. Even in the absence of errors, this is a necessary step for the elementary operations of extension, splitting, and preparation.

\begin{tcbexample}[$S_3$ charges and fluxes]
\label{ex: S3 charge and flux}

\vspace{.2cm}

In our next example, we describe the charges and fluxes of the $S_3$ GSC. This will be our recurring example to study the correction of charge and flux errors in Section~\ref{sec: spacetime logical blocks}. \\

We present the group $S_3$ as  
\begin{eqs}
    S_3=\langle r,s|r^3=s^2=1\,, srs=r^2 \rangle\,.
\end{eqs}
This implies that $S_3$ is a semidirect product: $S_3=\Z^r_3\rtimes \Z^s_2$, with a non-trivial action of $\Z_2^{s}$ on $\Z_3^r$. 
The group basis of the $S_3$ GSC is spanned by a qutrit and a qubit: $|\alpha,\beta\rangle = |r^{\alpha}s^{\beta}\rangle$ where $\alpha=0,1,2$ and $\beta=0,1$.
A concrete representation of qudit stabilizers for $S_3$ is given in Appendix~\ref{app: representation on qubits}.

\medskip \noindent {\bf Elementary violations:} The charges and fluxes correspond to violations of the vertex and plaquette stabilizers, respectively. 
It can be shown that there are 8 elementary syndrome types in the $S_3$ GSC corresponding to the anyons in the $S_3$ quantum double. 
These are labeled as $A,B,C,D,E,F,G,H$ \cite{cong2016boundaries}.
In general, these correspond to composites of charges and fluxes, but it will be sufficient for our purpose to discuss the pure charges and fluxes, see Appendix~\ref{app: anyon measurements} for the argument. 

\medskip \noindent {\bf Pure charges:} The pure charges (violations of the vertex stabilizers) are labeled by irreps $R$ of $G$. 
This can be understood as follows. 
Suppose the system is in the state $\ket{\Psi}$ and has a violated stabilizer at the vertex $v$. 
In this example $\ket{\Psi}$ is always in the physical Hilbert space, and we do not consider any ancillas.
This implies that $A_v^g \ket{\Psi} \neq \ket{\Psi}$, for some $g \in G$. In fact, the operators $A_v^g$ act as some non-trivial (possibly decomposable) representation of $G$ on the subspace spanned by the states $A_v^g \ket{\Psi}$. 
Therefore, the state in the vicinity of a vertex decomposes under the vertex operator action as a direct sum of irreps with multiplicity:
\begin{eqs}
   A_v^{g}= \bigoplus_{R}\left(R(g)\otimes 1_{\mathcal{M}_R}\right) ~.
\end{eqs}
Below, we will fix an irrep $R$ and ignore multiplicities $\mathcal{M}_R$. The state of a given charge of type $R$ and dimension $d_R$ is then labeled by an internal state spanned by the basis $\ket{j_R}$ where $1 \le j\le d_R$.\pagefootnotemark

These internal states are part of the physical Hilbert space. The term `internal' simply means they are not gauge-invariant under the $A_v^g$ action when $d_R > 1$; note that the case $d_R>1$ only occurs when $G$ is non-Abelian. 

Let us see how this works in the $S_3$ GSC. The irreps of $S_3$ are denoted $A,B,C$. 
$A$ corresponds to the trivial irrep, realized by the trivial action of $A_v^g$ on any code state. The $B$-type error corresponds to the sign irrep:
\begin{eqs}
    B(r^{\alpha}s^{\beta}) = (-1)^{\beta}~.
\end{eqs}
(As explained in Footnote~\ref{foot:GLV}, we use $R$ to denote both the irrep and its matrix representation in some fixed basis.) We have $d_B = 1$. The syndrome of a state $\ket{1_B}$ corresponding to $B$ at the vertex $v$ is: 
\begin{eqs}
    A_v^{s} \ket{1_B} = - \ket{1_B}~.
\end{eqs}
$C$ corresponds to the two-dimensional irrep, with matrix representation
\begin{align}
    C(r) = \begin{pmatrix}
        \omega & 0 \\ 0 & \omega^2
    \end{pmatrix}; \quad C(s) = \begin{pmatrix}
        0 & 1 \\ 1 & 0
    \end{pmatrix}, 
\end{align}
where $\omega = e^{2\pi i/3}$.
Note that, as a representation of the $\Z_3$ subgroup of $S_3$, $C$ splits into a direct sum of two irreps.
Physically these irreps correspond to $\Z_3$ charges valued 1 and 2 mod 3.
In the irrep $C$, $s$ acts off-diagonally  by permuting the $\Z_3$ representations, i.e. via the charge conjugation (outer) automorphism of ${\rm Rep}(\Z_3)$ .

The transformation of $C$ at the vertex $v$, spanned by the states $\ket{j_C},~j=1,2$ is 
\begin{eqs}
\label{eq:vertex ops on C charge}
    A_v^{r} \ket{j_C} &= \omega^j \ket{j_C}, \\
    A_v^s \ket{1_C} &= \ket{2_C}, \, A_v^s \ket{2_C} = \ket{1_C}.     
\end{eqs}
Since $A_{v}^s$ acts by permuting the internal space vectors, we may also decompose the $C$-representation space into irreps of $\Z_2^{s}$.
The state $\ket{1_C} + \ket{2_C}$ transforms trivially under the $\Z^s_2$ subgroup, while the state $\ket{1_C} - \ket{2_C}$ transforms as the sign irrep of $\Z_2^s$.

\medskip \noindent {\bf Pure fluxes:} Next, we consider the pure fluxes, corresponding to a violation of a plaquette stabilizer. A group-valued flux is labeled by the oriented product of group elements $\bar{g}_1 {g}_2 {g}_3 \bar{g}_4 = g$ around a plaquette $p$. Note that a gauge transformation $A_v^h$, where $v$ lies on the boundary of $p$, can change the value of $g$. Explicitly, if $v$ is the top left vertex of $p$, then applying $A_v^h$ takes $\bar{g}_1$ to $h\bar{g}_1$ and $\bar{g}_4$ to $\bar{g}_4\bar{h}$. This conjugates the group-valued flux: $g \rightarrow h g \bar{h}$. As a result, the gauge-invariant part of the flux is the conjugacy class $[g]$, where
\begin{equation}
    [g] := \{h g \bar{h}~| h \in G\}.
\end{equation}
In analogy with the case of charge errors, one can use the group element $g$ to label the internal state $\ket{g}_p$ of the flux at $p$. The dimension of the internal space corresponding to a flux of type $[g]$ equals the size of the conjugacy class, and can only be greater than 1 when $G$ is non-Abelian. As in the case of charges, we use the term `internal' because $\ket{g}_p$ can be scrambled by gauge transformations. 
There are three pure flux types in the $S_3$ GSC, corresponding to the three conjugacy classes of $S_3$. They are represented by the states
\begin{eqs}
    A &= [e] = \{\ket{e}_p\}, \\
    D &= [s] = \{\ket{s}_p,\ket{rs}_p,\ket{r^2s}_p\}, \\
    F &= [r] = \{\ket{r}_p,\ket{r^2}_p\}.
\end{eqs}

\end{tcbexample}
\pagefootnotetext{We note that, for simplicity, the notation used here differs from the notation for internal states in Appendix~\ref{app: charge and flux}.}

\subsubsection{Repeated measurements}

In order to ensure that quantum error correction with GSCs is fault tolerant, it is necessary to repeat the plaquette and vertex measurements.\footnote{One might be concerned about detecting dyons, which are neither pure fluxes nor pure charges. In Appendix~\ref{app: anyon measurements}, we prove that the plaquette and vertex measurements effectively split the dyons into constituent pure fluxes and charges. Therefore, the plaquette and vertex measurements are sufficient for decoding.} This results in a spacetime circuit, as detailed below. We also describe how fluxes and charges, which appear due to errors or the probabilistic nature of the operations, can be moved through the circuit using feedforward.

For simplicity, we consider a measurement schedule in which plaquette measurements are performed at every plaquette, followed by vertex measurements at every vertex. A pair of plaquette measurements at neighboring plaquettes can be depicted as:
\begin{eqs}
    \vcenter{\hbox{\includegraphics[scale=.27]{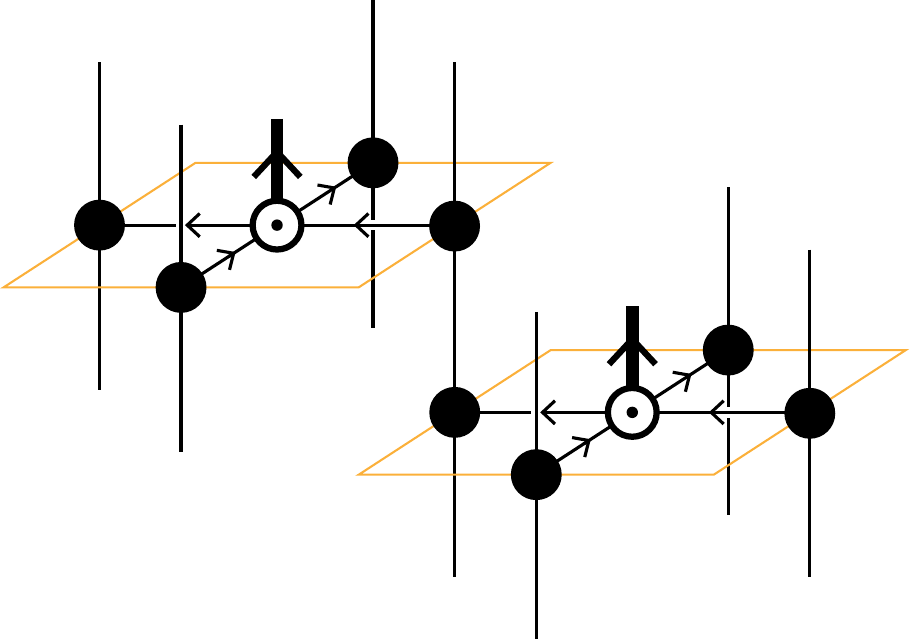}}}.
\end{eqs}
Note that we have used the merge and split property to simplify the decomposition in Eq.~\eqref{eq: decomposition of Bp}. Again,  using the merge and split property, we can contract the shared leg of the copy tensors. A layer of plaquette measurements can then be represented as:
\begin{eqs}
    \vcenter{\hbox{\includegraphics[scale=.27]{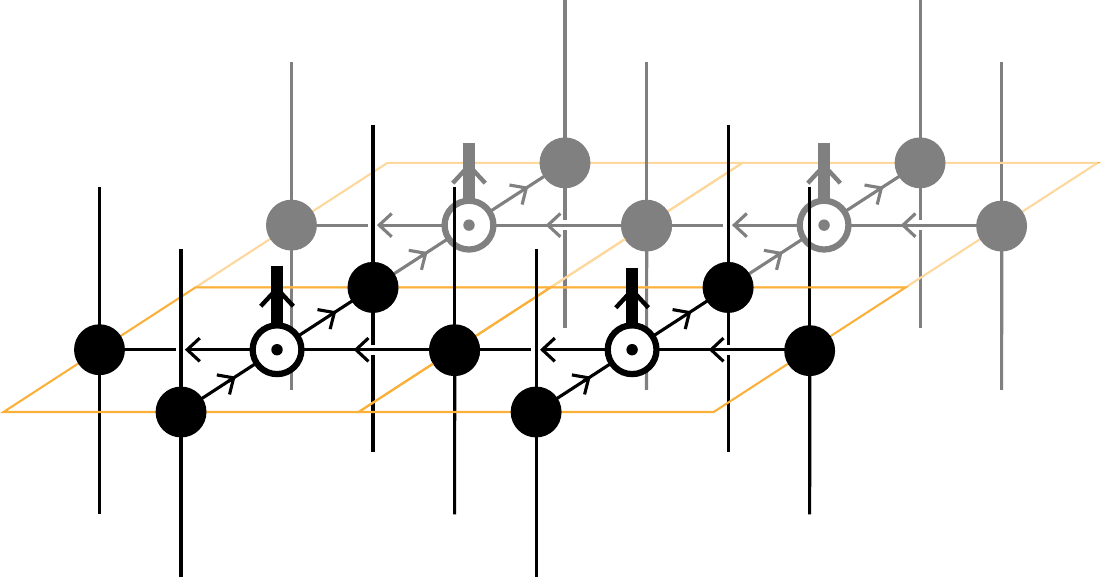}}}.
\end{eqs}
Similarly, a layer of vertex measurements can be depicted as:
\begin{eqs}
    \vcenter{\hbox{\includegraphics[scale=.27]{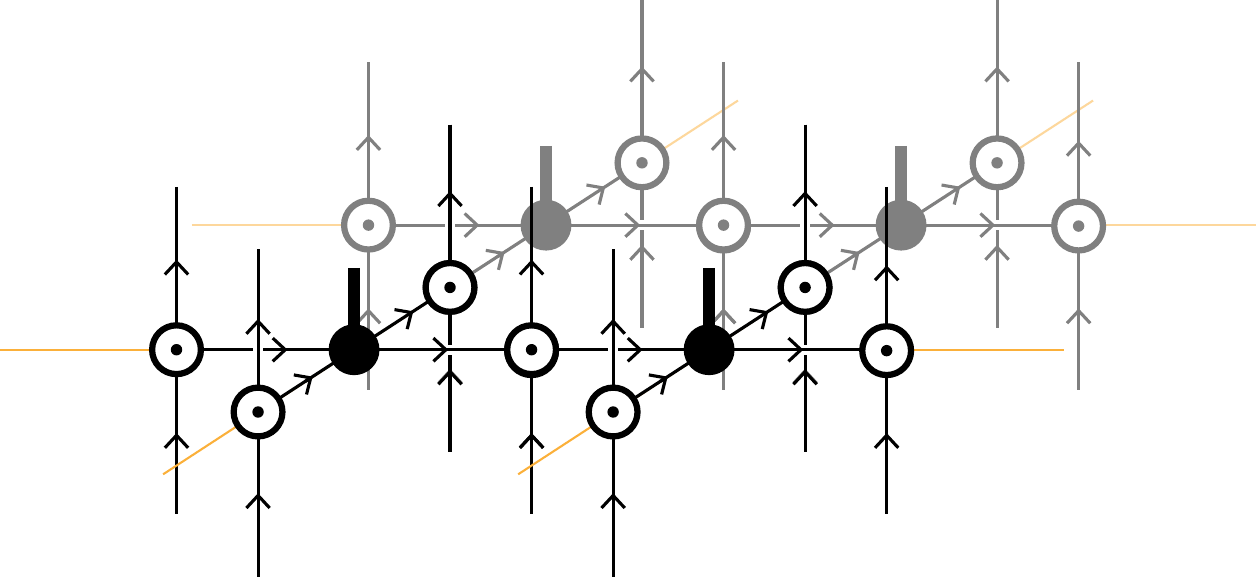}}}.
\end{eqs}
Putting together the layers of measurements gives us a spacetime circuit for performing error correction with a GSC. A single cube of the resulting circuit is:
\begin{eqs}\label{eq: flux detector cell}
    \vcenter{\hbox{\includegraphics[scale=.27]{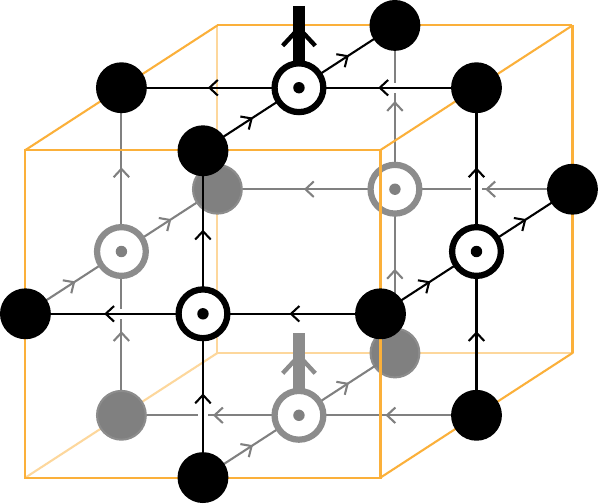}}},
\end{eqs}
while on the dual lattice, we have:
\begin{eqs} \label{eq: charge detector cell}
     \vcenter{\hbox{\includegraphics[scale=.27]{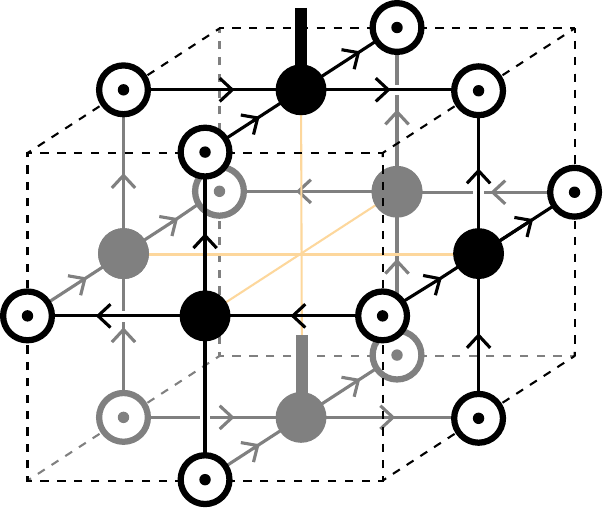}}}.
\end{eqs}

\subsubsection{Detector cells and movement operators} \label{sec: detectors and movement}

In the language of Refs.~\cite{Gidney2021stim,Gidney2021honeycomb,Davydova2023parentsubsystem,Kesselring2024condensation,Fu2025dynamical}, the diagram in Eq.~\eqref{eq: flux detector cell} forms a detector cell for the fluxes. In contrast to quantum error-correcting codes built from Pauli measurements, the group-valued measurement outcomes at the bottom and top of the detector cell do not need to agree. However, in the absence of errors, the two group-valued measurement outcomes must belong to the same conjugacy class,\footnote{Intuitively, the group-valued flux is not invariant under gauge transformations, while its conjugacy class is indeed gauge invariant.} i.e., the measurement outcomes may be:
\begin{eqs}\label{eq: flux detector}
    \vcenter{\hbox{\includegraphics[scale=.27]{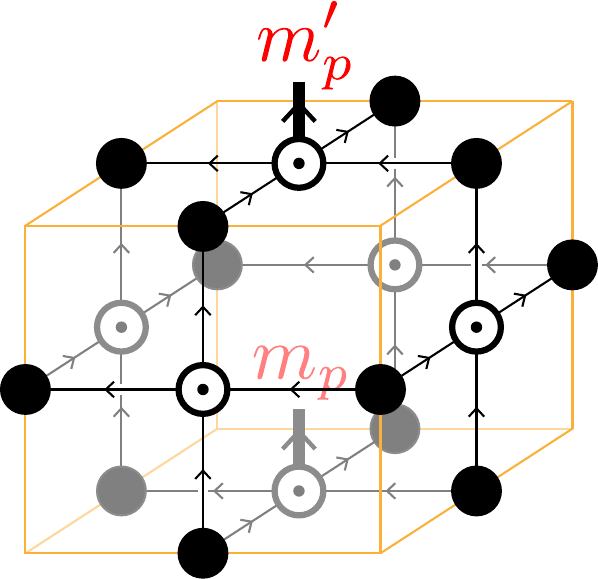}}},
\end{eqs}
where $m_p'=gm_p\bar{g}$, for some $g \in G$. We show this explicitly in Appendix~\ref{app: charge and flux}. Therefore, the repeated plaquette measurements define a detector for the GSC, which can be used to inform a decoder. 

\begin{figure*}[t]
        \centering
        \includegraphics[scale=.27]{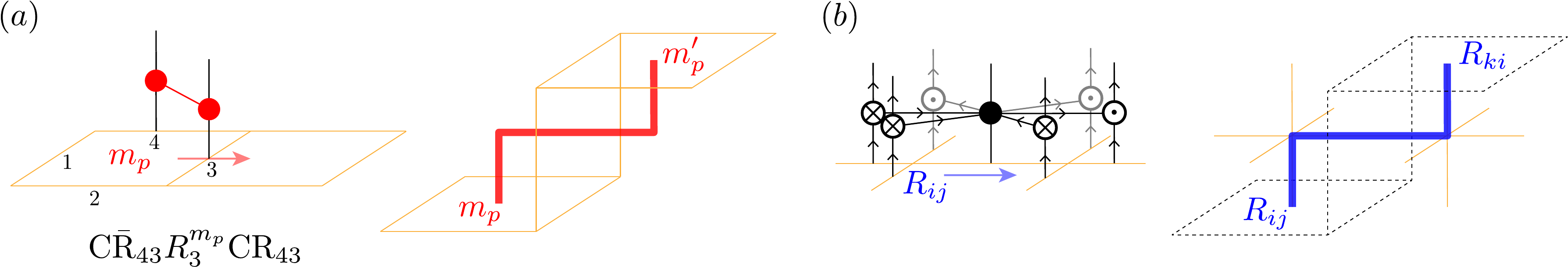}
            \caption{Example movement operators for the fluxes and charges. (a) A group-valued flux $m_p$ is moved to the right by applying the operator $\mathrm{C}\bar{\mathrm{R}}_{43}R_3^{m_p}\mathrm{C}{\mathrm{R}}_{43}$. The worldline of the flux is depicted to the right, with $m'_p=gm_p\bar{g}$, for some $g\in G$. (b) After the measurement outcome $R_{ij}$, the charge can be deterministically moved to an empty vertex by applying a product of controlled group actions. For further details, we refer to Appendix~\ref{app: charge and flux}.} 
    \label{fig: charge flux movement}
\end{figure*}

Depending on the outcomes of the detectors, we may need to move the fluxes to clean them up. For this purpose, the group-valued outcomes are valuable. In Appendix~\ref{app: charge and flux}, we show that a group-valued flux can be moved deterministically with feedforward and can be derived directly from the tensor network formalism. We depict a set of short spacetime ribbon operators in Fig.~\ref{fig: charge flux movement}, which move a flux to neighboring plaquettes. Note that if the group-valued measurement is faulty, it may be necessary to repeat the moving procedure.

Similarly for charges, the cube on the dual lattice in Eq.~\eqref{eq: charge detector cell} serves as a detector cell. As shown in Appendix~\ref{app: charge and flux}, in the absence of errors, a charge entering the cell from below is labeled by the same irrep as the charge exiting from the top of the cell:
\begin{eqs}\label{eq: charge detector}
    \vcenter{\hbox{\includegraphics[scale=.27]{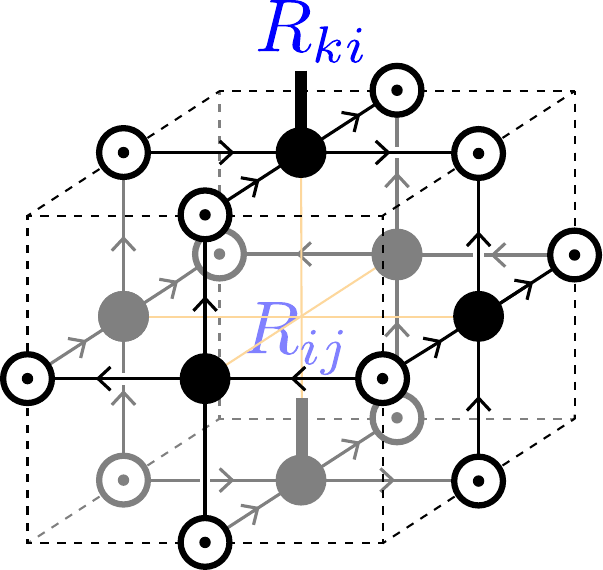}}}.
\end{eqs}
Note, however, that only the irrep type is consistent between the two measurement outcomes, and the matrix elements may fluctuate.\footnote{This captures the fact that the internal state of the charge is not gauge invariant, only the charge type. }

Even in the absence of errors, it is essential to be able to move the charges, as they appear throughout the elementary logical operations, as discussed in more detail in the next section. The charges can be moved deterministically using the short movement operator depicted in Fig.~\ref{fig: charge flux movement}. This movement operator assumes that the target vertex is empty, as described in Appendix~\ref{app: charge and flux}. When the target vertex hosts a charge, or the charge is being condensed into a boundary, the movement of charges is generally probabilistic, depending on the fusion outcomes. This means that a repeat-until-success approach is necessary to move them. 
It is also worth noting that if the group is nilpotent, the charges can be successfully moved in a constant number of attempts. 
We discuss the movement of charges in more detail in Appendix~\ref{app: charge and flux}.

\subsection{Spacetime partition function} \label{sec: partition function}

We now make an explicit connection to TQFTs by postselecting the measurement outcomes. Specifically, we postselect on outcomes that are $1\in G$ for the plaquette measurements and the trivial irrep for the vertex measurements. That is, we do not detect any fluxes or charges. This leaves us with a tensor network whose legs are contracted, and a portion of the tensor network looks like:
\begin{eqs}\label{eq: partition cube}
    \vcenter{\hbox{\includegraphics[scale=.27]{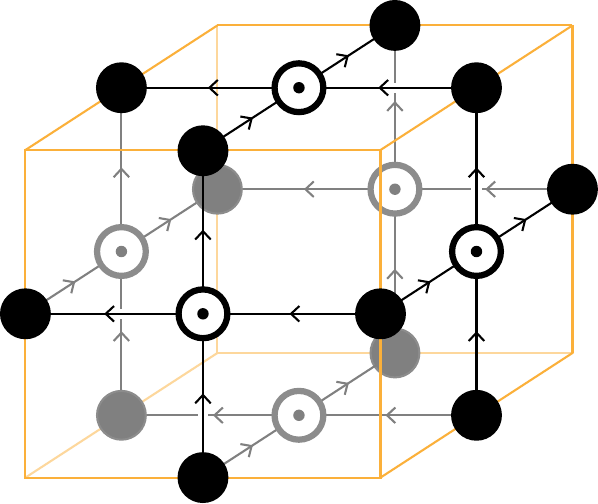}}}.
\end{eqs}
This tensor network is precisely the partition function of a $G$ topological gauge theory. 

To see this, note that each copy tensor includes a sum over group elements. The evaluation of the full tensor network can therefore be organized into a sum over configurations of group elements on the edges. In the language of TQFT, this assignment of group elements to edges defines a gauge field configuration. The multiplication tensors, then enforce trivial flux through the plaquettes, or in other words, that the gauge field is flat. We thus have a sum over all of the flat gauge field configurations, which is exactly the partition function of a $G$ topological gauge theory.

There is an equivalent formulation of this partition function in terms of configurations of membranes labeled by group elements. This is obtained by simply viewing the partition function on the dual lattice. The membranes cut through the edges of the direct lattice and carry the group label associated to the edge. They also inherit an orientation from the orientation of the edges. The membranes meet at plaquettes, and the multiplication tensors enforce that the membranes fuse according to the group laws. 

On the smooth boundaries, the membranes are allowed to terminate without violating any of the multiplication tensors. For reference, the tensor network on the front-side boundary looks like:
\begin{eqs}\label{eq: smooth partition cube}
    \vcenter{\hbox{\includegraphics[scale=.27]{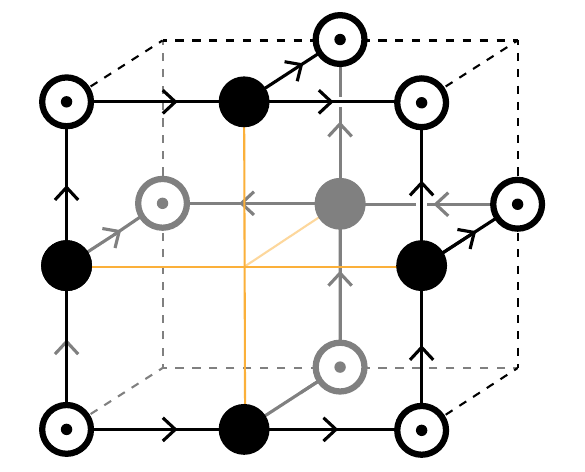}}}.
\end{eqs}
The fluctuating membranes on the smooth boundary can be interpreted as the fact that the fluxes can condense on the smooth boundary. These correspond to Neumann boundary conditions in the topological gauge theory. 

On the rough boundaries, the membranes cannot terminate, since that would lead to a violation of at least one multiplication tensor. Explicitly, a portion of the tensor network on the right-hand-side boundary is:
\begin{eqs}\label{eq: rough partition cube}
    \vcenter{\hbox{\includegraphics[scale=.27]{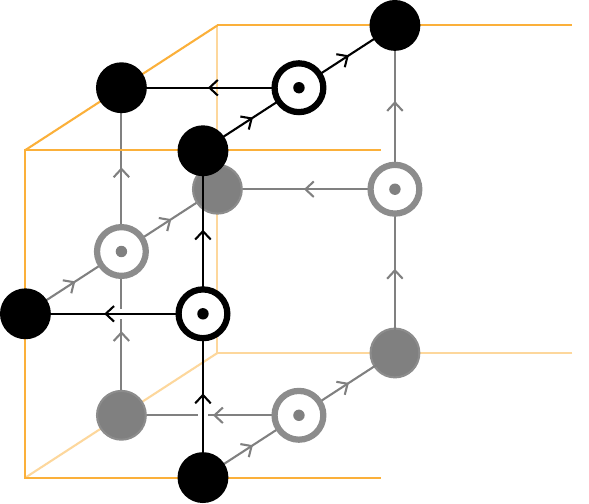}}}.
\end{eqs}
A membrane that terminates on the rough boundary would have to violate one of the three-legged multiplication tensors. This implies that the fluxes cannot condense on the rough boundaries. The charges can, however, condense on the rough boundary, and our elementary operations use this to remove charges from the system. The rough boundaries are Dirichlet boundary conditions in the topological gauge theory. 

Assuming that the measurement outcomes are not postselected to $1 \in G$ and the trivial irrep, we obtain a partition function with defects, corresponding to the worldlines of the charges and fluxes. In the absence of errors, these worldlines propagate parallel to the direction of time. The worldlines can be redirected with movement operators, as shown in Fig.~\ref{fig: charge flux movement}.

In the next section, we consider spacetime implementations of our elementary logical blocks, which can be built using the tensor-network formalism. We do not make their decomposition into tensors explicit, however. We also note that the elementary logical operations of extension and splitting require switching between degrees of freedom corresponding to different groups. We leave it implicit that additional tensors can be introduced for extending subgroups into a larger group. The key ingredient from this section is that the partition function can be understood in terms of a sum of membrane configurations. We find this particularly useful for describing the effects of the logical operations in spacetime.

\begin{table*}[t]
\centering

\begin{tblr}{
  colspec = {|Q[c,m,wd=0.22\textwidth]|
              Q[c,m,wd=0.30\textwidth]|
              Q[c,m,wd=0.28\textwidth]|
              Q[c,m,wd=0.10\textwidth]|},
  row{1} = {font=\bfseries},
  rowsep = 6pt,
  hlines, vlines,
  hline{2} = {1.1pt},
  cell{2-Z}{1-4} = {valign=m}, 
}

Elementary operation & Spacetime logical block & Logical operation & References \\
\vspace{.6cm}Identity & \includegraphics[width=0.11\textwidth]{Figures/Id.pdf}\vspace{-.6cm} & \vspace{.6cm} $|h\rangle_\mathsf{L} \xrightarrow[]{\mathrm{Id}_\mathsf{L}} |h\rangle_\mathsf{L}$ & \vspace{.6cm}Section~\ref{sec: partition function} \\

\vspace{-0.1cm}Left/right group action &
\includegraphics[width=0.12\textwidth]{Figures/Lg.pdf}\vspace{-.6cm}\,,\quad
\includegraphics[width=0.118\textwidth]{Figures/Rg.pdf}\vspace{-.6cm} &
\vspace{-.3cm}\begin{align*}|h\rangle_\mathsf{L} &\xrightarrow[]{L^g_\mathsf{L}} |gh\rangle_\mathsf{L}, \\ |h\rangle_\mathsf{L} &\xrightarrow[]{R^g_\mathsf{L}} |h\bar{g}\rangle_\mathsf{L}\end{align*} & \vspace{-.3cm} \begin{align*}&\text{Section~\ref{sec:GSC-TransversalGates}}\\ &\text{Section~\ref{sec: transversal logical gates spacetime}}\end{align*} \\

\vspace{.3cm}Group automorphism & \includegraphics[width=0.10\textwidth]{Figures/Uphi.pdf}\vspace{-.6cm} & \vspace{.4cm}$|h\rangle_\mathsf{L} \xrightarrow[]{U^\phi_\mathsf{L}} |\phi(h)\rangle_\mathsf{L}$ & \vspace{-0.7cm}\begin{align*}&\text{\!Section~\ref{sec:GSC-TransversalGates}}\\ &\text{\!Section~\ref{sec: transversal logical gates spacetime}}\end{align*} \\

\vspace{1.3cm}Extension & \includegraphics[width=0.18\textwidth]{Figures/extendHK2.pdf}\vspace{-1.5cm} & \vspace{-4cm}\begin{align*}|h\rangle_\mathsf{L}\otimes |k\rangle_\mathsf{L} &\xrightarrow[]{\mathrm{Extend}_{HK}} |hk\rangle_\mathsf{L}
\end{align*} \vspace{.1cm} & \vspace{-1.8cm}\begin{align*}&\text{\!Section~\ref{sec: extension and splitting}}\\ &\text{\!Section~\ref{sec: extension and splitting spacetime}}\end{align*} \\

\vspace{1cm}Splitting & \includegraphics[width=0.18\textwidth]{Figures/splitHK2.pdf}\vspace{-1.5cm} & \vspace{-1.2cm}\begin{align*}|hk\rangle_\mathsf{L} &\xrightarrow[]{\mathrm{Split}_{HK}} |h\rangle_\mathsf{L}\otimes |k\rangle_\mathsf{L}, \\ |hk\rangle_\mathsf{L} &\xrightarrow[]{\mathrm{Split}_{KH}} |h'\rangle_\mathsf{L} \otimes |k'\rangle_\mathsf{L},
\\
hk = &k' h'
\end{align*} & \vspace{-1.2cm}\begin{align*}&\text{\!Section~\ref{sec: extension and splitting}}\\ &\text{\!Section~\ref{sec: extension and splitting spacetime}}\end{align*} \\

\vspace{1.2cm}Preparation & \includegraphics[width=0.20\textwidth]{Figures/prepwhitespace.pdf}\vspace{-1.6cm}\vspace{0.5cm} & \vspace{1.2cm}Prepares $|1\rangle_\mathsf{L}$ and $\ket{+}_{\LL}$ & \vspace{-1.6cm}\begin{align*}&\text{\!Section~\ref{sec: preparation and readout}}\\ &\text{\!Section~\ref{sec: preparation and readout spacetime}}\end{align*} \\

\vspace{0cm}Readout & \vspace{.4cm}\includegraphics[width=0.12\textwidth]{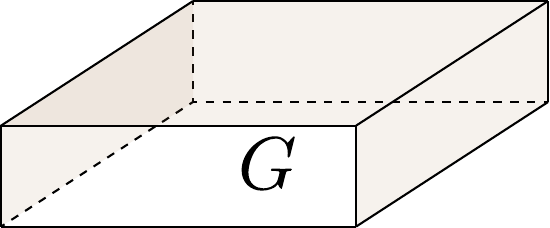}\vspace{-1cm} & \vspace{-.2cm}Measurement in group basis & \vspace{-.2cm}\begin{align*}&\text{\!Section~\ref{sec: preparation and readout}}\\ &\text{\!Section~\ref{sec: preparation and readout spacetime}}\end{align*} \\

\end{tblr}

\caption{Summary of elementary logical operations and spacetime logical blocks.}
\label{tab:logical_operations_summary}
\end{table*}

\section{Spacetime logical blocks}
\label{sec: spacetime logical blocks}

We now recast the elementary logical operations of Section~\ref{sec: elementary logical operations} as  spacetime logical blocks.  
This builds off of the work of Ref.~\cite{Bombin2023logicalblocks}, which formalized spacetime logical blocks for the $\Z_2$ surface code.
The set of logical operations are summarized in Table~\ref{tab:logical_operations_summary}.
These can be combined to generate a broad class of logical gates, as demonstrated in Section~\ref{sec: examples}. 

In this section, we emphasize the correspondence to TQFTs, as it provides a systematic framework for organizing logical gates and motivates natural generalizations. 
Specifically, we describe the operations in terms of topological defects and interfaces between topological gauge theories.

To get warmed up, we consider the spacetime logical block for the identity, which we depict as:
\begin{align} \label{eq: identity block}
    \vcenter{\hbox{\includegraphics[scale=.23]{Figures/Id.pdf}}}\,\,.
\end{align}
This block can be interpreted as the spacetime volume swept out by evolving a GSC for a finite time---through alternating plaquette and vertex measurements. Here, the rough boundaries are on the left and right and have been shaded, while the smooth boundaries on the front and back remain unshaded.

The diagram in Eq.~\eqref{eq: identity block} represents an operator, which can be made explicit by filling it with the tensor network circuit from the previous section. The legs at the bottom and top of the block are left uncontracted, and serve as the input and output of the operator. If the tensor network is postselected onto trivial outcomes (or error corrected) then the operator is a projector onto the code space and acts as the logical identity.

In what follows, we dress the logical identity with topological defects and interfaces which represent the elementary logical operations. 

\subsection{Transversal logical gates}\label{sec: transversal logical gates spacetime}

\subsubsection{Left and right group multiplication}

We start with the transversal logical operators $L^g_{\mathsf{L}}$ and $R^g_{\mathsf{L}}$, which implement left and right multiplication. 
The operators $L_{\mathsf{L}}^g$ and $R_{\mathsf{L}}^g$ are localized defects supported along rough boundaries, and depicted as: 
\begin{align*}
    \vcenter{\hbox{\includegraphics[scale=.23]{Figures/Lg.pdf}}}\,\,, \quad \vcenter{\hbox{\includegraphics[scale=.23]{Figures/Rg.pdf}}}.    
\end{align*}

Viewing the partition function as a sum over membranes, as described in Section~\ref{Sec:Lattice to continuum}, these correspond to defects on the boundary where the membranes can end. An allowed configuration of membranes is then: 
\begin{equation}
\begin{split}
\vcenter{\hbox{\includegraphics[scale=.23]{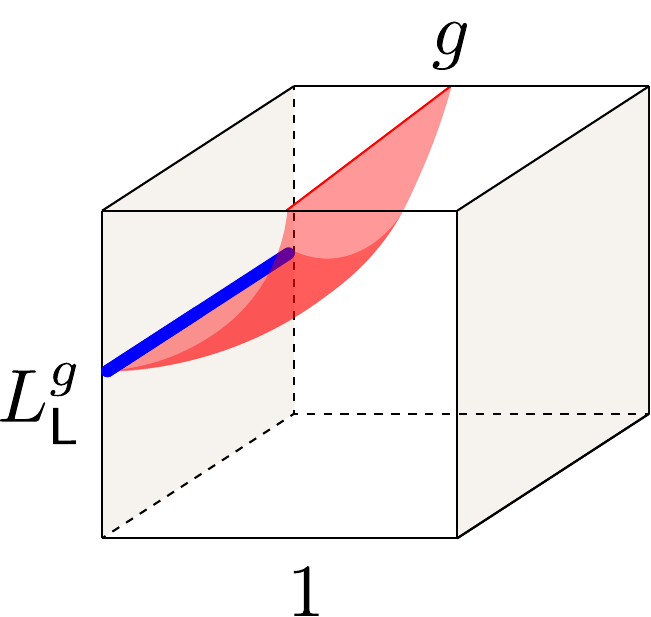}}}\,\,, \quad \vcenter{\hbox{\includegraphics[scale=.23]{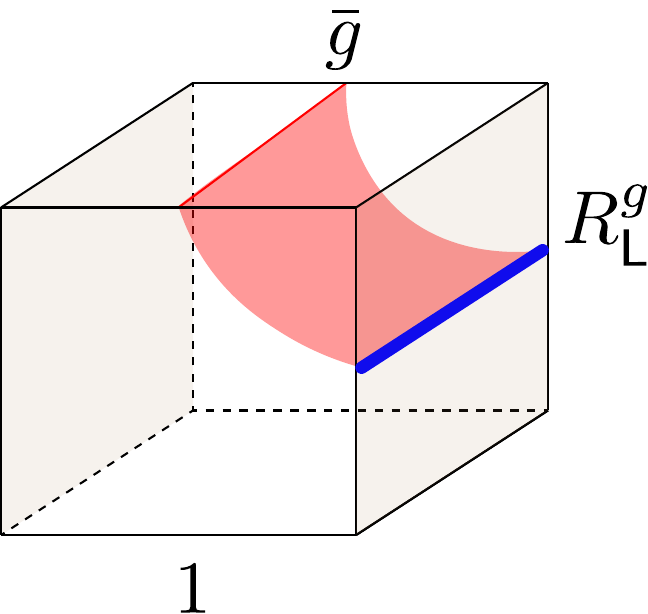}}}.   
\end{split}
\end{equation}

In the language of TQFT, the operators $L^g_{\LL}$ and $R^g_{\LL}$ are the nontrivial topological line defects on the rough boundary.
Explicitly, when a flux labeled by a conjugacy class is brought to a rough boundary, it decomposes into a direct sum of
boundary lines, which are labeled by the elements of the conjugacy
class.
These boundary topological lines are nothing but $L_{\mathrm{L}}^{g}$ and $R_{\mathrm{L}}^{g}$. \\

\noindent \textbf{Remark:} We can also consider logical operators obtained by projecting the bulk lines onto the smooth boundary. 
The flux lines become trivial while the charge lines become logical operators that are diagonal in the group basis.  
Such an operator is labeled by an irrep $R$ and a pair of indices $i,j\in \{1,\dots, d_{R}\}$ and acts on a logical basis state as:
\begin{equation}
\begin{split}
    Z_{\LL}^{R_{ij}}
|g\rangle_\mathsf{L} = {R}(g)_{ij}|g\rangle_\mathsf{L}.
\end{split}
\end{equation}
This logical operator generically requires a linear-depth circuit, and we save a discussion of its implementation for future studies.

\subsubsection{Group automorphisms}

The logical operator $U_{\rm L}^{\phi}$ is represented as a topological defect surface in the topological gauge theory:
\begin{equation}
\begin{split}
   \vcenter{\hbox{\includegraphics[scale=.23]{Figures/Uphi.pdf}}}\,\,. 
\end{split}
\end{equation}
A membrane labeled by $h$ gets transformed to the membrane $\phi(h)$ as it passes through the defect: 
\begin{equation}
\begin{split}
    \vcenter{\hbox{\includegraphics[scale=.23]{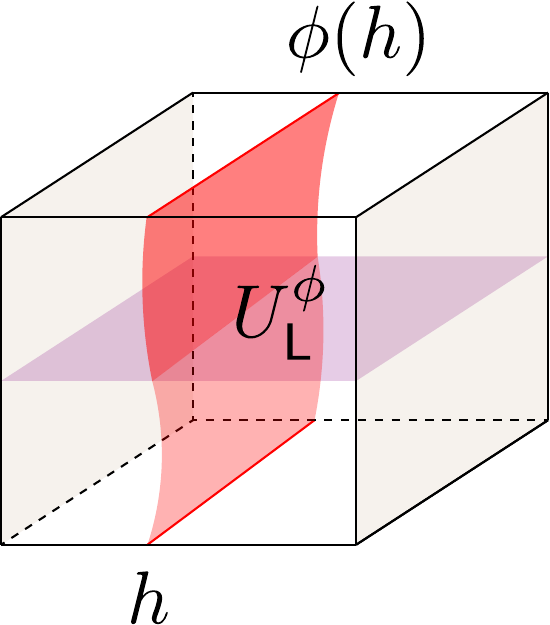}}}\,\,.      
\end{split}
\end{equation}
Under an inner automorphism, the $h$ membrane transforms to $gh\bar{g}$.
Therefore, in this case, $U_{\LL}^{\phi}$ can be represented as $L_{\mathsf{L}}^g R_{\mathsf{L}}^g$: 
\begin{equation}
\begin{split}
    \vcenter{\hbox{\includegraphics[scale=.23]{Figures/Uphimembrane2.pdf}}} \,\,= \,\, \vcenter{\hbox{\includegraphics[scale=.23]{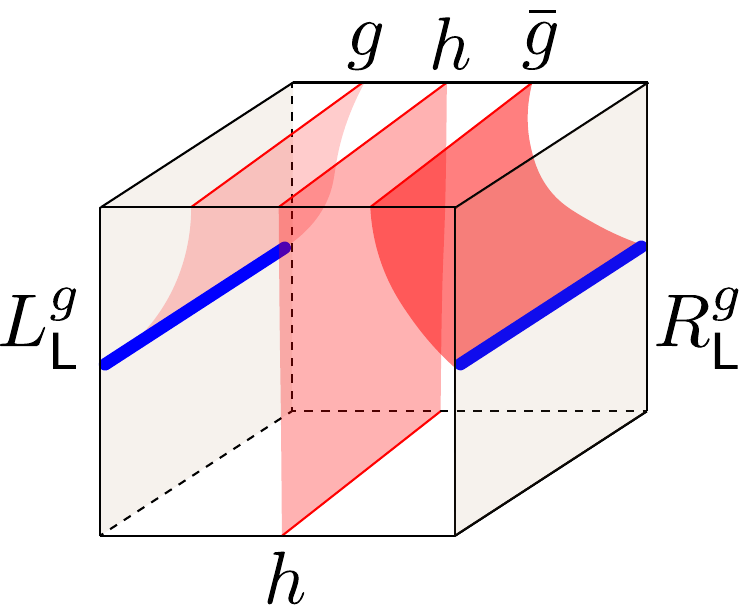}}}.        
\end{split}
\end{equation}

We note that there are other types of topological surface defects (0-form symmetries), such as those labeled by elements of $H^{2}(G,U(1))$ \cite{nikshych2013}. However, we do not consider these in this work, since they alter the rough and smooth boundaries and do not preserve the code space of the GSC. We refer to Ref.~\cite{li2026anyonpermutationsquantumdouble} for a discussion on implementations of other topological surface defects.

\begin{tcbexample}[$S_3$]

\vspace{.2cm}

    {\bf Left and right multiplication:} From the representation of the $S_3$ GSC on qubits and qutrits in Appendix~\ref{app: representation on qubits}, the action of the left and right group multiplication logicals on the group basis states is 
\begin{eqs}
    L_{\rm L}^{r}|\alpha,\beta\rangle_\mathsf{L} =& |\alpha+1, \beta\rangle_\mathsf{L},  \\
    L_{\rm L}^{s}|\alpha,\beta\rangle_\mathsf{L} =& |-\alpha, \beta+1\rangle_\mathsf{L}, \\
    R_{\rm L}^{r}|\alpha,\beta\rangle_\mathsf{L} =& |\alpha+2\beta+2, \beta\rangle_\mathsf{L},  \\
    R_{\rm L}^{s}|\alpha,\beta\rangle_\mathsf{L} =& |\alpha, \beta+1\rangle_\mathsf{L},  
\end{eqs}
where addition is taken $\mathrm{mod}\, 3$ on the first argument, and $\mathrm{mod}\, 2$ in the second argument.  

\medskip \noindent 
The charges condense at rough boundaries, while the fluxes become nontrivial boundary excitations.
Upon restriction to this boundary, the bulk fluxes $D$ and $F$ split into a direct sum of elementary boundary defects labeled by group elements in the conjugacy classes $[s]$ and $[r]$, respectively.
The fusion of these boundary defects reproduces the group multiplication of $S_3$. Tunneling a boundary defect along the rough boundary produces the left and right group multiplication on the logical state.

\medskip \noindent {\bf Group automorphisms:}
Next we consider the transversal logical operators corresponding to group automorphisms of $S_3$. 
Note that 
\begin{eqs}
    {\rm Aut}(S_3)\cong S_3\,,
\end{eqs}
with all the automorphisms being inner.
The corresponding logical operators $U_{\LL}^{\phi_g}$ implement a conjugation by the group element $g$. 
The action is read off to be
\begin{eqs}    U_{\rm L}^{\phi_r}|\alpha,\beta\rangle_\mathsf{L} =&|\alpha+2\beta, \beta\rangle_\mathsf{L},  \\
U_{\rm L}^{\phi_s}|\alpha,\beta\rangle_\mathsf{L} =& |-\alpha,\beta\rangle_\mathsf{L}.
\end{eqs}
Let us describe what happens to the fluxes and charges after acting with $U_{\rm L}^{\phi_h}$.
Under the action of $U_{\rm L}^{\phi_h}$, all the edge degrees of freedom transform under $\phi_h$. The flux state at a plaquette $p$ then simply transforms as
\begin{eqs}
    U_{\rm L}^{\phi_h}|g\rangle_p =|hg\bar{h}\rangle_p .  
\end{eqs}
Since all automorphisms of $S_3$ are inner, the conjugacy class itself does not change under the action of $U_{\rm L}^{\phi_{h}}$, but the internal flux state transforms.

\medskip \noindent Recall that pure charge errors are characterized by their transformation under the vertex stabilizers.
Consider a state $|i_R\rangle$ corresponding to a charge $R$ in the internal state $i$.
Under automorphism, this state transforms to $U_{\rm L}^{\phi}|i_R\rangle$ which transforms under the vertex stabilizers as
\begin{eqs}
\label{eq:action on charge post auto}
    A_v^{g}U_{\rm L}^{\phi}|i_R\rangle=&\;U_{\rm L}^{\phi}A_{v}^{\phi(g)}|R,i\rangle \\
    =&\; U_{\rm L}^{\phi} \sum_{j}R^R(\phi(g))_{ji}|j_R\rangle\,.
\end{eqs}
It can be seen that the automorphism does not change the irrep label, instead only acts on the internal space.
Automorphisms therefore act trivially on the one dimensional irrep $B$.
The states carrying a charge in the two-dimensional irrep initially transform under vertex stabilizers as in Eq.~\eqref{eq:vertex ops on C charge}. 
The action after implementing a group automorphism can then be deduced from Eq.~\eqref{eq:action on charge post auto}.
Under $U_{\rm L}^{\phi_r}$, we have:
\begin{eqs}
\label{eq:vertex ops on transformed C  charge}
    A_v^{r} U_{\rm L}^{\phi_r}\ket{j_C} &= \omega^j  U_{\rm L}^{\phi_r}\ket{j_C}\,, \\
    A_v^s  U_{\rm L}^{\phi_r} \ket{1_C} &= \omega^2 U_{\rm L}^{\phi_r}\ket{2_C}\,, \\  A_v^s  U_{\rm L}^{\phi_r}\ket{2_C} &= \omega  U_{\rm L}^{\phi_r}\ket{1_C}\,.
\end{eqs}
Meanwhile under $U_{\rm L}^{\phi_s}$:
\begin{eqs}
    A_v^{r} U_{\rm L}^{\phi_s}\ket{j_C} &= \omega^{-j}  U_{\rm L}^{\phi_s}\ket{j_C}\,, \\
    A_v^s  U_{\rm L}^{\phi_s} \ket{1_C} &=  U_{\rm L}^{\phi_s}\ket{2_C}\,, \\  A_v^s  U_{\rm L}^{\phi_s}\ket{2_C} &=   U_{\rm L}^{\phi_s}\ket{1_C}\,.
\end{eqs}
It is worth emphasizing that any charges and fluxes that could be corrected before can also be corrected after the action of automorphisms, whose only effect is to permute internal states.

\end{tcbexample}

\subsection{Extension and splitting} \label{sec: extension and splitting spacetime}

\subsubsection{Extension}
We now describe the spacetime logical block for the extension operation, which converts an $H$-GSC and a $K$-GSC into a $G = H \knit K$ GSC. We represent it graphically as:
\begin{equation}
\begin{split}
     \vcenter{\hbox{\includegraphics[scale=.2]{Figures/extendHK2.pdf}}},
\end{split}
\end{equation}
where the blue surfaces represent the interfaces between the GSCs. 

In Section~\ref{sec: extension}, we assumed that each step of the extension involved measuring projectors $A^G_v$ with $+1$ measurement outcomes. Here, we consider repeated vertex and plaquette measurements, as outlined in Section~\ref{sec: circuit implementation}, and pay closer attention to the possible measurement outcomes at each stage of the protocol. 

Recall that the extension begins by overlapping the rough boundary edges of an $H$- and $K$-GSC. Then, ancillary $K$ qudits are added to the $H$-GSC and ancillary $H$ qudits are added to the $K$-GSC. Instead of measuring the projector $A^G_v$, we perform $G$ vertex measurements on the overlapping vertices. 

Let us focus on the $G$ vertex measurements on the $H$-GSC side. 
The stabilizers $A_v^h$ are already satisfied for $h \in H$, since the ancillary $K$ qudits were initialized in the $\ket{1}$ state. The only violations of $A_v^g$ in the new patch of $G$-GSC are for $g=hk$, such that $k\neq 1$. These correspond to $K$ charges, in the sense described in Appendix~\ref{app: ResInd}. We show in Appendix~\ref{app: ResInd} that all of the $K$ charges can be condensed on the $H|G$ interface. 

Hence, the $K$ charges, which appear probabilistically as the $G$-GSC is grown, can always be removed by moving them to the $H|G$ interface. Analogously, the $H$ charges that appear on the side of the $K$-GSC can be removed by dumping them into the $G|K$ interface.
Within the spacetime TQFT description the error correction procedure is depicted as:
\begin{align}
    \vcenter{\hbox{\includegraphics[scale=.3]{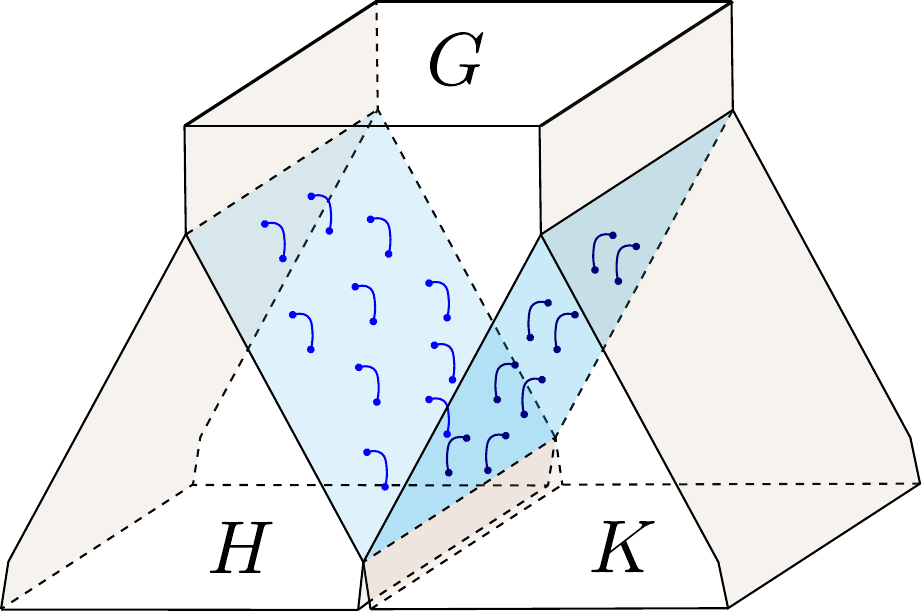}}}
\end{align}

Note that when a $K$ charge corresponding to the irrep $R$ is moved to the $H|G$ interface, it might not condense, for example, due to a physical error that occurs before moving the charge (see Example~\ref{eg: S3 extension} below). In this case, we end up with an irrep $R_H$ on the $H$ side of the interface, and another irrep $R'$ on the $G$ side. But if the internal state of $R$ transforms trivially under $H$, so does the internal state of $R_H \otimes R'$, implying that we can attempt to fuse these charges again and repeat until success.

\begin{tcbexample}[$S_3$ GSC extension]\label{eg: S3 extension}

\vspace{.2cm}

Here we consider the spacetime logical block that extends a $\Z_3$ and a $\Z_2$ surface code into an $S_3$ GSC.
For concreteness, we consider the $\Z_3$ GSC to be placed on the left and the $\Z_2$ GSC on the right.
In the process of extending into the $S_3$ GSC, we first overlap the right rough boundary of the $\Z_3$ GSC with the left rough boundary of the $\Z_2$ GSC as in Eq.~\eqref{eq:extension overlap}.
We then introduce additional qubits on one column to the left of the seam and qutrits on the right of the seam and perform $S_3$ vertex measurements on the vertices of the seam in Eq.~\eqref{eq:extension overlap}.

\medskip \noindent {\bf Charges from extending $\Z_3$ to $S_3$:} Let us consider the vertex stabilizers on the left of the seam. 
While the $\Z_3$ vertex terms remain satisfied, the $\Z_2$ terms are not necessarily satisfied.
Therefore, the extension step can only give rise to a $B$ charge in a state $\ket{1}_B$. 
This can be moved to the $\Z_3|S_3$ interface on the left, where $B$ condenses.

\medskip \noindent {\bf Charges from extending $\Z_2$ to $S_3$:}
More interestingly, on the right of the seam, we start with the $\Z_2$ stabilizers satisfied, and then perform the $S_3$ vertex measurements. 
Since the $\Z_2$ stabilizers corresponding to $A_v^s$ remain satisfied, the $B$ charge is not produced. 
However, a $C$ charge can be produced, as long as its internal state transforms trivially when restricted to $\Z_2$ (otherwise, there would be an $A_v^s$ stabilizer violation).
In the case where we have a single $C$ charge at some vertex $v$, the error \textit{must} be in the state $\ket{1_C} +\ket{2_C}$. 
The TQFT interpretation of this is that the $C$ anyon has a single condensation channel at the $\Z_2|S_3$ interface, corresponding to the internal state $\ket{1_C} +\ket{2_C}$. 

\medskip \noindent {\bf Fusing and splitting of charges:} Now, let us imagine that a qubit experiences a $Z$ error on some edge $e$ connecting the vertices $v,v'$, where $v$ carries the above $C$ charge.
If the system with errors is in the state $\ket{\Psi}$, then within the subspace spanned by $A_v^g \ket{\Psi}$ and $A_{v'}^g \ket{\Psi}$, the new internal state can be written as a linear combination of basis states $\ket{i_R} \otimes \ket{j_{R'}}$, where $R,R'$ are the charge types at $v,v'$. Initially, the system is in the state
\begin{equation}
    (\ket{1_C} +\ket{2_C}) \otimes \ket{1_A}~.
\end{equation}
Now the $Z$ error anticommutes with $A_v^s$, and therefore applies a $Z$ operation on the basis $\{\ket{j}_C\}$, which changes the internal state of the error at $v$ to $\ket{1_C} -\ket{2_C}$. Additionally, the error creates a $B$ charge at $v'$, since it also anticommutes with $A_{v'}^s$. The state after the $Z$ error can therefore be written as 
\begin{equation}
    (\ket{1_C} -\ket{2_C}) \otimes \ket{1_B}.
\end{equation}
If we now move the $C$ error at $v$ to the $\Z_2|S_3$ interface, it cannot condense, but instead turns into an $e$ error within the surface code. One resolution is to also move the $B$ error at $v'$ across the interface, where it will turn into a second $e$ error that can be annihilated with the first.
Of course, one could also fuse these charged errors within the $S_3$ GSC and then condense the resulting $C$ charge with the internal state $(\ket{1_C}+\ket{2_C}) $ on the $\Z_2|S_3$ interface.

\medskip \noindent {\bf Correcting charges:} We see that information about the internal state can be leveraged to condense $C$ charges at the $\Z_2|S_3$ interface.
Firstly, we measure the $C$ charge in the basis $\{\ket+_C\,, \ket{-_{C}}\}$, where $\ket{\pm_C}:= \ket{1_C} \pm \ket{2_C}$. 
Whenever we measure a $C$ charge in the internal state $\ket{+_C}$, we can condense it on the interface to $\Z_2$ by acting with a charge transport operator. 
On the other hand, when we measure $C$ in the internal state $\ket{-_C}$, we first convert it into $\ket{+_C} $ by pairing with a $B$ charge and then move it to the  interface to $\Z_2$ GSC.
The overall error is in a $\Z_2$ trivial internal state, therefore the remaining $B$ charges can always be annihilated. 
\end{tcbexample}

Another scenario involves charges on vertices of the $H$-GSC prior to performing the $G$ vertex measurements. These charges are then extended into the $G$-GSC. In this case, the initial $H$ charges must fuse to the identity. Therefore, although they get converted into $G$ charges after vertex measurements, the $G$ charges still have a fusion channel to the identity irrep \textit{when restricted to $H$}. We make this statement precise in Appendix~\ref{app: ResInd}. This implies that the remaining $G$ charges can also be condensed on the $H|G$ interface.

As the $G$-GSC is grown, the adjoining $H$ and $K$ patches are shrunk from the sides by sequentially measuring qudits on the outer rough boundaries in the group basis. Gauge transformations are then applied to map the measured qudits to $|1\rangle$ state. The goal of this step is to push any nontrivial holonomy on these patches into the $G$ patch. Any fluxes that appear in this step can be corrected in software (i.e. without physically moving fluxes) since we only need to identify the gauge transformations $A_v^g$ to apply on vertices $v$ located at the $H|G$ and $G|K$ interfaces. In the spacetime picture, this procedure is represented by rough boundaries that slope inward toward the central $G$-GSC block.

The action of the extension operation on logical states, described microscopically in Section~\ref{sec: elementary logical operations}, is transparent in the spacetime formulation.
The logical group basis states are represented by open membrane configurations assigning a nontrivial holonomy along a path connecting the rough boundaries. 
Membranes labeled by $h \in H$ and $k \in K$ in the $H$- and $K$-GSCs embed naturally as $H$- and $K$-subgroup membranes within the $G$-GSC. 
This embedding and the corresponding map on logical states is depicted by
\begin{eqs}
    \vcenter{\hbox{\includegraphics[scale=.23]{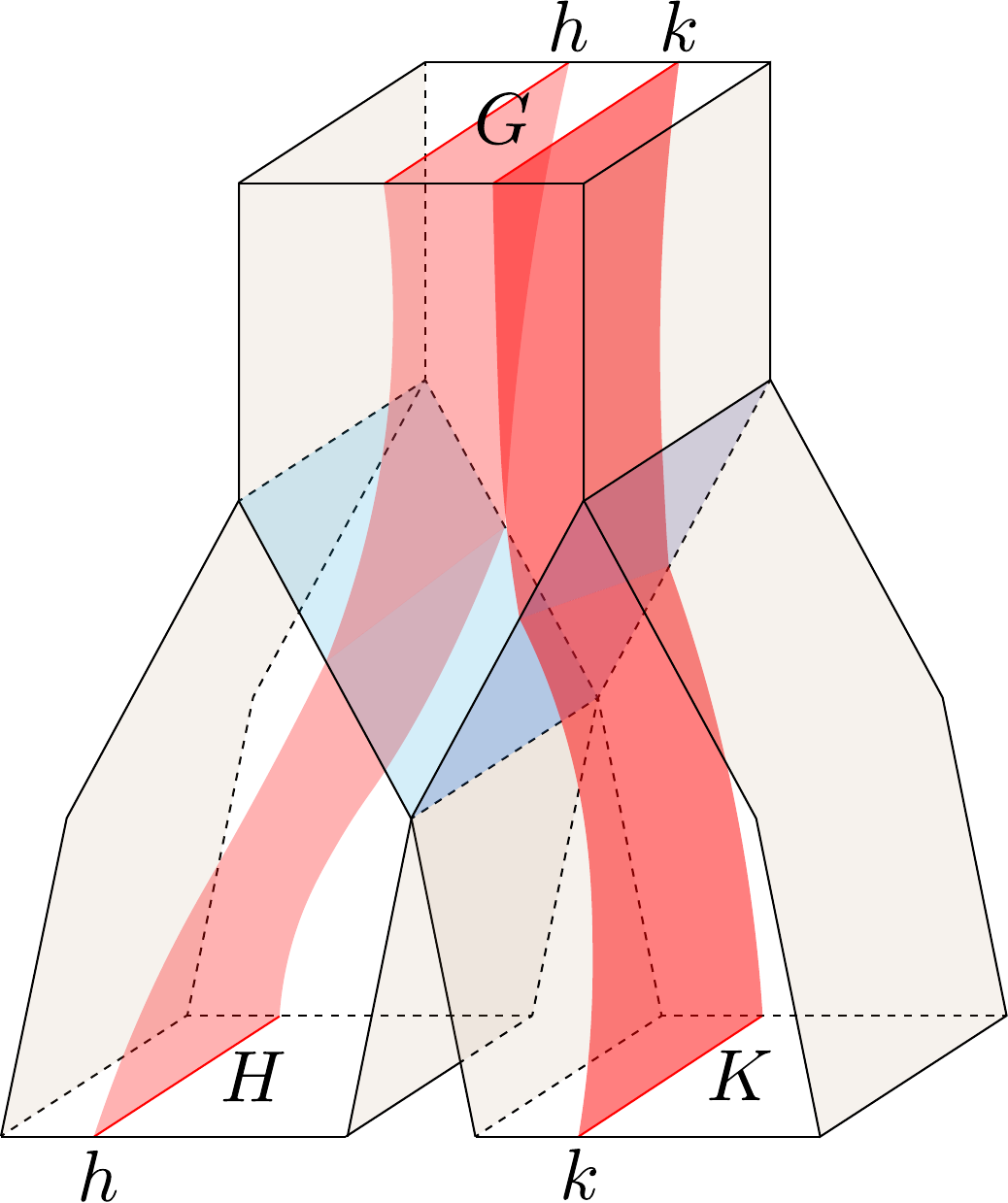}}} 
\end{eqs}
This reproduces the map on states given in Eq.~\eqref{eq: extension HK}.

\subsubsection{Splitting}

Next, we describe the spacetime logical block that corresponds to splitting a $G=H\knit K$ GSC into an $H$- and $K$-GSC. This process is depicted as 
\begin{equation}
\begin{split}
\includegraphics[height=12em]{Figures/splitHK2.pdf} 
\end{split}.
\end{equation}
The first step of splitting is to initialize $H$ and $K$ qudits in their respective $\ket{1}$ states to the left and right of the $G$ patch, and perform $H$- and $K$- vertex measurements,  respectively. The $H$ and $K$ charges arising in this step need to be transported to the outer rough boundaries, where they can be condensed. 

In the second step, all the qudits in the $G$-GSC are measured in the group basis. These measurements may produce fluxes, which are cleaned up in software, just like in the extension step. The crucial point is to apply suitable left and right multiplication to the degrees of freedom at the $H|G$ and $G|K$ interfaces so that the $G$ holonomy of the state in the original $G$ patch is transferred to the $H$- and $K$-GSCs.

The action of the splitting on states with definite $G$ holonomy can be immediately read off from the action on group membranes appearing in the TQFT partition function:
\begin{align}
\vcenter{\hbox{\includegraphics[scale=.2]{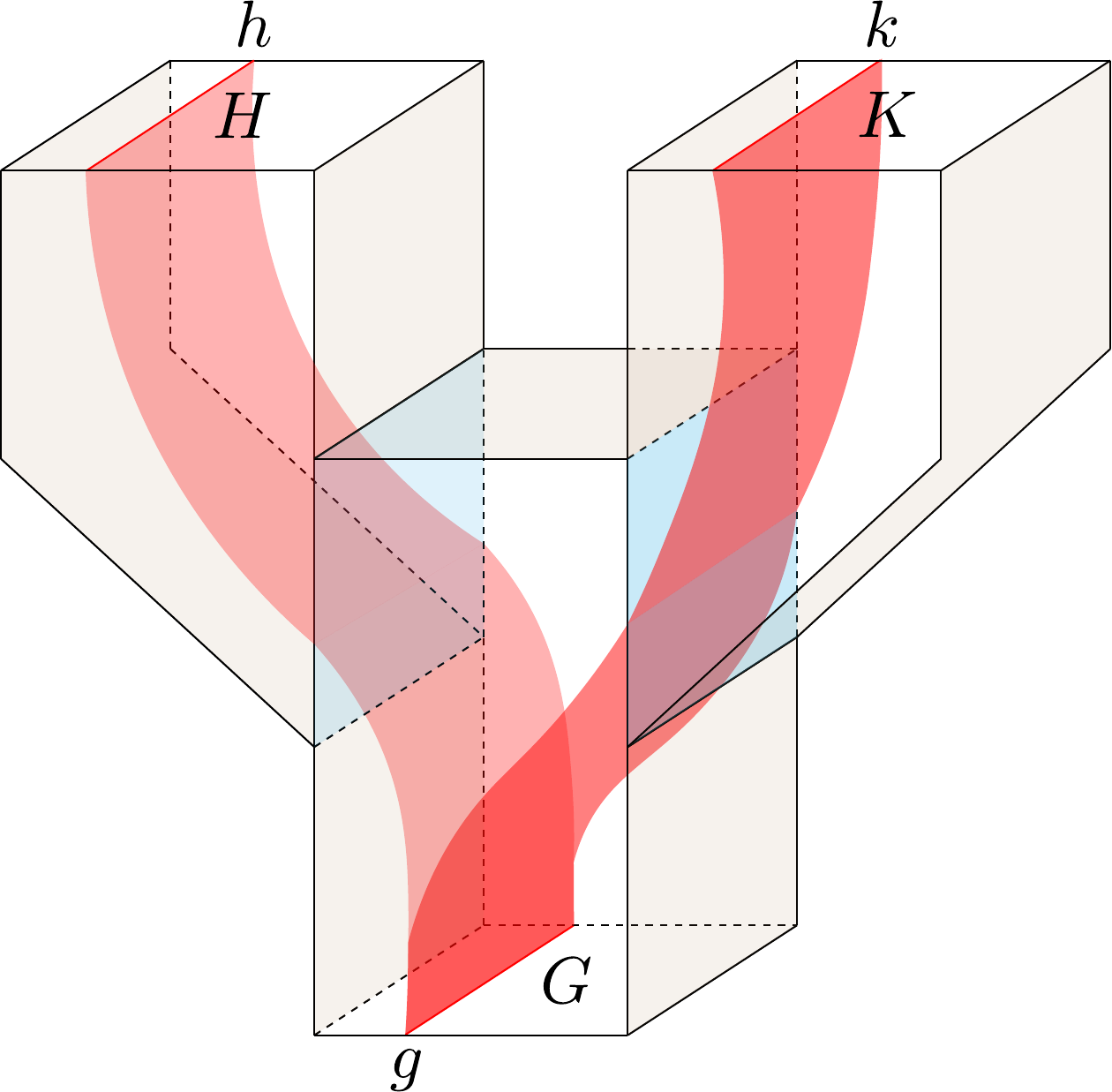}}} ,  
\end{align}
and for splitting in the other direction we have
\begin{align}
\vcenter{\hbox{\includegraphics[scale=.2]{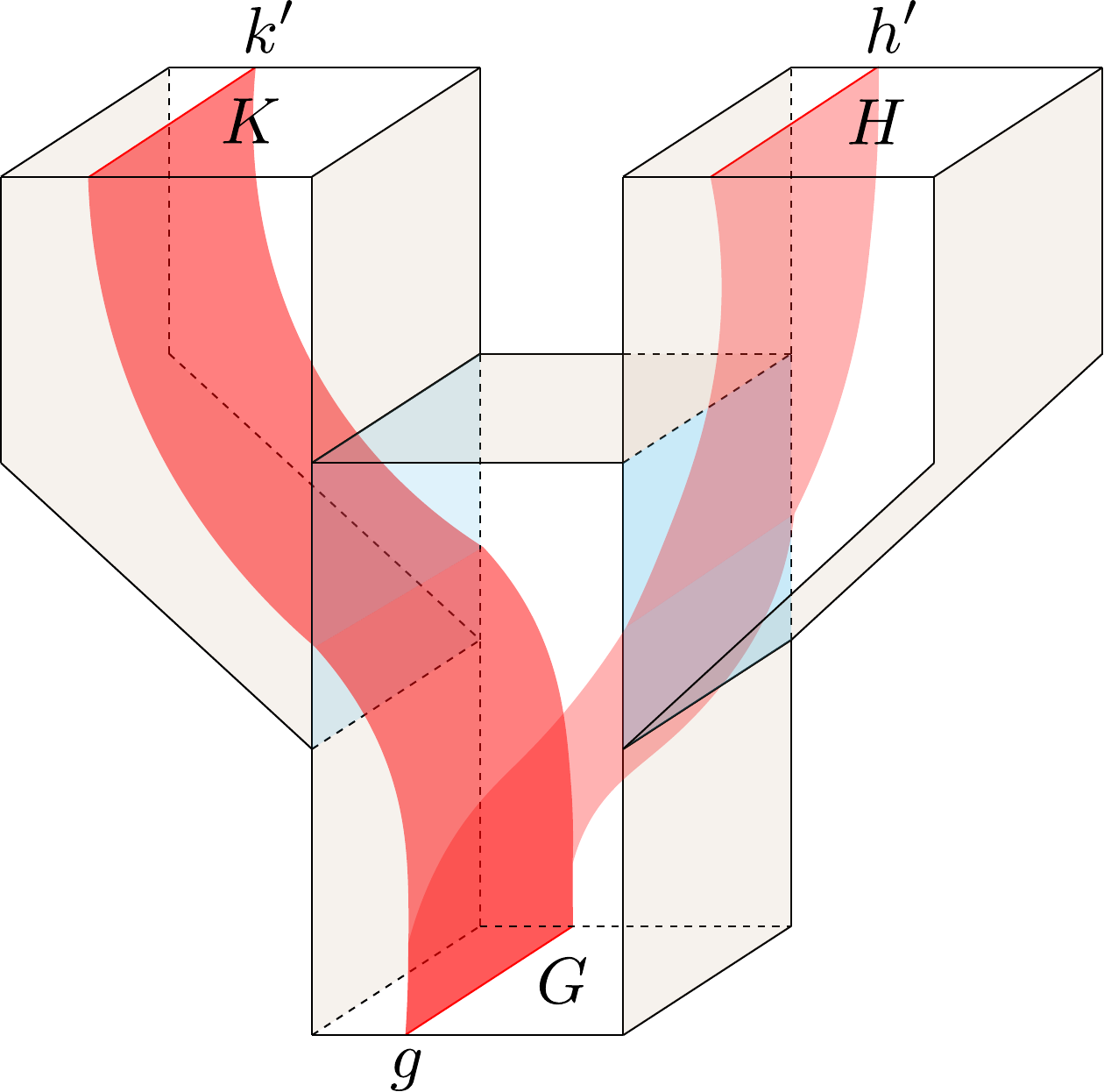}}},
\end{align}
where $h'$ and $k'$ are defined by $hk=k'h'$.

\subsection{Preparation and readout} \label{sec: preparation and readout spacetime}

\subsubsection{State preparation}

Recall from Section~\ref{sec: elementary logical operations} that the procedure to prepare the $\ket{1}_{\LL}$ state of a $G$ GSC begins by initializing a column of $G$-qudits in the $|1\rangle$ state. We then perform vertex measurements to generate a narrow strip of $G$-GSC. These two steps, namely adding ancilla and performing vertex measurements, are repeated to grow the $G$-GSC horizontally.
The spacetime diagram for this process is represented as:
\begin{equation}\label{eq:preparation}
\begin{split}
\includegraphics[scale=.2]{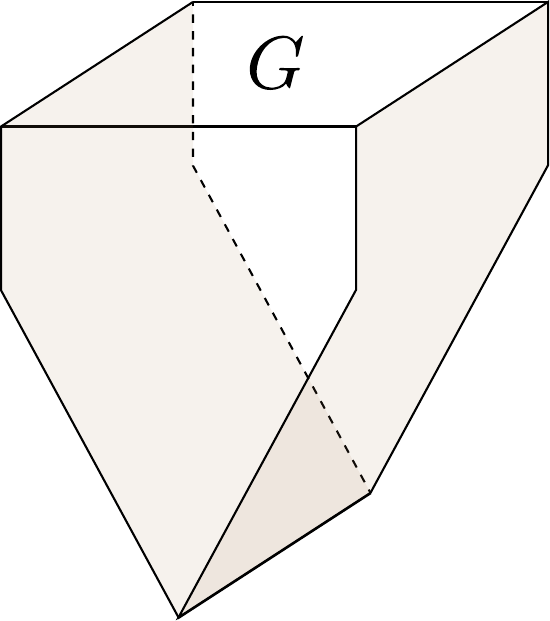}   
\end{split}
\end{equation}
The bottom of the diagram hosts a rough boundary, which corresponds to the fact that charges appear probabilistically from the vertex measurements. 

The $G$ charges detected by the vertex measurements are cleaned up by transporting them to the nearest rough boundary, where they can be condensed. 
Because the movement operators of charges require a circuit whose depth grows linearly with the distance they are moved, the preparation is organized sequentially in time, starting from a central column and expanding outward. This ensures that the charges typically only have to be moved a short distance to the nearest rough boundary.
We note that any other group basis state can then be prepared by simply acting with a $L^g_{\LL}$ or $R^g_{\LL}$, creating a membrane that ends on the rough boundary, as discussed above.

To prepare the logical $\ket{+}_\mathsf{L}$ state, we first prepare a patch with three smooth boundaries. This patch is grown along the rough boundary, so the charges appearing from the vertex measurements can be condensed on the nearby rough boundary. Once the patch is grown, we introduce a second rough boundary. We generically find fluxes at the newly introduced rough boundary. These can be moved to either of the smooth boundaries in linear time. The preparation is depicted in spacetime as:
\begin{equation}\label{eq:preparationplus}
\begin{split}
\includegraphics[scale=.2]{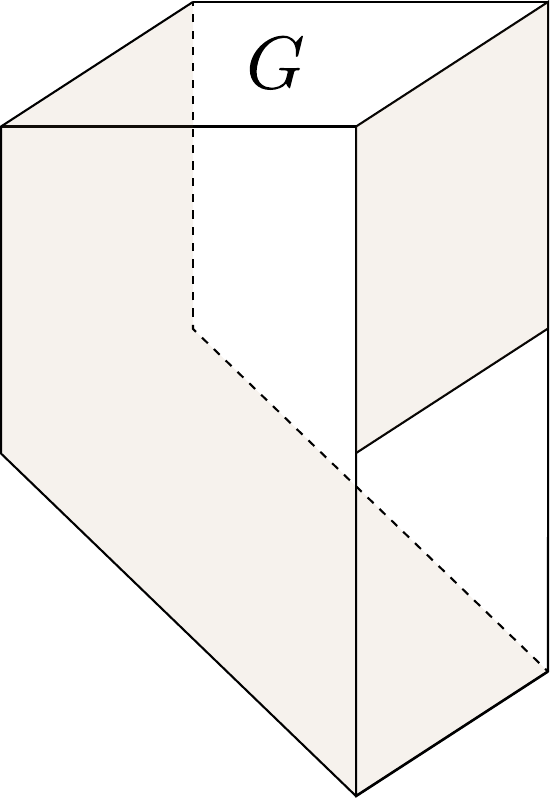}   
\end{split}
\end{equation}
This prepares the logical $\ket{+}_\mathsf{L}$ state, since the membranes are free to terminate on the smooth boundary on the bottom right.

\section{Putting it all together} \label{sec: examples}
In this section, we demonstrate several nontrivial applications of the spacetime logical blocks. This includes \textbf{(i)} performing non-Clifford gates through sliding GSCs, \textbf{(ii)} a scheme for magic state preparation, \textbf{(iii)} a transversal implementation of multi-controlled $X$ gates, \textbf{(iv)} a generalization to arbitrary classical reversible gates, and \textbf{(v)} a modification of GSCs, referred to as `coset surface codes', which are directly motivated by the spacetime logical blocks.

\subsection{Sliding}\label{sec:Egs-sliding}
Let us begin by considering the operation of sliding, introduced in Example~\ref{ex: extending and splitting D4}. In general, an $H$-GSC and $K$-GSC are extended to a $G=H\knit K$ GSC, then split back into a $K$- and $H$-GSC with their positions swapped. This allows for performing certain controlled gates, as described below.

We note that Ref.~\cite{davydova2025D4} previously showed how to realize a logical $\mathrm{CCZ}$ gate on three surface codes by sliding two of the surface codes across the third, with the intermediate code realizing a $\Z_2^3$ twisted quantum double model (topologically equivalent to the $D_4$ GSC). Ref.~\cite{sajith2025noncliffordS3} also showed that a logical controlled charge conjugation can be performed on $\Z_2$ and $\Z_3$ surface codes by sliding them across each other with an extension into an intermediate $S_3 = \Z_3 \rtimes \Z_2$ GSC.

The protocol starts with an $H$-GSC on the left and a $K$-GSC on the right. We then apply the two spacetime logical blocks shown below: 
\begin{equation}
\begin{split}
    \includegraphics[scale=0.35]{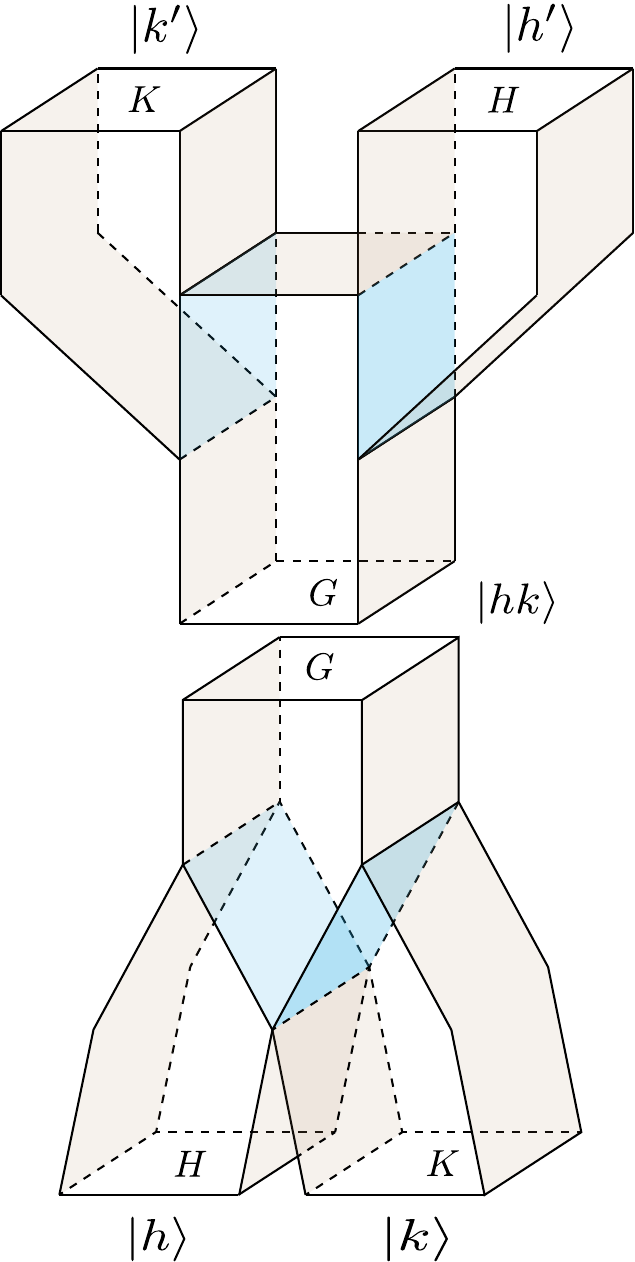}    
\end{split}.
\end{equation}
Here, we have assumed that the $H$- and $K$-GSCs are initialized in the logical basis states $\ket{h}_\mathsf{L}$ and $\ket{k}_\mathsf{L}$. Reading the diagram from bottom to top: \\

\noindent \textbf{(1)} We first perform $\mathrm{Extend}_{HK}$. This implements the logical operation:
        \begin{equation}
            \mathrm{Extend}_{HK}: \ket{h}_{\mathsf{L}} \otimes \ket{k}_{\mathsf{L}} \rightarrow \ket{hk}_{\mathsf{L}}.
        \end{equation} \\

\noindent \textbf{(2)}  We next perform $\mathrm{Split}_{KH}$. We emphasize that the $K$-GSC is now to the left of the $H$-GSC. The logical state transforms as:
    \begin{align}
            \mathrm{Split}_{KH}: \ket{hk}_{\mathsf{L}} &\rightarrow \ket{h'}_{\mathsf{L}} \otimes \ket{k'}_{\mathsf{L}},
        \end{align}
        where  $h' \in H$ and $k' \in K$ satisfy $hk = k' h'$. Note that for semidirect products, with $G = H \rtimes K$, we have $h' = \bar{k} h k$ and $k' = k$. In this case, the logical action can be interpreted as a controlled $K$-conjugation action on the $H$ logical qudit. In the more general case of knit products, each subgroup is transformed by a controlled action of the other. \\      

We recover Example~\ref{ex: extending and splitting D4} by taking $H$, $K$, and $G$ to be $\Z_2^a\times \Z_2^b$, $\Z_2^c$, and $D_4$, respectively. This differs from the sliding in Ref.~\cite{davydova2025D4} by rotating the $\Z_2^b$-GSC by an angle 90 degrees and applying a Hadamard to every site. Equivalently, the operations can be related to each other by mapping the $D_4$ GSC to the $\Z_2^3$ twisted quantum double model, as described in Appendix~\ref{app: representation on qubits}. The sliding described in Ref.~\cite{sajith2025noncliffordS3} can be obtained directly by taking $H=\Z_3$, $K=\Z_2$, and $G=H \rtimes K$. 

\subsection{Magic state preparation}\label{sec:Egs-magicstate}

Next, we show how the spacetime logical blocks can be used to prepare magic states. Our constructions are inspired by the magic-state preparation protocols recently described in Refs.~\cite{davydova2025D4,huang2025D4}.  In particular, Ref.~\cite{davydova2025D4} prepares a logical $\mathrm{CZ}$ state in a $\Z_2\times \Z_2$ surface code using a $\Z_2^3$ twisted quantum double model.
Ref.~\cite{huang2025D4} prepares a logical $T$ state in a $\Z_2$ surface code by initializing a $\Z_4$ surface code in a certain state, switching into a $D_4$ quantum double model (with specific boundary conditions), then switching into a $\Z_2$ surface code. 
Our examples are inspired by these constructions and employ the $D_4$ GSC. We expect that further examples can be developed by considering more general $G$-GSCs. 

\subsubsection{Preparing a $\mathrm{CX}$ state}

We first describe the preparation of a logical $\mathrm{CX}$ state, as introduced in Example~\ref{ex: preparation and readout in D4}, by recasting it in the language of spacetime logical blocks. The spacetime representation is: 
\begin{equation}
\begin{split}
        \includegraphics[scale=0.35]{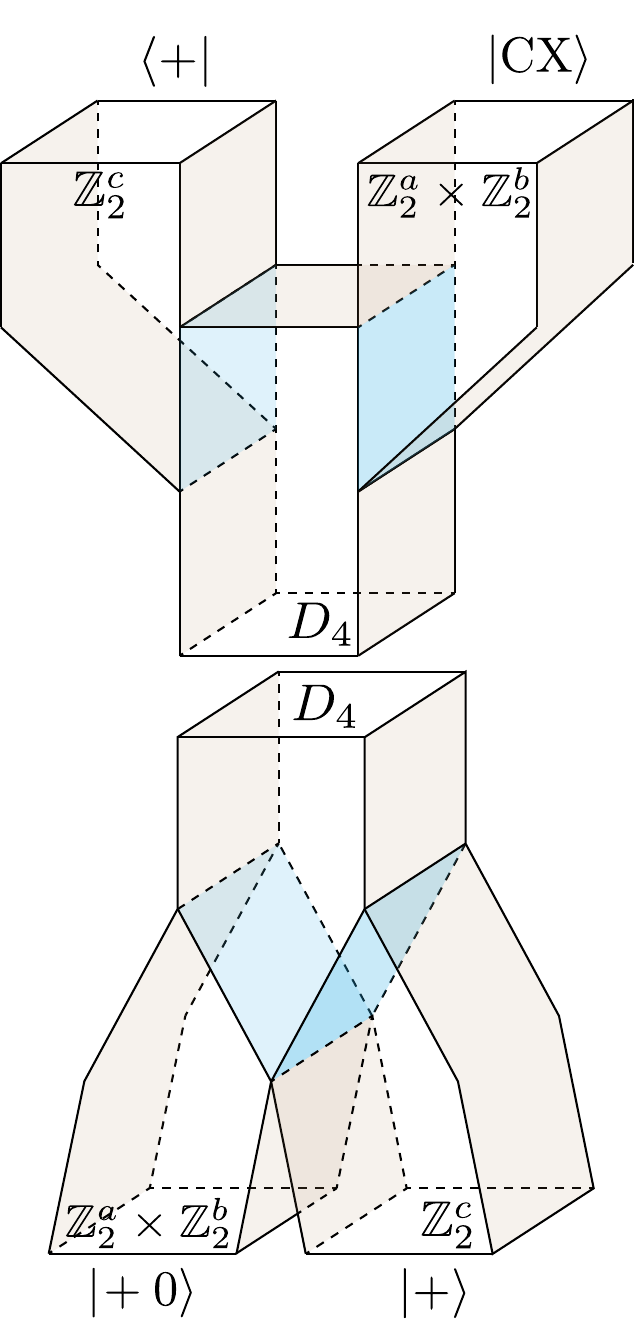}
\end{split}.
\end{equation}
To save space in the diagram, we have not included the logical blocks used for preparation. The protocol is as follows: \\

\noindent \textbf{(1)} We begin by initializing three surface codes in the state $|\!+0+\rangle_\mathsf{L}$. We label these as $\Z_2^a$, $\Z_2^b$, and $\Z_2^c$ GSCs, respectively. \\

\noindent \textbf{(2)} We slide the $\Z_2^a \times \Z_2^b$ GSC across the $\Z_2^c$ GSC. This produces the state:
    \begin{multline}
    \frac{1}{2\sqrt{2}}\left(2\ket{00}_\mathsf{L}+\ket{10}_\mathsf{L}+\ket{11}_\mathsf{L}\right)\otimes \ket{+}_\mathsf{L} \\
    +\frac{1}{2\sqrt{2}}\left(\ket{10}_\mathsf{L}-\ket{11}_\mathsf{L}\right)\otimes \ket{-}_\mathsf{L}.
    \end{multline} \\

\noindent \textbf{(3)}  We read out the $\Z_2^c$ GSC in the logical Pauli $X$ basis and postselect on a $+1$ outcome. This gives us the $\mathrm{CX}$ state:
    \begin{align}
        |\mathrm{CX}\rangle = \frac{1}{\sqrt{6}}\left( 2\ket{00}_\mathsf{L}+\ket{10}_\mathsf{L}+\ket{11}_\mathsf{L} \right).
    \end{align}

\subsubsection{Preparing a $\mathrm{T}$ state}
Before discussing the protocol, we first introduce the following alternative presentation of $D_4$:
\begin{equation}
    D_4 = \langle r, s\,|\, r^4 = s^2 = 1, s r s = r^3\rangle. 
\end{equation}
Here, $r$ generates the group $\Z_4^r$. The element $s$ generates the subgroup $ \Z_2^s$, and we have $D_4 = \Z_4^r \rtimes \Z_2^s$. 
The generators $r,s$ can be expressed in terms of $a,b,c$ as follows:
\begin{equation}
    a = s\,, \quad  b = r^2\,,\quad  c = sr.
\end{equation}
Note that the usual $a,b,c$ presentation realizes $D_4$ as a different semidirect product: $D_4 = (\Z_2^{a} \times \Z_2^{b}) \rtimes \Z_2^{c}.$ 

To prepare a T state, we combine an extension in the $r,s$ presentation with a split in the $a,b,c$ presentation. The corresponding spacetime logical blocks are shown below, with $1$ denoting the identity element:
\begin{equation}
\begin{split}
        \includegraphics[scale=0.35]{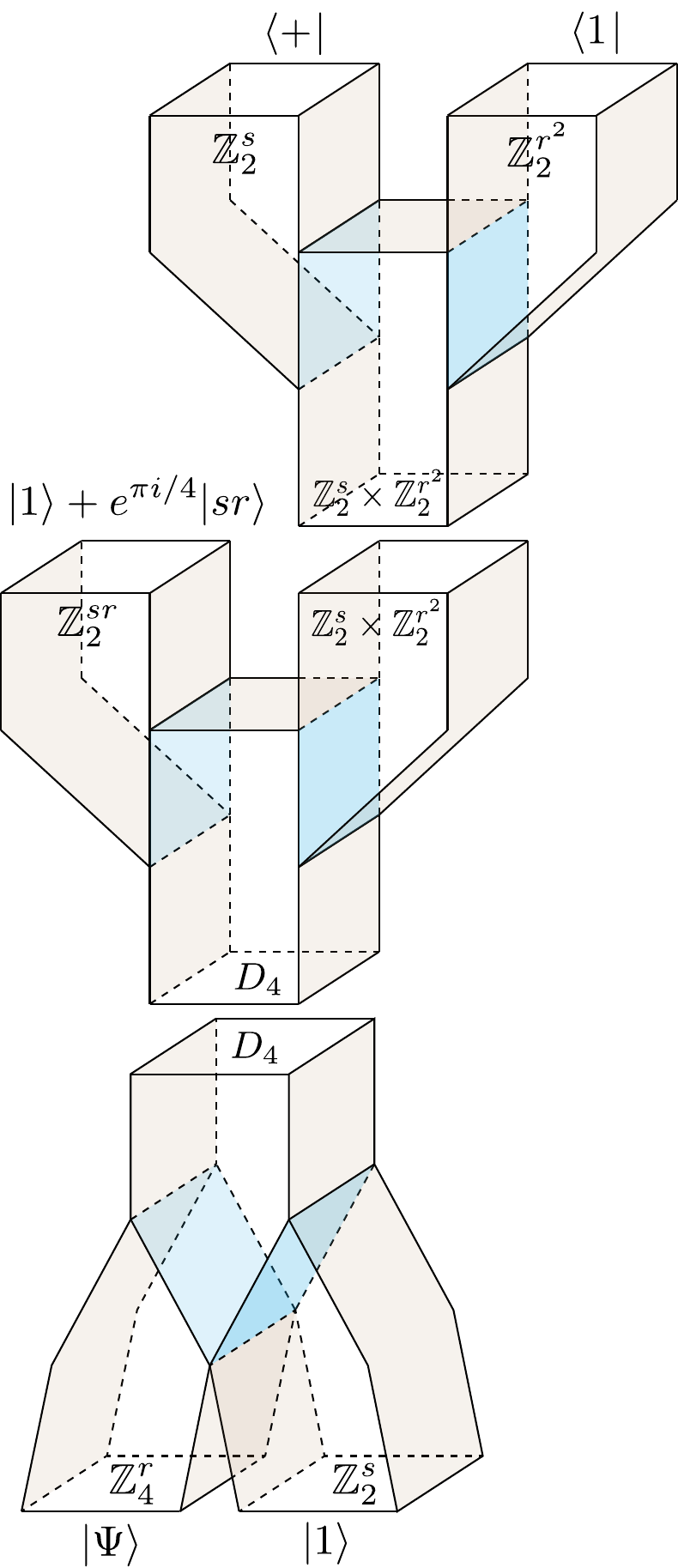}
\end{split}.
\end{equation}
In words, the protocol involves the following steps:\\

\noindent \textbf{(1)} We initialize a $\Z_4^r$ GSC in the stabilizer state
    \begin{eqs}
            \ket{\Psi}_\LL=\ket{1}_{\LL} &+ e^{i\pi/4} \ket{r}_{\LL} \\ &- \ket{r^2}_{\LL} + e^{i \pi/4} \ket{r^3}_{\LL},
    \end{eqs}
    which can be prepared with a fold-transversal gate, as noted in Ref.~\cite{huang2025D4}. 
    We also initialize a $\Z_2^s$ GSC in the state $\ket{1}_{\LL}$, where $1$ denotes the identity element of the group. \\

\noindent \textbf{(2)} We extend the $\Z_4^r$ and $\Z_2^s$ GSCs to a $D_4$ GSC. The logical state is $\ket{\Psi}_\LL$ with $r$ now treated as an element of $D_4$. \\

\noindent \textbf{(3)} We split the $D_4$ GSC into a $\Z_2^{sr}$ GSC on the left and a $\Z_2^{s} \times \Z_2^{r^2}$ GSC on the right. The logical state $\ket{\Psi}_\LL$ can be written using the elements of $\Z_2^{sr}$ and $\Z_2^{s} \times \Z_2^{r^2}$ as:
\begin{multline}
    \ket{\Psi}_\LL = 
     |1\rangle_\LL |1\rangle_\LL + e^{i \pi /4} |sr\rangle_\LL |s r^2\rangle_\LL \\
    -|1\rangle_\LL |r^2 \rangle_\LL + e^{i \pi /4} |sr\rangle_\LL |s\rangle_\LL.
\end{multline}\\

\noindent \textbf{(4)} We next split the $\Z_2^{s} \times \Z_2^{r^2}$ GSC into a $\Z_2^{s}$ GSC on the left and a $\Z_2^{r^2}$ GSC on the right. The logical state becomes:
\begin{multline}
    \ket{\Psi}_\LL = 
     |1\rangle_\LL |1\rangle_\LL |1\rangle_\LL + e^{i \pi /4} |sr\rangle_\LL |s \rangle_\LL |r^2\rangle_\LL \\
    -|1\rangle_\LL |1\rangle_\LL |r^2 \rangle_\LL + e^{i \pi /4} |sr\rangle_\LL |s\rangle_\LL |1\rangle_\LL.
\end{multline}
In anticipation of the subsequent step, we write the state in the $X$ basis on the $\Z_2^s$ GSC. This gives us:
\begin{eqs} \label{eq: T magic state}
    \ket{\Psi}_\LL = 
     &\frac{1}{\sqrt{2}}(|1\rangle_\LL + e^{\pi i/4}|sr\rangle_\LL)|+\rangle|1\rangle\\
    +&\frac{1}{\sqrt{2}}(|1\rangle_\LL - e^{\pi i/4}|sr\rangle_\LL)|-\rangle|1\rangle\\
    -&\frac{1}{\sqrt{2}}(|1\rangle_\LL - e^{\pi i/4}|sr\rangle_\LL)|+\rangle|r^2\rangle\\
    -&\frac{1}{\sqrt{2}}(|1\rangle_\LL + e^{\pi i/4}|sr\rangle_\LL)|-\rangle|r^2\rangle.
\end{eqs}
\\

\noindent \textbf{(5)} Finally, we read out the $\Z_2^s$ GSC in the $X$ basis and the $\Z_2^{r^2}$ GSC in the $Z$ basis. From Eq.~\eqref{eq: T magic state}, a measurement outcome of $|+\rangle|1\rangle$ or $|-\rangle|r^2\rangle$ results in a T state on the $\Z_2^{sr}$ GSC. If one of the other two measurement outcomes is obtained, then one can apply a Pauli $Z$ to map to a T state. We remark that a measurement in the $X$ basis is outside of our framework for GSCs, however, this is a well established operation for $\Z_2$ GSCs. \\

We note that this procedure differs slightly from the one presented in Ref.~\cite{huang2025D4}. In particular, a different set of boundary conditions was used for the $D_4$ quantum double in Ref.~\cite{huang2025D4}. 

\subsection{Multi-controlled $X$ gates} \label{sec:Egs-CCX}

\subsubsection{CCX}
In Example~\ref{ex: transversal gates D4}, we demonstrated that the logical left multiplication in a $D_4$ GSC implements a logical $\mathrm{CX}$ gate. Inspired by this connection, we consider a GSC based on a group $G=G_\mathrm{CCX}$ and show how left multiplication can implement a logical $\mathrm{CCX}$ gate. Here, $G_\mathrm{CCX}$ is the group generated by $X_1$, $X_2$, $X_3$, and $\mathrm{CCX}_{123}$ on three qubits. An over-complete but enlightening set of generators is given by:
\begin{align}
    G_\mathrm{CCX}=\langle X_1, X_2, X_3, \mathrm{CX}_{13}, \mathrm{CX}_{23}, \mathrm{CCX}_{123} \rangle.
\end{align}
Note that $\rm CX_{13}$ and $\mathrm{CX}_{23}$ appear when conjugating $X_1$ and $X_2$ by $\mathrm{CCX}_{123}$. This set of generators makes it clear that $G_\mathrm{CCX}$ can be expressed as:
\begin{align}\label{eq:GCCX}
    G_\mathrm{CCX} &= \bm{(}(\Z_2\times \Z_2 \times \Z_2)\rtimes (\Z_2 \times \Z_2)\bm{)}\rtimes \Z_2 \\
    &= (H \rtimes K_1) \rtimes K_2,
\end{align}
where the first three factors are generated by $X_1$, $X_2$, and $X_3$, the second two are generated by $\mathrm{CX}_{13}$ and $\mathrm{CX}_{23}$, and the final factor is generated by $\mathrm{CCX}_{123}$.\footnote{We remark that $G_\mathrm{CCX}$ is a 2-group. Given its explicit decomposition using semidirect products, this implies that $G_\mathrm{CCX}$ is both solvable and nilpotent. The associated anyon theory is likewise solvable and nilpotent.}

To avoid confusing the group elements of $G_\mathrm{CCX}$ with an action on the physical Hilbert or logical space, we label the group elements as
\begin{align*}
    a&=X_1, & d&=\mathrm{CX}_{13}, \\
    b&=X_2, & e&=\mathrm{CX}_{23}, \\
    c&=X_3, & f&=\mathrm{CCX}_{123}.
\end{align*}

Again, we provide an explicit representation of the $G_\mathrm{CCX}$ GSC on qubits in Appendix~\ref{app: representation on qubits}. Here, we consider only the code space and logical operators. The code space has dimension $|G_\mathrm{CCX}|=2^6$, and a generic element $g$ of $G_\mathrm{CCX}$ can be written as
\begin{align}
    g=a^\alpha b^\beta c^\gamma d^\delta e^\epsilon f^\eta,
\end{align}
with $\alpha,\beta,\gamma,\delta,\epsilon,\eta \in \{0,1\}$. Therefore, a logical basis state can be labeled as $|\alpha,\beta,\gamma,\delta,\epsilon,\eta \rangle$.

Similar to the case of $D_4$, the logical operators $L^a_\mathsf{L}$, $L^b_\mathsf{L}$, and $L^c_\mathsf{L}$ act as Pauli $X$ operators on the first three qubits:
\begin{eqs}
   L^a_\mathsf{L} |\alpha,\beta,\gamma,\delta,\epsilon,\eta \rangle &= |\alpha+1,\beta,\gamma,\delta,\epsilon,\eta \rangle, \\
   L^b_\mathsf{L} |\alpha,\beta,\gamma,\delta,\epsilon,\eta \rangle &= |\alpha,\beta+1,\gamma,\delta,\epsilon,\eta \rangle, \\
   L^c_\mathsf{L} |\alpha,\beta,\gamma,\delta,\epsilon,\eta \rangle &= |\alpha,\beta,\gamma+1,\delta,\epsilon,\eta \rangle.
\end{eqs}
Likewise, $L^d_\mathsf{L}R^d_\mathsf{L}$ and $L^e_\mathsf{L}R^e_\mathsf{L}$ act as $\mathrm{CX}$ operators on the first three qubits:
\begin{eqs}
    L^d_\mathsf{L}R^d_\mathsf{L} |\alpha,\beta,\gamma,\delta,\epsilon,\eta \rangle &= |\alpha,\beta,\gamma+\alpha,\delta,\epsilon,\eta \rangle, \\
    L^e_\mathsf{L}R^e_\mathsf{L} |\alpha,\beta,\gamma,\delta,\epsilon,\eta \rangle &= |\alpha,\beta,\gamma+\beta,\delta,\epsilon,\eta \rangle.
\end{eqs}
Following this trend, $L^f_\mathsf{L}R^f_\mathsf{L}$ applies a (transversal) $\mathrm{CCX}$ to the first three qubits. However, in addition to the $\mathrm{CCX}$, it entangles the first three qubits with the second two:
\begin{align}
    L^f_\mathsf{L}R^f_\mathsf{L}|\alpha,\beta,\gamma,\delta,\epsilon,\eta \rangle = |\alpha,\beta,\gamma+\alpha\beta,\delta+\beta,\epsilon+\alpha,\eta \rangle.
\end{align}
This is problematic if we imagine starting with three surface code qubits (labeled by $a,b,c$) and extending to a $G_{\mathrm{CCX}}$ GSC by adding three ancillas per edge and measuring new stabilizers. In this case, after we perform the transversal $L^f_{\mathsf{L}}R^f_{\mathsf{L}}$ operation, the logical information gets entangled with the ancillas, and will be destroyed once we split the $G_{\mathrm{CCX}}$ GSC to return to surface code patches. 

However, this problem is resolved if we initialize the $G_{\mathrm{CCX}}$ code in a $\ket{+}$ state of the $d,e$ and $f$ qubits. This is shown below, where we define $H = \Z_2^a \times \Z_2^b \times \Z_2^c$, $K_1 = \Z_2^d \times \Z_2^e$ and $K_2 = \Z_2^f$:
\begin{equation}\label{eq:GCCX-fig}
    \vcenter{\includegraphics[scale=0.35]{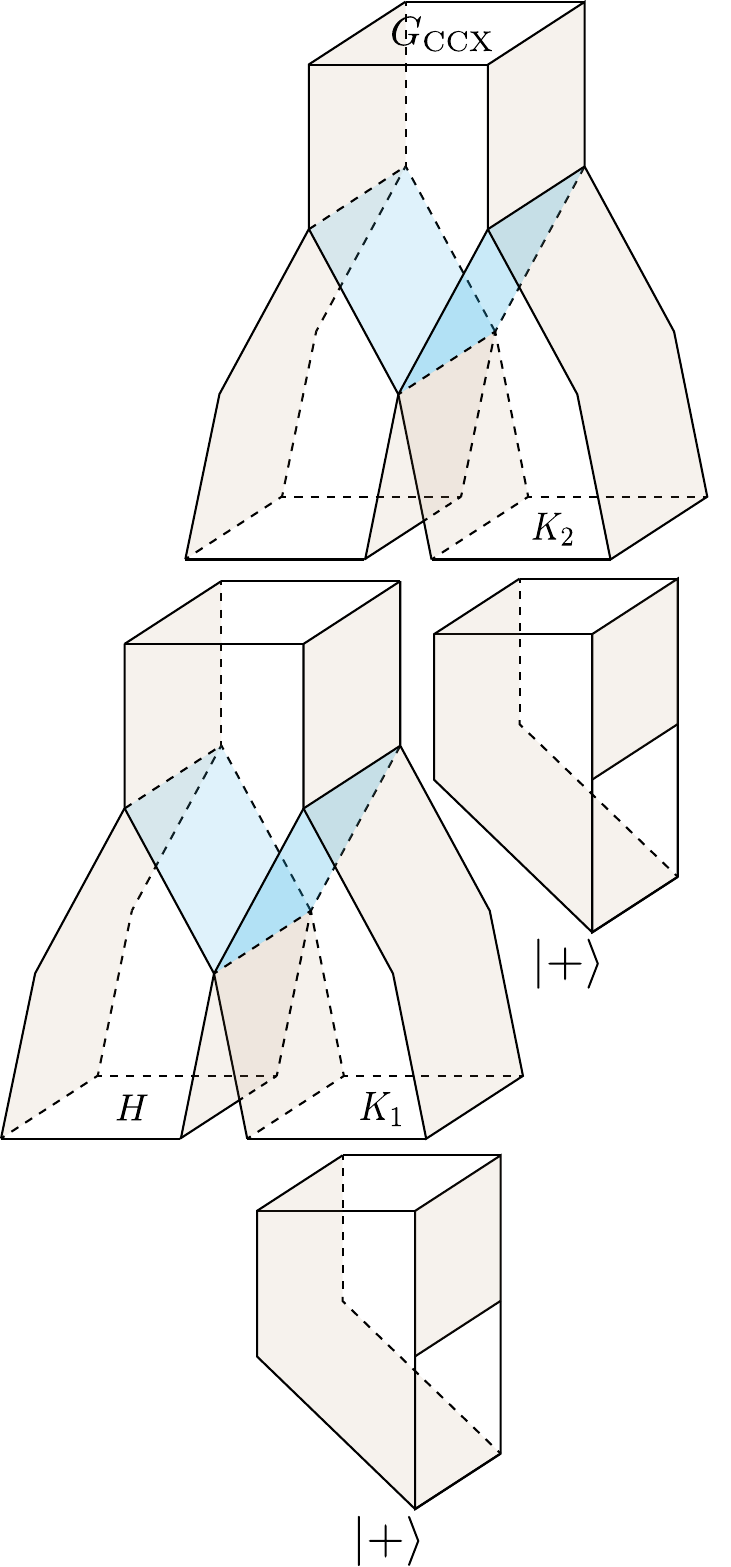}}.
\end{equation}
With this restriction, the remaining logical states are now labeled by the three qubits $\ket{\alpha,\beta,\gamma}_{\LL}$. The desired non-Clifford gate can be implemented through the logical left action:
\begin{equation}
    L_{\LL}^f\ket{\alpha,\beta,\gamma}_{\LL} = \ket{\alpha,\beta,\gamma + \alpha \beta}_{\LL},
\end{equation}
which leaves the $\ket{+}_{\LL}$ states of the $d,e,f$ qubits invariant. In our construction of $G_{\mathrm{CCX}}$, it is convenient that $\mathrm{CX}_{13},\mathrm{CX}_{23}$ and $\mathrm{CCX}_{123}$ all commute. This allows to perform the code extension from three surface code patches to a $G_{\mathrm{CCX}}$ patch in two steps as shown above, where $K_1 = \Z_2^d \times \Z_2^e$ and $K_2 = \Z_2^f$ are both Abelian groups. Note that this extension could also be performed in a single step.

\subsubsection{$\mathrm{C}^n\mathrm{X}$}

The natural generalization of $G_{\mathrm{CCX}}$ involves a group $G_{C^n X}$, which is generated by Pauli $X$ gates on $n+1$ qubits, and a multi-controlled $X$ gate with the first $n$ qubits as the control and the $n+1$ qubit as the target:
\begin{align}
    G_{C^n X} = \langle X_1, \dots, X_n, \mathrm{C^n X}_{n+1} \rangle,
\end{align}
where $\mathrm{C^n X}_{n+1} =\mathrm{CC \dots CX}_{1, \ldots n, n+1}$.
To understand the smallest group containing both $\mathrm{C^n X}_{n+1}$ and $X_1,\dots X_n$, we consider the following conjugation action (for $ \, i = 1,2, \dots , n)$:
\begin{align}
    \mathrm{C^n X}_{n+1} X_i \mathrm{C^n X}_{n+1} = X_i \mathrm{CX}_{\overline{i} ; n+1},
\end{align}
where the notation $\mathrm{C^n X}_{\overline{i}; n+1}$ means that $X_{n+1}$ is controlled on the first $n$ qubits {except} the $i$th. This shows that the group also includes at least the gates $\mathrm{CX}_{\overline{i}; n+1}$ for all $1 \leq i \leq n$, i.e., all controlled-$X$ gates with $n-1$ control qubits and target on the $n+1$ qubit. Similarly, conjugating $X_j$ by $\mathrm{CX}_{\overline{i}; n+1}$, with $i\neq j$ gives
\begin{align}
    \mathrm{CX}_{\overline{i}; n+1} X_j \mathrm{CX}_{\overline{i}; n+1} = X_j \mathrm{CX}_{\overline{i,j}; n+1}.
\end{align}
The notation $\overline{i,j}$ implies that the $X$ is controlled on the first $n$ qubits {except} the $i$th and $j$th, irrespective of their ordering, i.e., we need to include all controlled $X_{n+1}$ gates with $n-2$ control qubits. Iterating this conjugation action shows that the number of controlled gates with $k$ control qubits equals the binomial coefficient $\binom{n}{k}$ for each $1 \leq k \leq n-1$. As a result, the semidirect product structure is given by: 
\begin{align}
    G_{C^n X}=((\mathbb{Z}_2^{n+1} \rtimes \mathbb{Z}_2^{\binom{n}{1}}) \dots \rtimes \mathbb{Z}_2^{\binom{n}{n-1}})\rtimes \mathbb{Z}_2,
\end{align}
where each $\mathbb{Z}_2^{\binom{n}{k}}$ subgroup contains all controlled-$X$ operations with $k$ control qubits and target on qubit $n+1$. The examples with $G = \Z_2 \times \Z_2$, $D_4$ and $G_{\mathrm{CCX}}$ can be seen as special cases of the above formulation for $n=0,1,2$.

Due to the semidirect product structure of $G_{\mathrm{C}^n\mathrm{X}}$ we can straightforwardly compute the order of the group for any $n$:
\begin{align}
    \left | G_{C^n X} \right | &=2^{n + 1 + \sum_{i=1}^{n}\binom{n}{i}} = 2^{2^n +n}.
\end{align}
The local Hilbert space for $G_{\mathrm{C}^n\mathrm{X}}$ requires $2^n + n$ qubits per edge. Note that the group $G_{C^n X}$ is nilpotent (and hence solvable) for any $n$. 

Applying the same reasoning as in the example with $G_{\mathrm{CCX}}$, the gate $\mathrm{C^n X}_{n+1}$ can be performed on $n+1$ surface codes by switching to a $G_{\mathrm{C}^n\mathrm{X}}$ GSC, ensuring that all the logical qubits corresponding to non-Pauli gates are initialized in the $\ket{+}_{\LL}$ state. With this restriction, the ground states are labeled by elements of $\Z_2^{n+1}$. Then, performing a transversal $L_{\mathsf{L}}^{\mathrm{C}^n\mathrm{X}}$ achieves the desired logical gate.

The spacetime logical blocks used to switch into $G_{\mathrm{C}^n\mathrm{X}}$ also directly generalize those for $G_{\mathrm{CCX}}$, shown in Eq.~\eqref{eq:GCCX-fig}. It is convenient (although not strictly necessary) that all the multi-controlled gates generating this group commute with each other. Therefore, the code extension can be broken up into $n-1$ steps of the form $H \rtimes K_{i+1}$ for $i=2,3, \dots ,n$ in which $K_{i+1}$ is an Abelian group consisting only of gates of the form $\mathrm{C}^i\mathrm{X}$.

\subsection{Arbitrary reversible classical gates}\label{sec:group_engg}

The approach taken in Section~\ref{sec:Egs-CCX} can be further generalized to implement arbitrary reversible classical gates. Suppose we have $n$ patches of surface code, whose logical states are labeled in the computational basis as bit strings $\ket{\alpha_1, \alpha_2, \dots\alpha_{n}}_{\LL}$ with $\alpha_i = 0,1$. There are $2^n$ such logical bit strings, indexed from $0 \equiv 00 \dots 0$ to $2^n-1 \equiv 11 \dots 1$. Now an {arbitrary} reversible classical gate on $n$ qubits can be expressed as a permutation of the bit strings
\begin{eqs}
    \pi: [0,1,2,\dots , 2^n-1] \rightarrow [0,1,2, \dots , 2^n-1],
\end{eqs}
where $\pi \in S_{2^n}$. For instance, if the only action of $\pi$ is to swap $111 \dots 10$ with $111 \dots 11$, the corresponding logical operation is implemented by $\mathrm{C^{n-1} X}_n$. As another example, if $\pi$ takes $\alpha_1 \rightarrow \alpha_1 + 1$ mod 2 and leaves the other bits invariant, then logical operation is $X_1$. 

In this section, we show that given a desired set $\Pi = \{\pi_1, \pi_2, \dots , \pi_k\}$ of reversible classical gates, we can find a GSC based on a group $G_{\Pi}$ that can implement each $\pi_j$ transversally. We further show how to switch into the $G_{\Pi}$ code starting from $n$ patches of surface code, using our elementary logical blocks. This gives us the result that there is a transversal implementation of an arbitrary reversible classical gate on $n$ qubits using GSCs.

The crucial fact we will need is that any permutation $\pi \in S_{2^n}$ can be expressed {uniquely} as a product
\begin{equation}
    \pi = \sigma \circ \tau,
\end{equation}
where $\tau$ is a permutation that fixes the string $00 \dots 0$ and $\sigma$ is a permutation implemented by Pauli $X$ operators. Note that $\sigma \in \Z_2^n$ and $\tau \in S_{2^n-1}$ where $\Z_2^n$ and $S_{2^n-1}$ are both subgroups of $S_{2^n}$. Furthermore, the only permutation common to these subgroups is the Pauli $X$ operation that preserves $\ket{00\dots 0}$, which is the identity. As a result, $S_{2^n}$ can be written as a knit product:
\begin{equation}
    S_{2^n} = \Z_2^n \knit S_{2^n-1}.
\end{equation}
Note that $S_{2^n}$ is \textit{not} a semidirect product, because $\Z_2^n$ is not a normal subgroup for $n>2$. This is just the statement that the Pauli $X$ operations are not closed under conjugation by non-Clifford gates.

This result further implies that any subgroup $G_{\Pi}$ of $S_{2^n}$ which contains all the Pauli $X$ gates can also be expressed as a knit product:
\begin{equation}
    G_{\Pi} = \Z_2^n \knit K_{\Pi},
\end{equation}
where $K_{\Pi}$ is a subgroup of $S_{2^n-1}$ consisting of operations in $G_{\Pi}$ that fix $\ket{00 \dots 0}$. Intuitively, $K_{\Pi}$ consists of all the controlled operations in $G_{\Pi}$ that are controlled only on qubits in the $\ket{1}$ state. As mentioned above, $G_{\Pi}$ need not have a nested semidirect product structure, and for $n>2$ need not be solvable.

There are algorithms for computing the minimal subgroup of $S_{2^n}$ generated by a set of permutations. For concreteness, note that $G_{\Pi}$ can be determined in simple cases using a generalization of the procedure illustrated previously for $G_{\mathrm{CCX}}$. Without loss of generality, we assume that each desired permutation $\pi_j$ preserves $\ket{00\dots 0}$, implying that $\pi_j \in K_{\Pi}$. (If not, we can simply replace $\pi_j$ with $\sigma_j \pi_j$, where $\sigma_j$ is the unique Pauli permutation such that $\sigma_j \pi_j$ preserves $\ket{00\dots 0}$.) 

Now fix some $\pi_m$ and define $O_{\pi_m}$ as the operator that implements $\pi_m$ on the logical bit strings. We first conjugate $X_1, X_2, \dots , X_n$ by $O_{\pi_m}$ to obtain new permutation operators $O_{\pi_m}^{(i)} = O_{\pi_m} X_iO_{\pi_m}^{\dagger}$. These permutations should be included in $K_{\Pi}$ after left multiplying by suitable Pauli operators if necessary. We then conjugate each $X_i$ by $O_{\pi_m}^{(j)}$ to obtain another set of permutations $O_{\pi_m}^{(ij)} = O_{\pi_m}^{(j)} X_i O_{\pi_m}^{(j)\dagger}$. Note that if $O_{\pi_m}$ is in the $n$th level of the Clifford hierarchy, then $O_{\pi_m}^{(i)}$ is in the $(n-1)$th level, $O_{\pi_m}^{(ij)}$ is in the $(n-2)$th level, and so on. Iterating this procedure gives us the minimal group $K_{\pi_m}$ that includes the desired permutation $O_{\pi_m}$. The group $K_{\Pi}$ is the minimal subgroup of $S_{2^{n}-1}$ which contains each $K_{\pi_m}$. 

Having determined $K_{\Pi}$, the general protocol consists of the following steps:
\begin{enumerate}
    \item Begin with $n$ patches of surface code initialized in an arbitrary logical state.
    \item Extend to a $G_{\Pi} = \Z_2^n \knit K_{\Pi}$ GSC, in which the $K_{\Pi}$ logical qudits are in the $\ket{+}_\mathsf{L}$ state.
    \item Perform the desired transversal operations in $G_{\Pi}$ by left multiplication. This permutes the $\Z_2^n$ logical qubits, leaving the $K_{\Pi}$ qudits invariant.
    \item Split into $\Z_2^n$ and $K_{\Pi}$ GSCs. The $K_{\Pi}$ GSCs are unentangled with the $\Z_2^n$ GSCs and can be discarded.
\end{enumerate}

\subsection{Coset surface codes}\label{Sec:Egs-CSC}

The general construction above involves preparing an $H \knit K$ GSC with the logical $H$ qubits inherited from surface codes and the logical $K$ qudits initialized in the $\ket{+}_\mathsf{L}$ state. The $K$ qudits are initialized in the $\ket{+}_\mathsf{L}$ state to avoid becoming entangled with the $H$ qubits through the action of the transversal logical gate. In this section, we introduce coset surface codes (CSCs), which neatly capture the paradigm of initializing the $K$ qudits in the $\ket{+}_\mathsf{L}$ state.  

To begin, we consider the spacetime diagram for the extension of an $H$- and $K$-GSC with the $K$ logical qudits initialized in the $\ket{+}_\mathsf{L}$ state. This is pictured on the left-hand side of Eq.~\eqref{eq: CSC spacetime}. We then imagine ``pushing'' the leg associated to the $K$-GSC into the diagram:
\begin{equation} \label{eq: CSC spacetime}
    \vcenter{\includegraphics[scale=0.35]{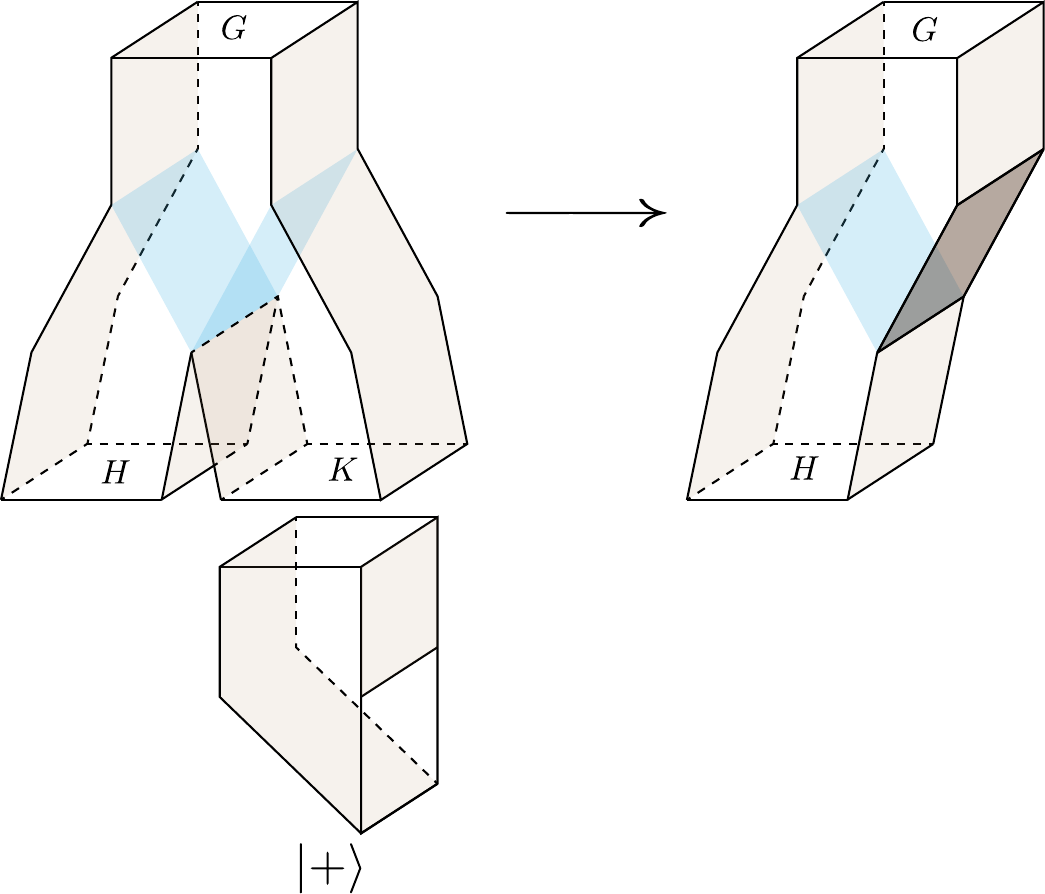}}.
\end{equation}
This allows us to reinterpret the initialization of the $K$-GSC in the $\ket{+}_\mathsf{L}$ state as a boundary condition on the right-hand side of the $H\knit K$ GSC, shaded dark brown in the diagram.

The shaded boundary corresponds to a boundary in which the fluxes of $K$ are able to condense. Such a boundary is prepared by modifying the boundary vertex and plaquette stabilizers, as discussed for instance in Ref.~\cite{beigi2011boundaries}. With this modified boundary, it is no longer a $G$-GSC, strictly speaking. Instead, it has logical basis states with the property that $\ket{g}_{\LL} \equiv \ket{gk}_{\LL}$ for any $k \in K$, as can be seen in the left gauge. Therefore, the logical states are in fact labeled by \textit{cosets} $g K$, and the dimension of the logical subspace is $|G|/|K|$. For this reason we refer to the modified code as a CSC of $G$.

\section{Discussion}

In this paper, we introduced group surface codes (GSCs), generalizing the $\Z_2$ surface code to arbitrary finite groups $G$. After showing that the dimension of their code space is the order of the group $|G|$ (implying that code states can be labeled in the `computational basis' by group elements), we identified a number of transversal logical gates which perform left and right multiplication, or enact group automorphisms on the code states (see Table~\ref{tab:Summary_logical_gates}). For suitable choices of $G$, these give transversal non-Clifford gates in arbitrary levels of the Clifford hierarchy. 

We then defined a set of elementary logical operations (summarized in Table~\ref{tab:logical_operations_summary}) for transversal gates, extending and splitting, and preparation and readout. We constructed spacetime circuits using tensor networks, and explained how each logical operation can be viewed as a spacetime logical block decorated with topological defects and interfaces in topological quantum field theory. Importantly, combining these spacetime logical blocks provides a concrete protocol to start with $k$ $\Z_2$ surface codes, switch to a specific GSC, implement the desired non-Clifford operation, and switch back to $\Z_2$ surface codes. Besides the transversal operations in the GSC, our methods give a unified understanding of various schemes to perform non-Clifford operations and prepare magic states (see Table~\ref{tab:Summary_logical_gates}), which have appeared recently in the literature~\cite{huang2025D4,davydova2025D4,sajith2025noncliffordS3}. 

One key perspective emerging from our work is the idea of engineering a particular non-Abelian topological order to transversally implement a chosen gate set. In this work, we showed that there exists a modified GSC that can transversally implement an arbitrary set of reversible classical gates, and we showed how to switch into such a code starting from surface code patches.  

Below we discuss possible extensions and generalizations of this work. One natural direction is to expand the list of elementary logical operations for GSCs. Some important examples include:
\begin{itemize}
    \item Lattice surgery, which includes enacting logical measurements through merging and splitting, in contrast to the purely unitary operations we have considered in this paper.
    \item Converting a rough or smooth boundary of the GSC into other boundary conditions. We saw an example of such a boundary transformation in Section~\ref{Sec:Egs-CSC} in the context of CSCs, and anticipate that other boundary types can further enlarge the available toolkit of logical operations. Working with different boundaries could potentially allow us to perform phase gates, which are not straightforward to implement in the current framework of GSCs.
    \item The `logical antipode' operation, introduced in Ref.~\cite{cowtan2025homologyhopfalgebrasquantum}, which takes $\ket{g}_{\LL} \rightarrow \ket{\bar{g}}_{\LL}$ but is not, in general, a group automorphism. It requires a combination of a group element inversion on each site and a spatial rotation of the entire code by 180 degrees.
    \item Charge tunneling processes. In principle, these give access to phase gates associated with the Fourier transform of $G$. Our tensor network picture for moving charges in spacetime (Section~\ref{sec: spacetime logical blocks}) can be viewed as a first step in this direction.
\end{itemize}
   Further generalizations of this work include developing a connection to twisted quantum doubles, considering topological codes based on solvable anyon theories or Hopf algebra quantum doubles, and studying analogs of GSCs in higher dimensions or on more general lattices as in Ref.~\cite{mcdonough2026groupCSS}.  

Finally, we mention some important future directions related to error correction and fault tolerance, which we did not explore in depth in this work. An immediate goal is to use the ideas in Refs.~\cite{davydova2025D4,lyons2026FT} to establish an error correction threshold for quantum computation with GSCs in the presence of local stochastic noise. Another is to consider codes tailored to biased noise by inserting defects as in Ref.~\cite{Tiurev2024domainwallcolorcode}. Analogous to Refs.~\cite{Bombin2024unifyingflavorsof,Bauer2024topologicalerror}, it would also be interesting to study whether, by appropriately replacing tensors with measurements, the circuits described in Section~\ref{Sec:Lattice to continuum} can be used to define Floquet codes and measurement-based quantum error correcting codes.   

\vspace{0.2in}
\noindent{\it Acknowledgments} -- TDE is grateful to Andrew Landahl and Margarita Davydova for conversations that inspired this work. TDE thanks Julio C. Magdalena de la Fuente and Andreas Bauer for valuable discussions about the path-integral approach to QEC. TDE and VM also thank Yabo Li for insightful conversations about Ref.~\cite{li2026anyonpermutationsquantumdouble}. VM thanks Matthias Fl\'or for useful discussions. NM thanks Alex May for a discussion on reversible classical gates. 
NM and TDE are also grateful to Jacob Bridgeman and Meng Cheng for conversations about the boundaries of the $S_3$ quantum double.
AT thanks Ingo Runkel and David Hofmeier for discussions.  
Research at Perimeter Institute is supported in part by the Government of Canada through the Department of Innovation, Science and Economic Development and by the Province of Ontario through the Ministry of Colleges and Universities. This research was supported in part by grant no. NSF PHY-2309135 to the Kavli Institute for Theoretical Physics (KITP). This research was supported in part by the International Centre for Theoretical Sciences (ICTS) for participating in the program - `Generalized symmetries and anomalies in quantum phases of matter 2026' (code: ICTS/ GSYQM2026/01). AT is funded by Villum Fonden Grant no. VIL60714.

\begin{appendix}

\section{Representation on qubits} \label{app: representation on qubits}

In this appendix, we provide explicit representations of $L^g$ and $R^g$ on qubits (and qutrits) for the groups $D_4$, $G_{\mathrm{CCX}}$ (see Section~\ref{sec:Egs-CCX}) and $S_3$. For $D_4$, in particular, we also write the vertex and plaquette stabilizers on qubits. We further show that the $D_4$ GSC stabilizers can be mapped to those of the $\Z_2^3$ twisted quantum double (TQD) model in Ref.~\cite{kobayashi2025cliffordhierarchystabilizercodes}.

\subsubsection{$D_4$ GSC on qubits}

The physical Hilbert space on each edge of the $D_4$ GSC is $|D_4| = 8$ dimensional. This allows us to parameterize the group elements of $D_4$ in terms of three qubits. To do so, we find it convenient to use the following presentation of $D_4$:
\begin{equation}
    D_4 = \langle a,b,c| a^2 = b^2 = c^2 = 1, cac = ab \rangle.
\end{equation}
Any element of $D_4$ can be written as $a^{\alpha}b^{\beta}c^{\gamma}$ with $\alpha,\beta,\gamma \in\{0,1\}$. We can then define the computational basis states at an edge as $\ket{\alpha,\beta,\gamma}$, where we interpret $\alpha, \beta, \gamma$ as labeling the group element $a^{\alpha}b^{\beta}c^{\gamma}$.

The operators $L^g$ and $R^g$ map the state $\ket{a^{\alpha}b^{\beta}c^{\gamma}}$ into the states $\ket{ga^{\alpha}b^{\beta}c^{\gamma}}$ and $\ket{a^{\alpha}b^{\beta}c^{\gamma}\bar{g}}$ respectively. This gives us the following representation on qubits:

\begin{equation} \begin{aligned} L^a \ket{a^{\alpha}b^{\beta}c^{\gamma}} &= \ket{a^{\alpha+1}b^{\beta}c^{\gamma}} &&\!\Longrightarrow\! &L^a &= X_1, \\ R^a \ket{a^{\alpha}b^{\beta}c^{\gamma}} &= \ket{a^{\alpha+1}b^{\beta+\gamma}c^{\gamma}} &&\!\Longrightarrow\! &R^a &= X_1\mathrm{CX}_{32}, \\ L^b \ket{a^{\alpha}b^{\beta}c^{\gamma}} &= \ket{a^{\alpha}b^{\beta+1}c^{\gamma}} &&\!\Longrightarrow\! &L^b &= X_2, \\ R^b \ket{a^{\alpha}b^{\beta}c^{\gamma}} &= \ket{a^{\alpha}b^{\beta+1}c^{\gamma}} &&\!\Longrightarrow\! &R^b &= X_2, \\ L^c \ket{a^{\alpha}b^{\beta}c^{\gamma}} &= \ket{a^{\alpha}b^{\beta+\alpha}c^{\gamma+1}} &&\!\Longrightarrow\! &L^c &= X_3\mathrm{CX}_{12}, \\ R^c \ket{a^{\alpha}b^{\beta}c^{\gamma}} &= \ket{a^{\alpha}b^{\beta}c^{\gamma+1}} &&\!\Longrightarrow\! &R^c &= X_3, \end{aligned} \end{equation}
where we have labeled the $\alpha$, $\beta$, and $\gamma$ qubits as 1, 2, and 3 respectively.

With this, we can immediately construct the vertex stabilizers $A_v^g$, for $g = a,b,c$:
\begin{align}\label{eq: D4 Av 1}
A_v^a =     \vcenter{\hbox{\includegraphics[scale=.3]{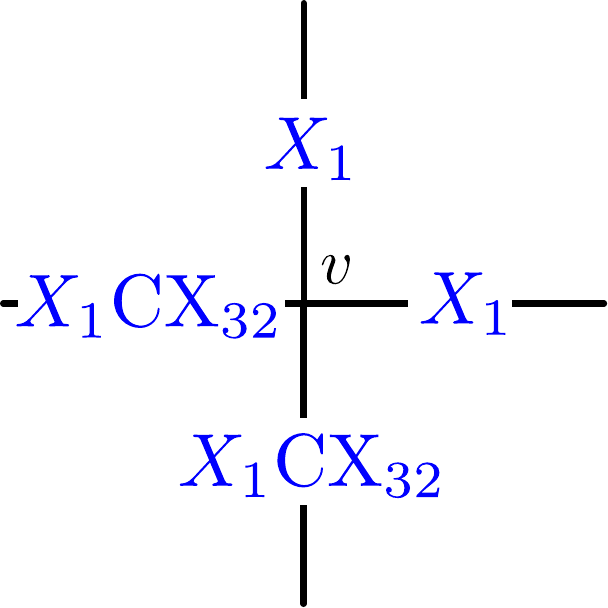}}}~,
\end{align}
\begin{align}\label{eq: D4 Av 2}
A_v^b =     \vcenter{\hbox{\includegraphics[scale=.3]{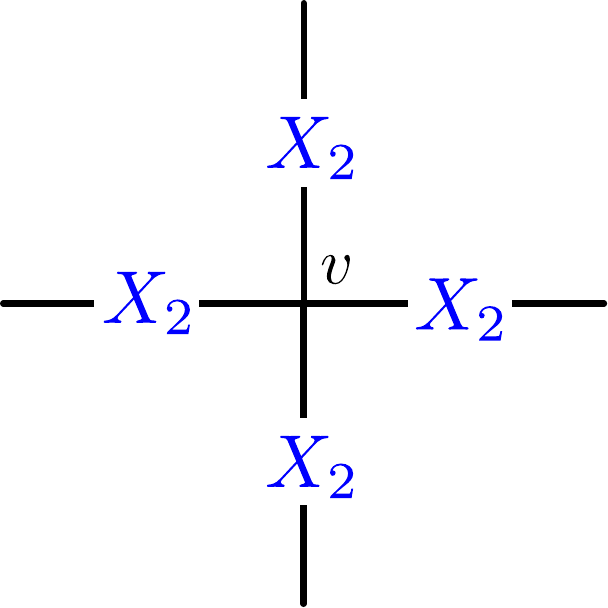}}}~,
\end{align}
\begin{align} \label{eq: D4 Av 3}
\,\, \,\,\,\, A_v^c =      \vcenter{\hbox{\includegraphics[scale=.3]{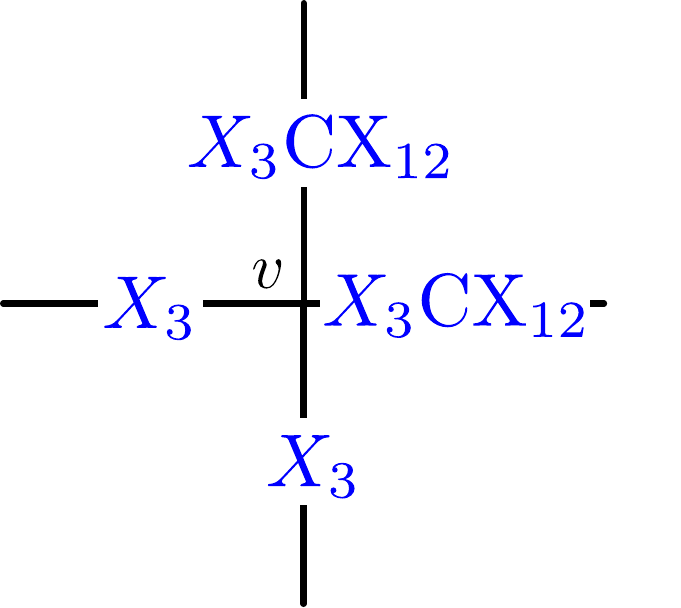}}}~.
\end{align}

To represent the plaquette stabilizers, we first write $g = a^{\alpha} b^{\beta} c^{\gamma}$ and note that $\overline{g} = a^{\alpha} b^{\beta + \alpha \gamma} c^{\gamma}$. We then consider the product of four elements around a plaquette $p$ whose edges are labeled $1,2,3,4$ (not to be confused with the numbering of the qubits at an edge):
\begin{eqs}\label{eq:D4-flux}
\overline{g}_1 g_2 g_3 \overline{g}_4 
&= a^{\alpha_1+\alpha_2+\alpha_3+\alpha_4} \\
&\times b^{\beta_1+\beta_2+\beta_3+\beta_4 + S(\alpha,\gamma)} \\
&\times c^{\gamma_1+\gamma_2+\gamma_3+\gamma_4},
\end{eqs}
where $S(\alpha,\gamma)$ is
\begin{multline}
    S(\alpha,\gamma) 
    = \alpha_4 \gamma_1  + \alpha_3 \gamma_2   \\
    + \gamma_1 (\alpha_1 + \alpha_2+\alpha_3+\alpha_4) \\
    + (\gamma_1+\gamma_2+\gamma_3+\gamma_4) \alpha_4.
\end{multline}
If the flux at the plaquette is trivial, the exponents of $a,b,c$ in Eq.~\eqref{eq:D4-flux} should all vanish. This implies 
\begin{eqs} \label{eq: flux free a and c}
    \alpha_1+\alpha_2+\alpha_3+\alpha_4 &= 0 \text{ mod }2, \\ \gamma_1+\gamma_2+\gamma_3+\gamma_4 &= 0 \text{ mod }2,
\end{eqs}
which can be enforced by ordinary $\Z_2$ surface code plaquette stabilizers $B_p^a, B_p^c$ acting on the $1$ and $3$ qubits at each edge. When Eq.~\eqref{eq: flux free a and c} holds, then the exponent of $b$ simplifies. The no flux condition can be enforced by the stabilizers
\begin{align}~\label{eq:D4_Bp1}
B_p^a &=     \vcenter{\hbox{\includegraphics[scale=.3]{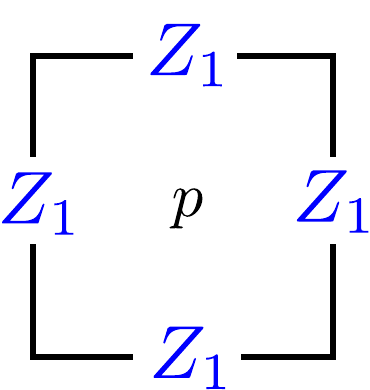}}}~,
\end{align}
\begin{align}\label{eq:D4_Bp2}
B_p^b &=     \vcenter{\hbox{\includegraphics[scale=.3]{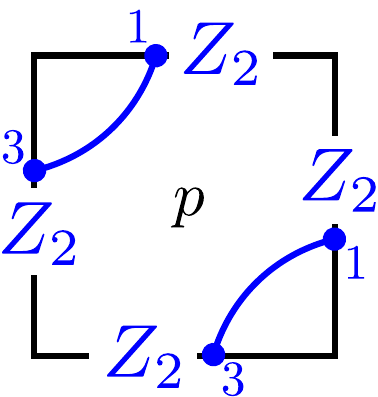}}}~,
\end{align}
\begin{align}\label{eq:D4_Bp3}
B_p^c &=     \vcenter{\hbox{\includegraphics[scale=.3]{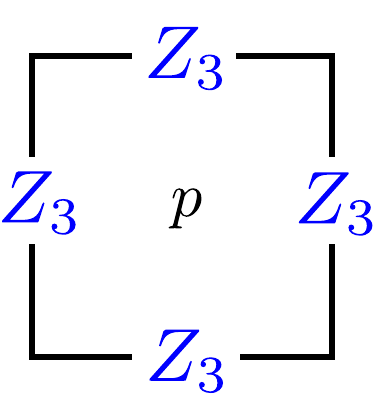}}}~,
\end{align}
where the blue arc denotes a $\mathrm{CZ}_{13}$. Note that this expression for $B_p^b$ only holds within the flux-free subspace.

To perform the vertex measurements described in Section~\ref{Sec:Lattice to continuum}, we use controlled left action operators, represented schematically as $\mathrm{CL}$. Let us express the action of $\mathrm{CL}$ operators on qubits for the case of $D_4$. The operator $\mathrm{CL}$ then acts on two 8-dimensional Hilbert spaces, which we decompose as three qubits each, as above. Suppose the qubits are labeled $1,2,3$ and $1',2',3'$, respectively. Suppose the group basis states are labeled $\ket{a^{\alpha} b^{\beta}c^{\gamma}}$ and $\ket{a^{\delta} b^{\epsilon} c^{\eta}}$. Then, the $\mathrm{CL}$ operator acts on the qubits $1',2',3'$ controlled on qubits $1,2,3$ as:
\begin{equation}
\begin{split}
    \mathrm{CL} \ket{a^{\alpha} b^{\beta}c^{\gamma}; a^{\delta} b^{\epsilon} c^{\eta}} &= \ket{a^{\alpha} b^{\beta}c^{\gamma};  a^{\alpha} b^{\beta}c^{\gamma}a^{\delta} b^{\epsilon} c^{\eta}} \\
    &= \ket{a^{\alpha} b^{\beta}c^{\gamma}; a^{\alpha + \delta} b^{\beta + \epsilon + \gamma \delta}c^{\gamma + \eta}}.
\end{split}
\end{equation}
From the second line, it follows that $\mathrm{CL}$ is equivalent to:
\begin{equation}
 \mathrm{CL} = \mathrm{CX}_{33'}\mathrm{CX}_{22'}\mathrm{CX}_{11'}\mathrm{CCX}_{31'2'}.   
\end{equation}

\subsubsection{Mapping to $\Z_2^3$ twisted quantum double model}

It is well known that the quantum double model for $D_4$, i.e., the bulk of the $D_4$ GSC, has the same underlying topological order as the $\Z_2^3$ twisted quantum double (TQD) model~\cite{propitius1995topological}.  Here, we make this explicit by mapping the stabilizers of the $D_4$ quantum double model in Eqs.~\eqref{eq: D4 Av 1}-\eqref{eq: D4 Av 3} and \eqref{eq:D4_Bp1}-\eqref{eq:D4_Bp3} to those of the $\Z_2^3$ TQD model on a square lattice in Ref.~\cite{kobayashi2025cliffordhierarchystabilizercodes}.

The mapping is implemented by the following two transformations: 
\begin{enumerate}[label=(\roman*)]
    \item We translate all the $b$ qubits, labeled by 2, by the lattice vector $(1/2,1/2)$.
    \item We apply a Hadamard to each of the $b$ qubits. 
\end{enumerate}
This maps the vertex stabilizers $A_v^a$ and $A_v^c$ to:
\begin{align}
    \tilde{A}_v^a =\vcenter{\hbox{\includegraphics[scale=.3]{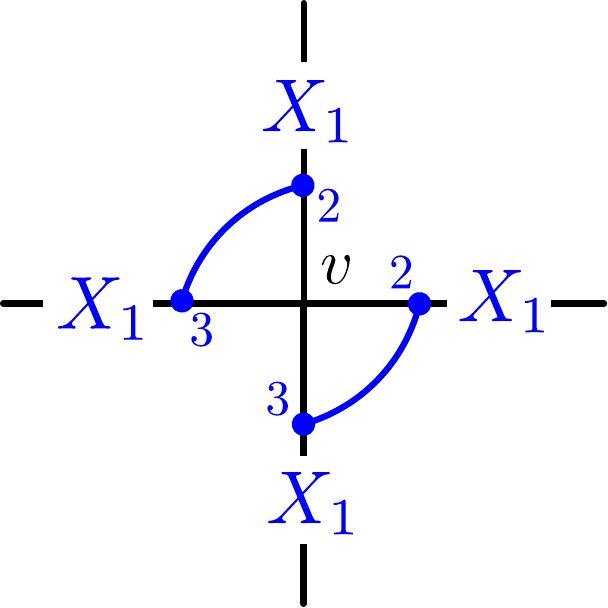}}},
\end{align} 
\begin{align}
    \tilde{A}_v^c = \vcenter{\hbox{\includegraphics[scale=.3]{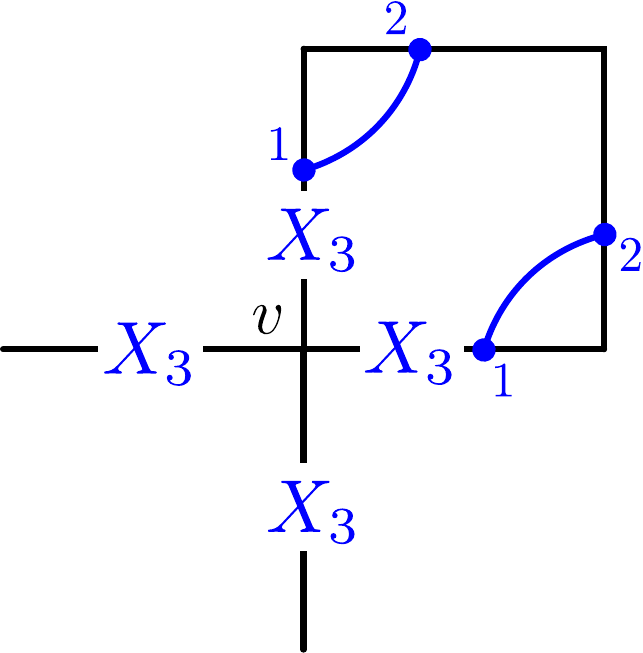}}}~.
\end{align}
Note that the factors of $\mathrm{CX}$ become $\mathrm{CZ}$, due to the Hadamard gates.

The plaquette terms $\tilde{B}_p^a, \tilde{B}_p^c$ are unchanged:
\begin{align}
    \tilde{B}_p^a &=\vcenter{\hbox{\includegraphics[scale=.3]{Figures/Bpa.pdf}}}~, \\  \tilde{B}_p^c &= \vcenter{\hbox{\includegraphics[scale=.3]{Figures/Bpc}}}~.
\end{align}

There is a more nontrivial effect on the stabilizers $A_v^b$ and $B_p^b$. Due to the shift of the $b$ qubits, the vertex stabilizer is mapped to a plaquette stabilizer and vice-versa. The stabilizers $A_v^b$ and $B_p^b$ map to $\tilde{B}_p^b$ and $\tilde{A}_v^b$, respectively:
\begin{align}
    \tilde{A}_v^b &= \vcenter{\hbox{\includegraphics[scale=.3]{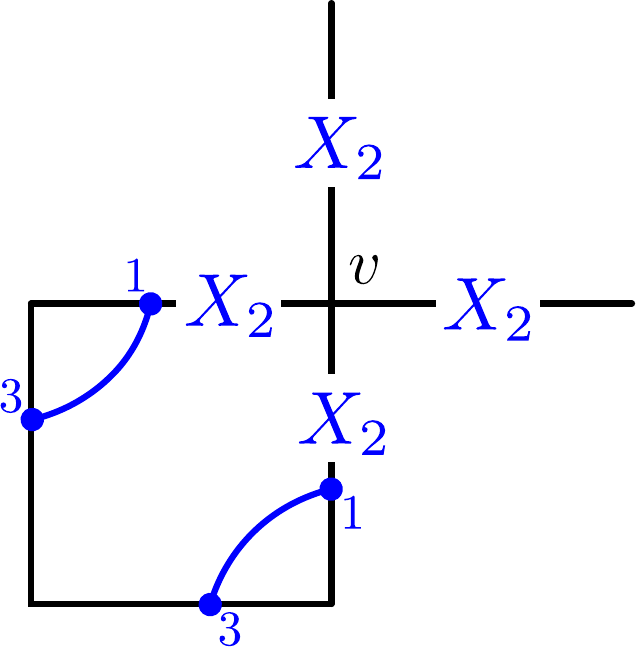}}}~, \\
    \tilde{B}_p^b &=  \vcenter{\hbox{\includegraphics[scale=.3]{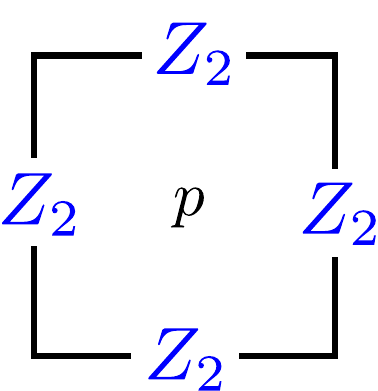}}}.
\end{align}

If the labels 1, 2, 3 are replaced with the colors $\textcolor{red}{r},\textcolor{blue}{b},\textcolor{green}{g}$, we obtain precisely the $\Z_2^3$ TQD model of Ref.~\cite{kobayashi2025cliffordhierarchystabilizercodes}. First of all, the transformation between the models implies that the fault-tolerance proof in Ref.~\cite{davydova2025D4} carries over to the $D_4$ GSC presented in this work. Second, the transformation, when applied to a surface code on the $b$ qubits, implements an automorphism of the anyon theory that permutes the $e$ and $m$ anyons. This explains the difference in the boundary conditions between our work and Ref.~\cite{davydova2025D4}. It also explains why the sliding operation in Ref.~\cite{davydova2025D4} enacts a logical $\mathrm{CCZ}$, as compared to the $\mathrm{CCX}$ in this work. 

\subsubsection{$G_{\mathrm{CCX}}$ GSC on qubits}

Let us now give explicit representations of $L^g$ and $R^g$ on qubits for the group $G_{\mathrm{CCX}}$. 
We label the generators of $G_{CCX}$ using the notation from the main text:
\begin{align*}
    a&=X_1, & d&=\mathrm{CX}_{13}, \\
    b&=X_2, & e&=\mathrm{CX}_{23}, \\
    c&=X_3, & f&=\mathrm{CCX}_{123}.
\end{align*}
These generators satisfy the following non-trivial commutation relations:
\begin{align*}
    dad&=ac, & ebe&= bc, \\
    faf&=e, & fbf&=d.
\end{align*}
A general group basis state may be written as $\ket{g} = \ket{a^{\alpha}b^{\beta}c^{\gamma}d^{\delta}e^{\epsilon}f^{\eta}} \equiv \ket{\alpha,\beta,\gamma,\delta,\epsilon,\eta}$. The left and right action of each generator can be determined through its action on the group basis state. For example,
\begin{eqs}
    L^f \ket{\alpha,\beta,\gamma,\delta,\epsilon,\eta} &= \ket{\alpha,\beta,\gamma+\alpha\beta,\delta+\beta,\epsilon+\alpha,\eta+1} \\
    &\implies L^f = X_6 \mathrm{CX}_{24}\mathrm{CX}_{15}\mathrm{CCX}_{123}. 
\end{eqs}
Repeating this calculation, we find:
\begin{equation} \begin{aligned} L^a &= X_1, &\,\, R^a &= X_1\mathrm{CX}_{43}\mathrm{CX}_{65}, \\ L^b &= X_2, & R^b &= X_2\mathrm{CX}_{53}\mathrm{CX}_{64}, \\ L^c &= X_3, & R^c &= X_3, \\ L^d &= X_4\mathrm{CX}_{13}, & R^d &= X_4, \\ L^e &= X_5\mathrm{CX}_{23}, & R^e &= X_5, \\ L^f &= X_6\mathrm{CX}_{24}\mathrm{CX}_{15}\mathrm{CCX}_{123}, & R^f &= X_6. \end{aligned} \end{equation}

\subsubsection{$S_3$ GSC on qutrits and qubits}

Let us denote group elements of $S_3$ as $g = r^{\alpha} s^{\beta}$, where $r^3 = s^2 = 1$. We then define qutrit and qubit Pauli $X$ operators ${X}_1$ and $X_2$ on the qutrit and the qubit, respectively. Denoting a group basis state by $\ket{\alpha, \beta}$, the action of these operators are:
\begin{align}
    {X}_1 \ket{\alpha, \beta} &= \ket{\alpha+1, \beta}~; \\
    X_2 \ket{\alpha, \beta} &= \ket{\alpha, \beta+1}~.
\end{align}
We also define the charge conjugation operator $\mathcal{C}_1$, which implements the action
\begin{align}
    \mathcal{C}_1 \ket{\alpha, \beta} = \ket{-\alpha, \beta}~.
\end{align}
We can now straightforwardly express the multiplication in $S_3$ in terms of the above operators:
\begin{equation} \begin{aligned} L^{r}\ket{\alpha,\beta} &= \ket{\alpha+1,\beta} &&\!\Longrightarrow\! &L^r &= {X}_1, \\ L^{s}\ket{\alpha,\beta} &= \ket{-\alpha,\beta+1} &&\!\Longrightarrow\! &L^s &= \mathcal{C}_1 X_2, \\ R^{r}\ket{\alpha,\beta} &= \ket{\alpha+2\beta+2,\beta} &&\!\Longrightarrow\! &R^r &= {X}^2_1 \mathrm{CX}^2_{21}, \\ R^{s}\ket{\alpha,\beta} &= \ket{\alpha,\beta+1} &&\!\Longrightarrow\! &R^s &= X_2. \end{aligned} \end{equation}

\section{Detectors and movement operators}
\label{app: charge and flux}

In this appendix, we describe the detectors associated to fluxes and charges, i.e., a series of measurements of fluxes and charges that are deterministic in the absence of errors. In Section~\ref{Sec:Lattice to continuum}, we saw that the plaquette measurement outcomes are group valued, while the vertex measurements outcomes are an irrep type with two indices. Here, we show that only the conjugacy class and irrep type have deterministic outcomes. The internal states, that is, the group value and the indices, are scrambled by the vertex measurements. 

We also show in this appendix how the fluxes and charges can be moved using feedforward. For fluxes, we show that their movement can be done deterministically. For charges, their movement is probabilistic in the general case, although knowledge of the internal state can aid in reducing the likelihood of leftover charges.

\subsubsection{Flux detectors}

The goals of this section are to \textbf{(i)} show that in the absence of errors, conjugacy classes remain invariant on each plaquette, and \textbf{(ii)} fluxes can be moved deterministically through controlled actions and feedforward.

The basic building block, representing a projection onto the trivial flux subspace at two successive timesteps, is the following:
\begin{eqs}
    \vcenter{\hbox{\includegraphics[scale=.4]{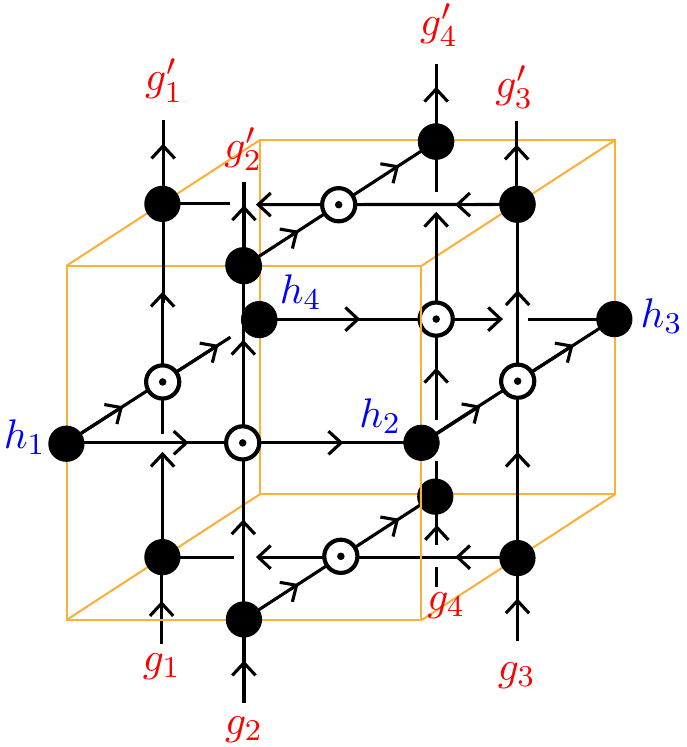}}}.
\end{eqs}
Here, there is a sum over the $h_i$ labels, corresponding to different intermediate gauge transformations.

The plaquette operators in the $xy$ plane implement post-selected projection to the flux-free subspace,
according to Eq.~\eqref{eq: decomposition of Bp}.
We want to show that, if we carry out the plaquette measurements instead of postselecting, as in Eq.~\eqref{eq: plaquette measurement}, the conjugacy classes of the $xy$ plaquettes remain unchanged, assuming no errors have occurred.

Arbitrary plaquette measurement outcomes are represented by modifying the multiplication tensors on the $xy$ plane by adding an extra, uncontracted leg. The uncontracted leg carries the classical information from the measurement of the ancilla in the group basis. We introduce the following notation to simplify the tensor networks and to emphasize the fixed index on the tensor:
\begin{eqs}
    \vcenter{\hbox{\includegraphics[scale=.4]{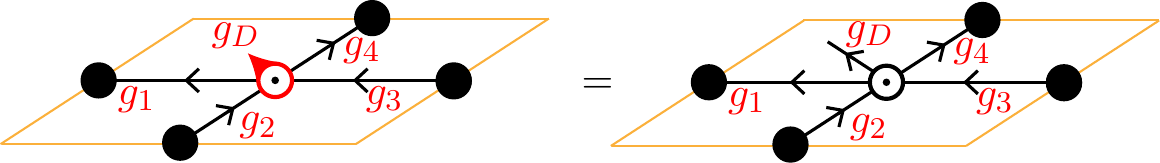}}}.
\end{eqs}
The tensor configuration with nontrivial flux outcomes $g_U$ and $g_D$, respectively, on the up and down plaquettes is given by:
\begin{eqs}
\label{eq: flux cell with up and down modified}
    \vcenter{\hbox{\includegraphics[scale=.4]{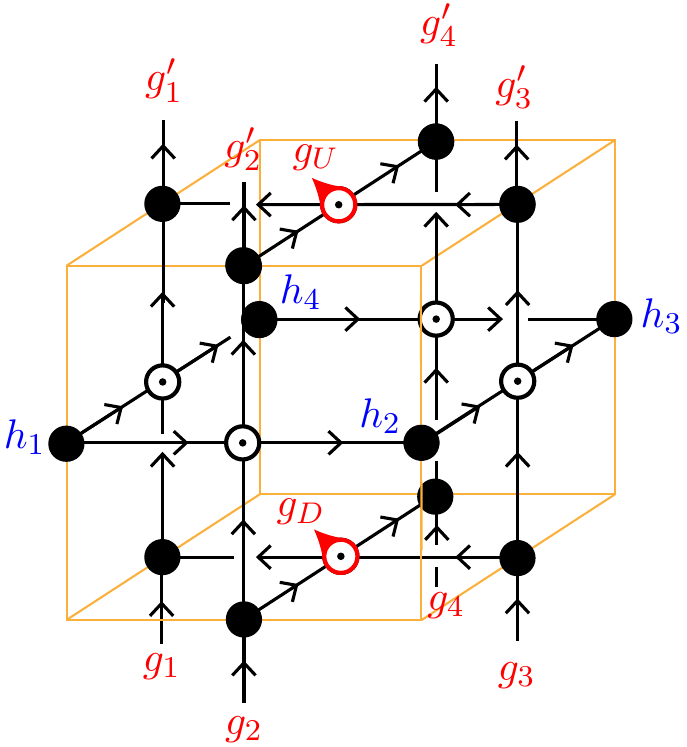}}}.
\end{eqs}

Note that the orientation of the uncontracted leg is arbitrary. However, once chosen, it determines a site, which is a pair $s=(v,p)$ consisting of a plaquette and one of the vertices on its boundary. The vertex in the site determines the starting point for the product of group elements at the plaquette. Choosing a different starting point corresponds to a cyclic permutation of the group elements in the product, resulting in a conjugation action on the flux. If the flux is trivial, the starting point does not matter since $ge\bar{g}= e$ for all $g$. 

We now show that the measurement outcomes $g_U$ and $g_D$ must be in the same conjugacy class. To do so, we consider the constraints imposed by the multiplication tensors:
\begin{equation}
\begin{aligned}
\label{Flux constraint equations }
    \bar{g}_1g_2 g_3 \bar{g}_4 &= g_D,  &  \bar{h}_4\bar{g}_1'  h_1 g_1 &= 1, &  g_2 \bar{h}_2 \bar{g}_2' h_1 &= 1, \nonumber \\ \bar{h}_3 \bar{g}_3' h_2 g_3 &= 1, &
    g_4 \bar{h}_3\bar{g}_4' h_4 &= 1, & \bar{g}_1'g_2' g_3' \bar{g}_4' &= g_U.
\end{aligned}
\end{equation}
Rearranging the constraints coming from the side faces shows that these are responsible for transporting the group elements on each physical edge between successive timesteps, i.e., from $g_i$ to $g_i'$:
\begin{equation}
\begin{aligned}
\label{eq: flux side faces constraints for no errors}
    g_1' &= h_1 g_1 \bar{h}_4, & g_2' &= h_1 g_2 \bar{h}_2, \\
    g_3' &= h_2 g_3 \bar{h}_3 , & g_4' &= h_4 g_4 \bar{h}_3.
\end{aligned}
\end{equation}
These equations show that each edge is in the support of two vertex operators, as expected.

Using Eq.~\eqref{eq: flux side faces constraints for no errors} to simplify $\bar{g}_1' g_2' g_3' \bar{g}_4'$, we obtain:
\begin{align}
\label{eq: flux stationary in time result}
    g_U = h_4 g_D \bar{h}_4.
\end{align}
This implies that the group values $g_U$ and $g_D$ must be in the same conjugacy class. The conjugation by $h_4$ is due to the vertex measurement at the intermediate timestep. 
Therefore, repeated plaquette measurements are able to detect errors, when there is a change in the conjugacy class at a plaquette. We note that the orientation of the $g_U$ and $g_D$ legs does not affect the conclusion, since a different choice of orientation only changes the measurement outcome by conjugation with some group element.

\subsubsection{Movement of fluxes}

We now describe how the fluxes can be moved one plaquette per timestep using feedforward based on the plaquette measurements. The main result is the tensor network description of the movement operator in Eq.~\eqref{eq: tensor representation of moving flux to right}. We make the connection to ribbon operators explicit in the next section. 
We proceed by modifying the multiplication tensor on the side of the cube shown in Eq.~\eqref{eq: flux cell with up and down modified}. For concreteness, consider moving the flux to the right by allowing the group element on the right face to be nontrivial. We argue below that the correct choice of group element leads to the flux being transported to the adjacent plaquette on the next timestep.

Let us denote by $g_D$ and $g_U$ the fluxes on the down and up plaquettes as before, and $g_R$ the uncontracted leg on the right multiplication tensor:
\begin{equation}
\label{eq: flux cell with up and down and right modified}
    \vcenter{\hbox{\includegraphics[scale=.4]{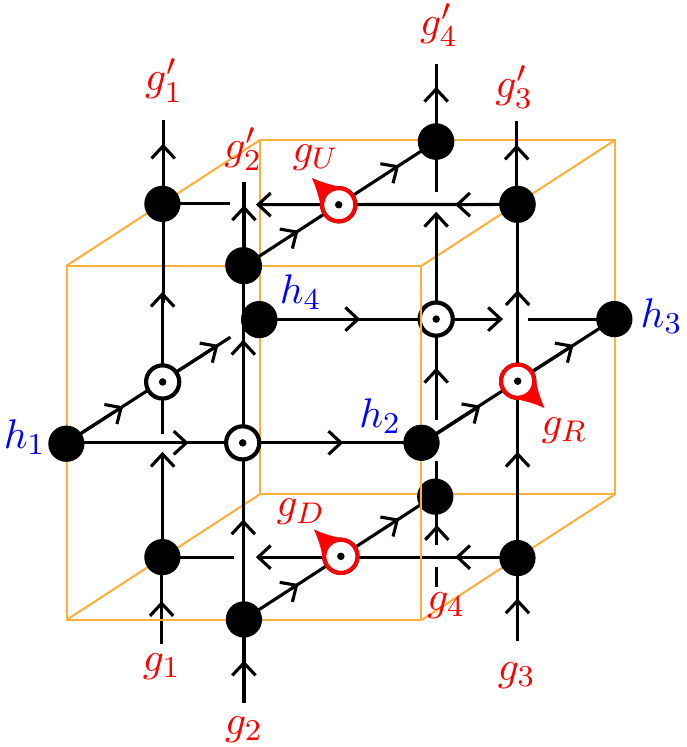}}}.
\end{equation}
All $g_i'$ edges on the top plaquette have the same value as in Eq.~\eqref{eq: flux side faces constraints for no errors} apart from $g_3'$, since the modified multiplication tensor on the right face now reads:
\begin{eqs}
    g_3' = h_2 g_3 \bar{g}_R \bar{h}_3.
\end{eqs}
Thus, the flux on the upward facing plaquette becomes:
\begin{eqs}
   g_U = h_4 \bar{g}_1 g_2 g_3 \bar{g}_R \bar{g}_4 \bar{h}_4.
\end{eqs}
For a trivial flux $g_U$ on the up plaquette, which would imply that the flux has been moved to a different plaquette, we need:
\begin{eqs}
\label{eq: no flux left on top}
    g_R = \bar{g}_4 g_D g_4 \implies g_U = h_4 \bar{g}_1 g_2 g_3 \bar{g}_4 \bar{g}_D \bar{h}_4 = 1.
\end{eqs}
Let us now consider the flux $g_U^{(R)}$ on the up plaquette on the adjacent cube to the right with $g_L^{(R)} = g_R$, and $g_D^{(R)}=1$. The diagram on the cube to the right is:
\begin{equation}
\label{eq: left flux cell with up and left and up modified}
    \vcenter{\hbox{\includegraphics[scale=.4]{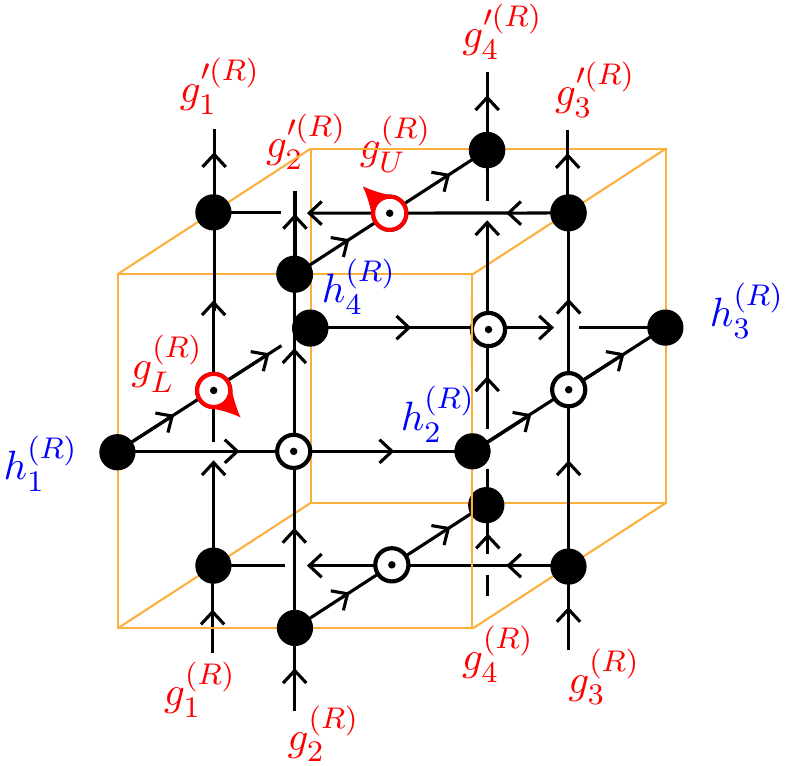}}}.
\end{equation}
The group elements on the legs at the top again satisfy the same relations as in Eq.~\eqref{eq: flux side faces constraints for no errors}, except for $g_1'^{(R)}$:
\begin{eqs}
    g_1'^{(R)} = h_1^{(R)} g_1^{(R)} \bar{g}_L^{(R)} \bar{h}_4^{(R)}.
\end{eqs}
Therefore, the flux on the up plaquette is:
\begin{eqs}
\label{eq: succesful movement of flux to right}
    g_U^{(R)} =& h_4^{(R)} g_L^{(R)} \bar{g}_1^{(R)} g_2^{(R)} g_3^{(R)} \bar{g}_4^{(R)} \bar{h}_4^{(R)} \\
    =& h_4^{(R)} g_R \bar{h}_4^{(R)} \\
    =& h_4^{(R)} \bar{g}_4 g_D g_4 \bar{h}_4^{(R)}.
\end{eqs}
This shows that $g_U^{(R)}$ is in the same conjugacy class as $g_D$. Together with the fact that $g_U$ is trivial, we see that the flux has been successfully moved to the right, although the internal state (the group value) has changed.

Now that we know the modification to the multiplication tensor that is needed to move the flux, we can convert the modification into an operation implemented by controlled actions on the physical qudits.
Specifically, we note the equivalence of the following three tensor configurations:
\begin{eqs}
\label{eq: equivalence of tensor right flux }
    \vcenter{\hbox{\includegraphics[scale=.4]{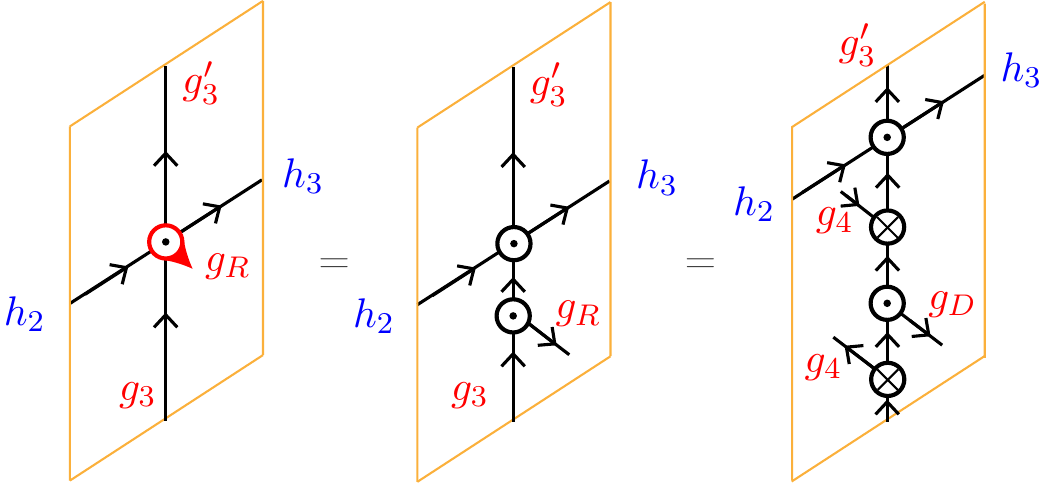}}},
\end{eqs}
which hold for $g_R = \bar{g}_4 g_D g_4$.
The rightmost configuration can be physically implemented by a combination of controlled operations and feedforward, providing a deterministic circuit level implementation of flux transport between adjacent plaquettes.
The explicit operation is $g_3 \rightarrow g_3 \bar{g}_4 \bar{g}_D g_4$, with tensor representation:
\begin{eqs}
\label{eq: tensor representation of moving flux to right}
    \vcenter{\hbox{\includegraphics[scale=.4]{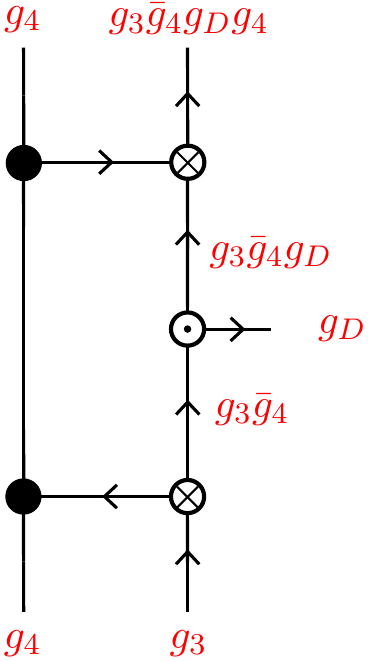}}}.
\end{eqs}
Movement in the three other directions can be computed similarly, yielding expressions for the uncontracted leg on each face:
\begin{eqs}
\label{Other bottom to side fluxes}
    g_L = \bar{g}_D, \, \, 
     g_B = \bar{g}_D, \, \, 
     \, \,  g_F = g_1 g_D \bar{g}_1.
\end{eqs}
The corresponding fluxes on the up plaquettes of the adjacent cubes are:
\begin{eqs}
\label{eq: other moved fluxes from conservation}
    g_T^{(B)} &= {\bar{g}_4}^{'(B)} h_4 g_D \bar{h}_4 {g_4}^{'(B)}, \\
    g_T^{(F)} &= h_4^{(F)} g_1 g_D \bar{g}_1 \bar{h}_4^{(F)}, \\ g_T^{(L)} &= {g_4}^{'(L)} h_4 g_D \bar{h}_4 {\bar{g}_4}^{'(L)}.
\end{eqs}
This gives explicit controlled operations and feedforward for moving fluxes. Note that the success of moving the flux is deterministic, assuming that the measurement of $g_U$ is correct.

A concern might be that, in moving the fluxes, charges are created along the path. We show that the flux movement operators do not, in fact, create any other syndromes. We accomplish this by showing that the movement operators commute with all of the vertex stabilizers $A_v^h$. We do this explicitly for the movement operator to the right. We compute the commutation with $A_v$ for vertices $v_1,v_2$ in the following layout:
\begin{eqs}
    \label{eq: checking flux commutation with A}
    \vcenter{\hbox{\includegraphics[scale=.6]{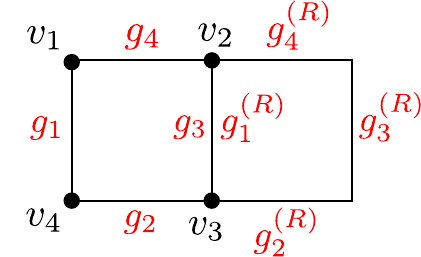}}}.
\end{eqs}
Note that $A_{v_4}$ has no intersecting support and does not affect $g_D$. $A_{v_3}$ acts on $g_3$ by left multiplication and does not affect $g_D$, so it commutes automatically. As a result, we only need to consider the commutation with $A_{v_1}$ and $A_{v_2}$.

For this computation, it is useful to rewrite the movement operator as a (controlled) right multiplication by $\bar{g}_4 g_D g_4$:
\begin{eqs}
    R^{\bar{g}_4 g_D g_4} = \sum_{g_3} \ket{g_3 \bar{g}_4 \bar{g}_D g_4}\bra{g_3}.
\end{eqs}
For the $A_{v_i}^h$ vertex operators, we only consider their action on the $g_1,g_2,g_3,g_4$ legs, where the support intersects that of $R^{\bar{g}_4 g_D g_4}$. 

We now proceed by direct computation, considering the action on the state $\ket{g_1}\ket{g_2}\ket{g_3}\ket{g_4}$ for $A_{v_1}$ and $A_{v_2}$. For $v_1$, we have: 
\begin{eqs}
    &R^{\bar{g}_4 g_D g_4} A_{v_1}^h \ket{g_1}\ket{g_2}\ket{g_3}\ket{g_4} \\ =& R^{\bar{g}_4 \bar{h} h g_D \bar{h} h g_4} \ket{g_1 \bar{h}}\ket{g_2}\ket{g_3}\ket{h g_4} \\=&
    \ket{g_1 \bar{h}}\ket{g_2}\ket{g_3 \bar{g}_4 \bar{g}_D g_4}\ket{h g_4}; \\\\
    & A_{v_1}^h R^{\bar{g}_4 g_D g_4} \ket{g_1}\ket{g_2}\ket{g_3}\ket{g_4} \\
    =& A_{v_1}^h \ket{g_1}\ket{g_2}\ket{g_3 \bar{g}_4 \bar{g}_D g_4}\ket{g_4} \\
    =& \ket{g_1 \bar{h}}\ket{g_2}\ket{g_3 \bar{g}_4 \bar{g}_D g_4}\ket{h g_4},
\end{eqs}
where we used the fact that $A_{v_1}^h$ transforms $g_D \mapsto h g_D \bar{h}$ and $g_4 \mapsto h g_4$, so that the movement operator changes accordingly (although these transformations cancel out). Therefore, we have:
\begin{eqs}
    R^{\bar{g}_4 g_D g_4} A_{v_1}^h = A_{v_1}^h R^{\bar{g}_4 g_D g_4}.
\end{eqs}
Similarly, for $v_2$ we compute:
\begin{eqs}
    &R^{\bar{g}_4 g_D g_4} A_{v_2}^h \ket{g_1}\ket{g_2}\ket{g_3}\ket{g_4} \\ =&
    R^{h \bar{g}_4 g_D g_4\bar{h}} \ket{g_1}\ket{g_2}\ket{g_3\bar{h}}\ket{g_4\bar{h}} \\
    =&\ket{g_1}\ket{g_2}\ket{g_3 \bar{g}_4 \bar{g}_D g_4\bar{h}}\ket{g_4\bar{h}}; \\\\
    &A_{v_2}^h R^{\bar{g}_4 g_D g_4} \ket{g_1}\ket{g_2}\ket{g_3}\ket{g_4} \\ 
    =& A_{v_2}^h \ket{g_1}\ket{g_2}\ket{g_3 \bar{g}_4 \bar{g}_D g_4} \ket{g_4} \\
    =& \ket{g_1}\ket{g_2}\ket{g_3 \bar{g}_4 \bar{g}_D g_4\bar{h}}\ket{g_4\bar{h}},
\end{eqs}
where we have used the $A_{v_2}^h$ action $g_4 \mapsto g_4 \bar{h}$, which changes the movement operator. As before, this shows:
\begin{eqs}
    R^{\bar{g}_4 g_D g_4} A_{v_2}^h = A_{v_2}^h R^{\bar{g}_4 g_D g_4}.
\end{eqs}
As a result, the right flux movement operator does not generate any charges, as it commutes with all vertex stabilizers. Similar calculations show that this holds for flux movement in all directions, so that the flux movement operator does not create violated stabilizers along its path, as claimed.

\subsubsection{Charge detectors}

Similar to the conjugacy class for fluxes, we show that the irrep type of a vertex measurement is invariant throughout the measurement dynamics, assuming that there are no errors. This implies that the measurements of the irrep types form detectors. In this section, we also give a prescription for implementing charge movement through feedforward. Unlike the flux case, this transport is not deterministic in general. In the special case that the charge is being moved to an empty site, we give a deterministic protocol for moving the charge.

The charge detection cell projecting onto the trivial charge subspace is represented as:
\begin{eqs}
\label{charge detector fig}
    \vcenter{\hbox{\includegraphics[scale=.4]{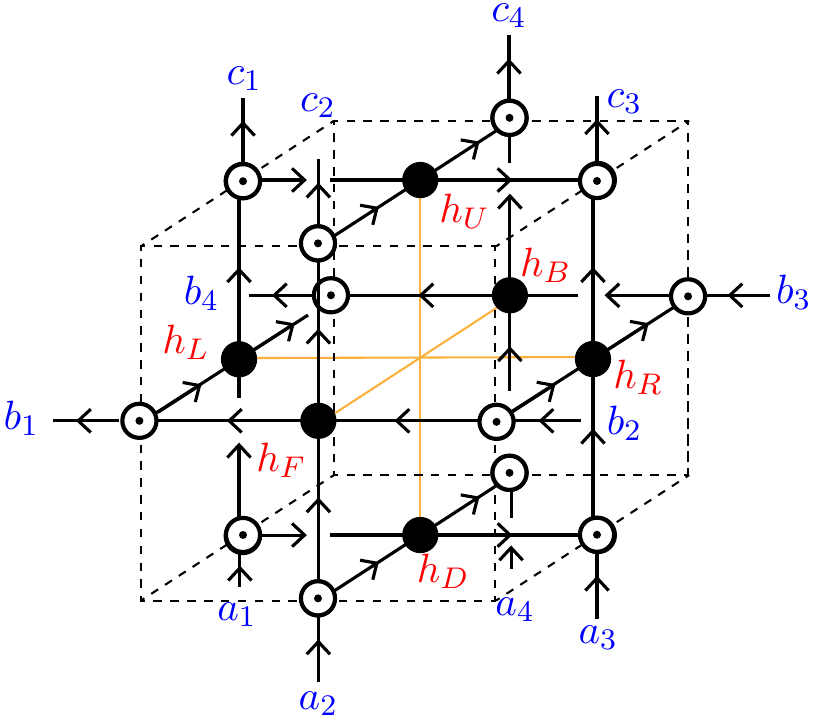}}}. 
\end{eqs}
We want to show that, in the absence of errors, if we measure a nontrivial charge (at the bottom face of the cube), the same charge type will be present on the subsequent measurement round (top face of the cube), although the internal state may change.

To start the discussion, we clarify the meaning of the internal state of a charge. The vertex stabilizers $A_v^g$ at a vertex $v$ form a representation of the group $G$, since they satisfy $A_v^gA_v^h=A_v^{gh}$. This implies that the physical Hilbert space can be decomposed as:
\begin{align}\label{eq: internal state decomposition}
    \mathcal{H} = \bigoplus_{R \in \mathrm{Irr}(G)} \mathcal{H}_{R,v} \otimes \mathcal{M}_{R,v},
\end{align}
where the sum is over irreps $R$ of $G$, $\mathcal{H}_{R,v}$ is the $d_R$-dimensional Hilbert space supporting the irrep $R$, and $\mathcal{M}_{R,v}$ is the so-called multiplicity space. The action of $A_v^g$ in the $R$ sector is $R(g)\otimes I_{\mathcal{M}_{R,v}}$. Such a decomposition is always possible for a representation of a finite group acting on a finite-dimensional subsystem~\cite{serre1977linear}.

Intuitively, if a state is in the sector labeled by $R$, then there is a charge of type $R$ at the vertex $v$. The internal state of the charge then corresponds to the specific state on the subsystem $\mathcal{H}_{R,v}$. More precisely, we can write a basis for the physical Hilbert space, according to Eq.~\eqref{eq: internal state decomposition}. We let the states $\ket{i}_{R,v}$ label a basis for $\mathcal{H}_{R,v}$, with $i \in \{ 1,\ldots, d_R\}$, and let $\ket{\alpha}_{R,v}$ label a basis for $\mathcal{M}_{R,v}$. Then a basis for the physical Hilbert space is:
\begin{align}\label{eq: internal state basis}
    \{\ket{i}_{R,v}\otimes \ket{\alpha}_{R,v} \}_{R,i,\alpha}.
\end{align}
We say that the state $\ket{i}_{R,v}\otimes \ket{\alpha}_{R,v}$ hosts a charge $R$ in the internal state $\ket{i}_{R,v}$ at the vertex $v$. When clear from context, we suppress the multiplicity state to simply write $\ket{i}_{R,v}$.

With this perspective on internal states, we now study the effect of performing a vertex measurement at $v$. We let $\ket{\psi}$ be an arbitrary state written in the basis of Eq.~\eqref{eq: internal state basis}:
\begin{align}
    \ket{\psi} = \sum_{S,\ell,\alpha}c_{S,\ell,\alpha}\ket{\ell}_{S,v}\otimes \ket{\alpha}_{S,v}.
\end{align}
Here, $S$, $\ell$, and $\alpha$ index the irreps, and basis states for the internal space and multiplicity space, respectively.
Since the multiplicity state plays no role in what is to follow, we suppress it for notational convenience.

To perform the vertex measurement, we first add an ancilla in the $\ket{+}_{\mathrm{anc}}$ state to obtain:
\begin{align}
  \frac{1}{\sqrt{|G|}}\sum_{h}  \sum_{S,\ell}c_{S,\ell}\ket{\ell}_{S,v}\otimes \ket{h}_{\mathrm{anc}},
\end{align}
where the first sum is over all $h \in G$.
We then apply controlled group actions according to Eq.~\eqref{eq: Avdecomposition} to produce the state:
\begin{align}
  \frac{1}{\sqrt{|G|}}\sum_{h}  \sum_{S,\ell}c_{S,\ell}A_v^h\ket{\ell}_{S,v}\otimes \ket{h}_{\mathrm{anc}}.
\end{align}
Using that $A_v^h$ acts on $\mathcal{H}_{S,v}$ as $S(h)$, we arrive at:
\begin{align}\label{eq: before substitute in irrep}
  \frac{1}{\sqrt{|G|}}\sum_{h}  \sum_{S,\ell}c_{S,\ell}\sum_k S(h)_{k,\ell}\ket{k}_{S,v}\otimes \ket{h}_{\mathrm{anc}}.
\end{align}

The ancilla is then measured in the irrep basis. To make this explicit, we first write $\ket{h}_{\mathrm{anc}}$ in the irrep basis as:
\begin{align}\label{eq: h in irrep basis}
     \ket{h}_{\mathrm{anc}} = \sum_{R} \sqrt{\frac{d_R}{|G|}} \sum_{i,j} R(h)^*_{ij} \ket{R;i,j}_{\mathrm{anc}}.
\end{align}
The next step is to insert Eq.~\eqref{eq: h in irrep basis} into Eq.~\eqref{eq: before substitute in irrep} and project onto a measurement outcome $|R;i,j\rangle_{\mathrm{anc}}$. Ignoring the normalization, this gives us the state:
\begin{align}
   \sum_{h} \sum_{S,\ell}c_{S,\ell}\sum_k  S(h)_{k,\ell}R(h)^*_{ij}\ket{k}_{S,v}\otimes\ket{R;i,j}_{\mathrm{anc}}.
\end{align}
The sum over $h\in G$ can be carried out using Schur orthogonality:
\begin{align}
    \sum_{h} S(h)_{k,\ell}R(h)^*_{ij} \propto \delta_{R,S}\delta_{k,i}\delta_{\ell,j}.
\end{align}
This leaves us with the (unnormalized) state $\ket{\psi_{R;i,j}}$, which, ignoring the ancilla, is:
\begin{align} 
   \ket{\psi_{R;i,j}} \propto \sum_\alpha c_{R,j,\alpha} \ket{i}_{R,v} \otimes \ket{\alpha}_{R,v} .
\end{align}
This calculation shows us that a measurement outcome of $\ket{R;i,j}$ implies that there is a definite charge $R$ in the internal state $\ket{i}_{R,v}$ at the vertex $v$. The second index tells us that the internal state transitioned from $\ket{j}_{R,v}$ to $\ket{i}_{R,v}$. We note that the measurement outcome of $\ket{R;i,j}$ is equivalent to acting with the operator $\Pi^R_{ij}$ on the state $\ket{\psi}$, where $\Pi^R_{ij}$ is:
\begin{align} \label{eq: transition operator}
    \Pi^R_{ij} = \ket{i}_{R,v}\bra{j}_{R,v} \otimes I_{M_{R,v}}.
\end{align}
We now return to the detector cell in Eq.~\eqref{charge detector fig} and consider consecutive vertex measurements with outcomes $\ket{R^D;i_D,j_D}$ and $\ket{R^U;i_U,j_U}$. This can be represented graphically as:
\begin{eqs}
\label{fig: charge cube top bottom}
    \vcenter{\hbox{\includegraphics[scale=.4]{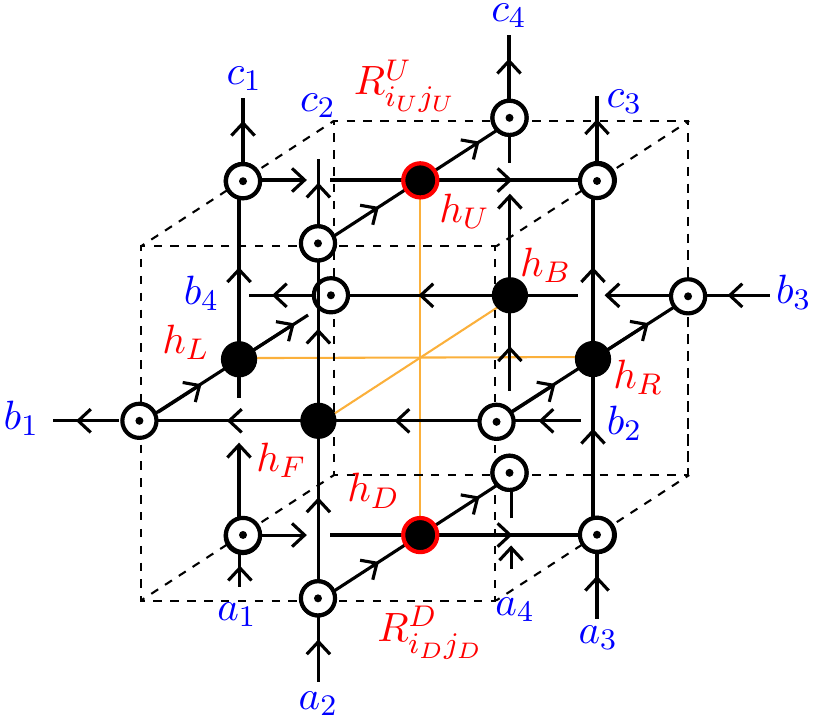}}}, 
\end{eqs}
where the modified copy tensors on the bottom and top faces of the cube are weighted by the respective matrix elements. 

The consecutive measurements are equivalent to applying the operator $\Pi^{R^D}_{i_Dj_D}$ followed by $\Pi^{R^U}_{i_Uj_U}$.\footnote{Here, we assume that there are no fluxes at the associated plaquette.} From the definition in Eq.~\eqref{eq: transition operator}, these satisfy:
\begin{align}
    \Pi^{R^U}_{i_Uj_U}\Pi^{R^D}_{i_Dj_D}=\delta_{R^U,R^D} \delta_{j_U,i_D} \Pi^{R^D}_{i_Uj_D}.
\end{align}
Therefore, the charge type $R^U$ in the second measurement must be $R^D$, assuming no errors have occurred. We also see that the index $i_U$ is not determined by the first measurement outcome. This tells us that only the irrep type will be unchanged in the absence of errors.

\subsubsection{Deterministic charge movement in the bulk}
\label{subsection: deterministic charge movement in bulk}

In this section, we show that it is possible to deterministically move a charge in the bulk by one edge, as long as the target vertex does not already contain a charge. To describe the movement operator, we first build intuition with charge ribbons.

We first consider a charge and its dual joined by a charge line operator.
The charge line operator is built from the group basis ribbon, defined by: 
\begin{eqs}
\label{group basis ribbon for charge}
    F^{1,g}_{u,v} |\bm{h_{e}}\rangle= \delta_{g, g_{\bm{h_e}}}|\bm{h_{e}}\rangle,
\end{eqs}
where $g_{\bm{h_e}}$ indicates the oriented product of group elements along the path connecting vertices $u$ and $v$ in the group configuration $|\bm{h_{e}}\rangle$. 
The state describing the system with two charges, one at vertex $u$ and the other at $v$, with irreps $R^*$ and $R$ in states $\ket{j}$ and $\ket{j'}$ respectively, is then obtained by applying the charge ribbon $F^{R,j,j'}_{u,v}$ on a code state $\ket{[\bm{s_e}]}$ (see Eq.~\eqref{eq: code state}):
\begin{eqs}
\label{two charge state}
    F^{R,j,j'}_{u,v}\ket{[\bm{s_e}]} = \sum_{g\in G} R(g)_{jj'} F^{1,g}_{u,v} \ket{[\bm{s_e}]},
\end{eqs}
where we have decomposed the irrep basis charge ribbon $F^{R,j,j'}_{u,v}$ in terms of the group basis ribbons from Eq.~\eqref{group basis ribbon for charge}. Pictorially, this yields:
\begin{eqs}
\label{two charge configuration}
    \vcenter{\hbox{\includegraphics[scale=.5]{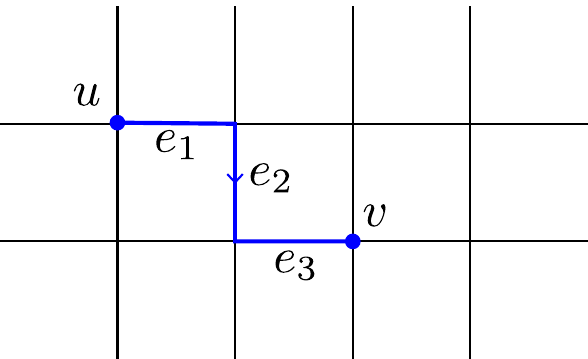}}}.  
\end{eqs}
The action of the charge ribbon on a given group configuration is especially simple:
\begin{eqs}
\label{eq: irrep charge ribbon on group config state}
    F^{R,j,j'}_{u,v}\ket{\bm{h_e}} = & \sum_{g\in G} R(g)_{jj'} F^{1,g}_{u,v} \ket{\bm{h_e}} \\
    = & \sum_{g\in G} R(g)_{jj'}\delta_{g, g_{\bm{h_e}}} \ket{\bm{h_e}} \\
    = & R(g_{\bm{h_e}})_{jj'} \ket{\bm{h_e}}.
\end{eqs}

It is useful to rewrite the two-charge state in a more direct way. 
Explicitly applying a charge ribbon to the code state, we obtain:
\begin{eqs}
\label{eq: two charge state in group config basis}
    F^{R,j,j'}_{u,v} \ket{[\bm{s_e}]} \propto & \sum_{\ket{\bm{h_e}} \in [\bm{s_e}]}   F^{R,j,j'}_{u,v} |\bm{h_e}\rangle \\
    =& \sum_{\ket{\bm{h_e}} \in [\bm{s_e}]}  R(g_{\bm{h_e}})_{jj'} |\bm{h_e}\rangle,
\end{eqs}
using Eq.~\eqref{eq: irrep charge ribbon on group config state}.  This shows that the charge ribbon is a diagonal operator in the group configuration basis $\{ \ket{\bm{h_e}} \}$, and assigns a different complex number to each configuration in the superposition. 
More specifically, it assigns an amplitude $R(g_{\bm{h_e}})_{jj'}$ to configuration $\ket{\bm{h_e}}$, where the irrep $R$ depends on the charge type, the indices $j,j'$ depend on the internal charge states, and the group element $g_{\bm{h_e}}$ is the holonomy on the path connecting the vertices $u,v$.

To develop a strategy for moving the charge, we consider the action of $A_{v}^k$ on the two-charge state:
\begin{eqs}
\label{A acting on 2 charge state}
    A_{v}^k F^{R,j,j'}_{u,v}|[\bm{s_e}]\rangle &= \sum_{g\in G} R(g)_{jj'} A_{v}^k F^{1,g}_{u,v} |[\bm{s_e}]\rangle.
\end{eqs}
The above expression can be simplified using the commutation relations for $A_{v}^k$ and $F^{1,g}_{u,v}$ (see \cite{cui2018topological}):
\begin{eqs}
    A_{v}^k F^{1,g}_{u,v} = F^{1,g\bar{k}}_{u, v} A_{v}^k.
\end{eqs}
This allows us to simplify Eq.~\eqref{A acting on 2 charge state} as:
\begin{eqs}
\label{eq: simplifying A action on 2-charge state}
    &\sum_{g\in G} R(g)_{jj'} A_{v}^k F^{1,g}_{u,v} |[\bm{s_e}]\rangle \\
    = &\sum_{g\in G} R(g)_{jj'} F^{1,g\bar{k}}_{u,v} A_{v}^k |[\bm{s_e}]\rangle \\
    =& \sum_{g\in G} R(g)_{jj'} F^{1,g\bar{k}}_{u,v} |[\bm{s_e}]\rangle \\
    \propto & \sum_{g\in G}  \sum_{\ket{\bm{h_e}} \in [\bm{s_e}]} R(g)_{jj'} F^{1,g\bar{k}}_{u,v} |\bm{h_e}\rangle \\
    = & \sum_{g\in G}  \sum_{\ket{\bm{h_e}} \in [\bm{s_e}]} R(g)_{jj'} \delta_{g\bar{k}, g_{\bm{h_e}}}|\bm{h_e}\rangle \\
    = & \sum_{\ket{\bm{h_e}} \in [\bm{s_e}]} R(g_{\bm{h_e}}k)_{jj'} |\bm{h_e}\rangle,
\end{eqs}
where we used that $A_{v}^k$ is a stabilizer with trivial action on $|[\bm{s_e}]\rangle$, and that $\delta_{g\bar{k}, g_{\bm{h_e}}} = \delta_{g, g_{\bm{h_e}}k}$.
The $A_{v}^k$ action is then easy to interpret: it adds $k$ to the holonomy. 

Naively, we could move the charge by acting with $A_v^{k_{\langle vw\rangle}}$, i.e., the vertex term controlled on the group element at $\langle vw \rangle$. This would give us:
\begin{eqs}
    \sum_{\ket{\bm{h_e}} \in [\bm{s_e}]} R(g_{\bm{h_e}}k_{\langle vw \rangle})_{jj'} |\bm{h_e}\rangle,
\end{eqs}
thereby extending the holonomy from $g_{\bm{h_e}}$ to $g_{\bm{h_e}}k_{\langle vw \rangle}$.
Pictorially, the extended holonomy is:
\begin{eqs}
\label{charge moved from v1 to v2}
    \vcenter{\hbox{\includegraphics[scale=.45]{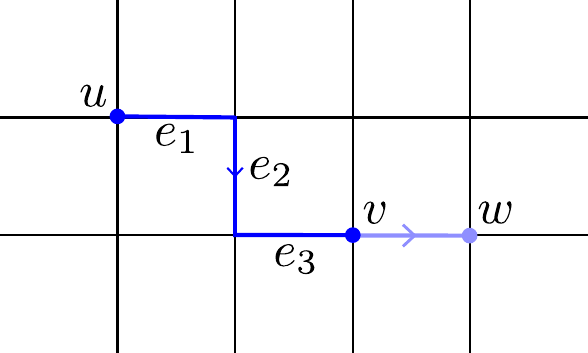}}}.  
\end{eqs}

Although conceptually simple, there is a problem with this approach to moving the charge. The operator $A_v^{k_{\langle vw \rangle}}$ is controlled on the group element at $\langle vw\rangle$ while also acting upon it. On the edge $\langle vw\rangle$, it maps $k_{\langle vw\rangle}$ to $k^2_{\langle vw \rangle}$. In general, this is not a unitary map. Therefore, we must find a way to apply the $k_{\langle vw \rangle}$-controlled action without including the edge $\langle vw \rangle$ as one of its targets.

One solution comes from noting that, for any group element $k$ and vertex $v$, $A_v^k$ is the same as a closed $k$-flux ribbon exclusively around vertex $v$, as shown in Ref.~\cite{lo2025universalquantumcomputations3}. 
The correspondence between flux loops and vertex operators is not limited to the single vertex case. Take a flux loop for $\ell \in G$ around two neighboring vertices $v$ and $w$: it can be shown that its action on group configuration $\ket{\bm{h_e}}$ is $A_v^\ell A_w^{k_{\langle vw \rangle} \ell \bar{k}_{\langle vw \rangle}}$, where $k_{\langle vw \rangle}$ is the group element on the edge connecting $v$ and $w$ in $\ket{\bm{h_e}}$.\footnote{Here, we assume the layout is that of Eq.~\eqref{charge moved from v1 to v2}, with the starting site for the closed flux loop on the bottom left of the $v$ vertex, to agree with the ordering for the single vertex operator $A_v^g$. The action is defined similarly for any other layout and starting site.} 
The reason we want to consider this larger loop is that the common edge between the two vertices is not acted upon, offering a solution to the issue of the common edge being both control and target in the discussion above.

We can therefore use the two-vertex loop to move the charge from vertex $v$ to vertex $w$, under the assumption that $w$ hosts no charge of its own. More precisely, for the $\ket{\bm{h_e}}$ configuration, a $k_{\langle vw \rangle}$ loop around $v$ and $w$ will implement the unitary $A_{v}^{k_{\langle vw \rangle}} A_{w}^{k_{\langle vw \rangle}k_{\langle vw \rangle}\bar{k}_{\langle vw \rangle}} = A_{v}^{k_{\langle vw \rangle}} A_{w}^{k_{\langle vw \rangle}}$.
Giving explicit $g_i$ labels to all edges incident on vertices $v$ and $w$, the action of the 2-vertex flux loop controlled on the edge connecting $v$ and $w$ is given pictorially by:
\begin{multline}
\label{loop around two vertices}
    F^{C,k_{\langle vw \rangle}, k_{\langle vw \rangle}}\Bigg |\vcenter{\hbox{\includegraphics[scale=.5]{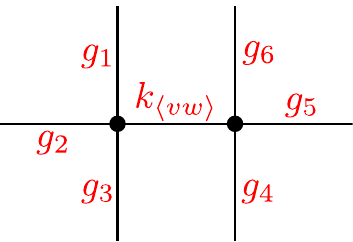}}}\Bigg \rangle \\
   = \Bigg |\vcenter{\hbox{\includegraphics[scale=.5]{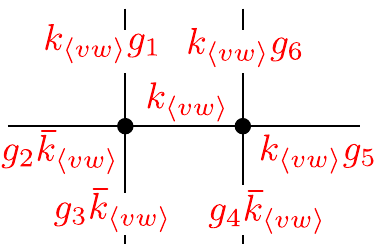}}}\Bigg \rangle.
\end{multline}
The resulting operation has the following implementation using controlled group action gates:
\begin{eqs}
\vcenter{\hbox{\includegraphics[scale=.5]{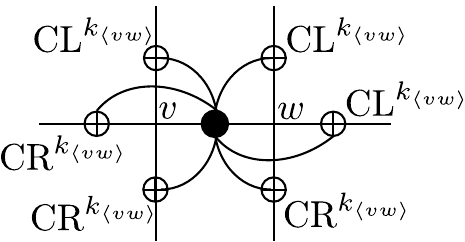}}}.
\end{eqs}

We now show that this operator, which we will denote $W_{\langle vw \rangle}$, correctly moves a charge from vertex $v$ to vertex $w$, while retaining its charge type and internal state, assuming that $w$ is initially charge-free. Before applying $W_{\langle vw \rangle}$, we have:
\begin{eqs}
\label{eq: initial state before controlle charge movement}
    A_v^G \ket{\psi} = 0, \quad A_w^G \ket{\psi} = \ket{\psi},
\end{eqs} 
where $A_v^G$ is the projector onto the trivial charge sector defined in Eq.~\eqref{eq: Av}, indicating the presence of a charge at $v$ and absence thereof at $w$. We first want to show that, after applying $W_{\langle vw \rangle}$, the position of the charge has changed:
\begin{eqs}
\label{eq: state after controlled charge movement}
    A_v^G W_{\langle vw \rangle} \ket{\psi} = W_{\langle vw \rangle}\ket{\psi}, \quad A_w^G W_{\langle vw \rangle} \ket{\psi} = 0.
\end{eqs}

We proceed by first computing $A_v^g W_{\langle vw \rangle}$ and $W_{\langle vw \rangle} A_w^\ell$:
\begin{eqs}
    A^g_v W_{\langle vw \rangle} \ket{\bm{h_e}} &= A_v^g A_v^{k_{\langle vw \rangle}} A_w^{k_{\langle vw \rangle}} \ket{\bm{h_e}} \\
    &= A_v^{g {k_{\langle vw \rangle}}} A_w^{k_{\langle vw \rangle}} \ket{\bm{h_e}}, \\
    W_{\langle vw \rangle} A_w^\ell \ket{\bm{h_e}} & = A_v^{{k_{\langle vw \rangle}}\bar{\ell}} A_w^{{k_{\langle vw \rangle}}\bar{\ell}} A_w^\ell \ket{\bm{h_e}} \\
    &= A_v^{{k_{\langle vw \rangle}}\bar{\ell}} A_w^{{k_{\langle vw \rangle}}} \ket{\bm{h_e}},
\end{eqs}
where on the third line we use the fact that $A_w^\ell$ acts as ${k_{\langle vw \rangle}} \mapsto {k_{\langle vw \rangle}}\bar{\ell}$, therefore changing the control value for the vertex operations. For $A^g_v W_{\langle vw \rangle}$ to be equal to $W_{\langle vw \rangle} A_w^\ell$, we need:
\begin{eqs}
    \ell = {{\bar{k}_{\langle vw \rangle}}} \bar{g} {k_{\langle vw \rangle}}.
\end{eqs}
This gives us the equality:
\begin{eqs}
    A_v^g W_{\langle vw \rangle} = W_{\langle vw \rangle} A_w^{{{\bar{k}_{\langle vw \rangle}}} \bar{g} {k_{\langle vw \rangle}}}.
\end{eqs}
Given that $W_{\langle vw \rangle}$ does not depend on $g$, and that the map taking $g$ to ${{\bar{k}_{\langle vw \rangle}}} \bar{g} {k_{\langle vw \rangle}}$ is a bijection on the group elements, we can sum over $g$ on both sides of the equation to get:
\begin{eqs}
\label{eq: U Av and Aw relations}
    A_v^G W_{\langle vw \rangle} = W_{\langle vw \rangle} A_w^G.
\end{eqs}
The above shows that the position of the charge is swapped from $v$ to $w$ by $W_{\langle vw \rangle}$, proving Eq.~\eqref{eq: state after controlled charge movement}. Moreover, since $W_{\langle vw \rangle}$ is a local operator, it cannot change the charge type, so that the charge type is conserved. It remains to show that the internal state is also conserved.

To argue about the internal state, we first point out that $W_{\langle vw \rangle}$ commutes $A_v^g A_w^g$, for all $g\in G$. For an arbitary configuration state $\ket{\bm{h_e}}$, we have:
\begin{eqs}
    W_{\langle vw \rangle} A_v^g A_w^g \ket{\bm{h_e}} =& A_v^{g {k_{\langle vw \rangle}} \bar{g}} A_w^{g{k_{\langle vw \rangle}} \bar{g}} A_v^g A_w^g \ket{\bm{h_e}} \\ =& A_v^{g{k_{\langle vw \rangle}}} A_w^{g{k_{\langle vw \rangle}}}\ket{\bm{h_e}}, \\
   A_v^g A_w^g W_{\langle vw \rangle} \ket{\bm{h_e}} =& A_v^g A_w^g A_v^{k_{\langle vw \rangle}} A_w^{k_{\langle vw \rangle}} \ket{\bm{h_e}}\\
    =& A_v^{g{k_{\langle vw \rangle}}} A_w^{g{k_{\langle vw \rangle}}}\ket{\bm{h_e}}, 
\end{eqs}
showing that they indeed commute. 

We now use the above calculation to argue that the internal state is coherently moved to the vertex $w$. We start by considering a vertex operator $A_v^g$ acting on a state $\ket{\psi}$ satisfying Eqs.~\eqref{eq: initial state before controlle charge movement} and~\eqref{eq: state after controlled charge movement}:
\begin{eqs}
    W_{\langle vw \rangle} A_v^g \ket{\psi} =& W_{\langle vw \rangle} A_v^g A_w^g \ket{\psi} \\ 
    =& A_v^g A_w^g W_{\langle vw \rangle} \ket{\psi} \\
    =& A_w^g W_{\langle vw \rangle} \ket{\psi}.
\end{eqs}
In the last line, we used that Eq.~\eqref{eq: state after controlled charge movement} implies
\begin{eqs}
A_v^g W_{\langle vw \rangle}\ket{\psi} = W_{\langle vw \rangle}\ket{\psi}, \quad \forall g \in G.
\end{eqs}

Since the operator $\Pi_{ij,v}^R$ is a linear combination of $A_v^g$ terms, we have:
\begin{eqs}
    W_{\langle vw \rangle} \Pi^R_{ij,v}\ket{\psi} = \Pi^R_{ij,w} W_{\langle vw \rangle} \ket{\psi},
\end{eqs}
finally proving that $W_{\langle vw \rangle}$ moves the charge from $v$ to $w$, preserving both its charge type and internal state, contingent on $w$ being initially charge-free.

We caution that the movement operator above does not naively generalize to a movement operator that condenses a charge on a rough boundary. This is because in the bulk, moving a charge from $v$ to $w$ requires a flux loop around $v$ and $w$, but when $w$ is on the the rough boundary there is no local flux loop encompassing both $v$ and $w$. Here, we interpret $w$ as a fictitious vertex attached to all of the rough edges of a boundary component. The movement operator would require a flux loop supported on the entire rough boundary component.

\section{Anyon measurements} \label{app: anyon measurements}

Beyond fluxes and charges, GSCs can host composite excitations, known as dyons. The collection of fluxes, charges, and dyons constitute the anyons of the underlying quantum double model. One might wonder whether it is necessary to perform measurements that fully resolve the anyon types in order to perform error correction. In this appendix, we argue that the plaquette and vertex measurements are sufficient for resolving the effects of errors, even though they are not able to directly determine the anyon types. 

The first thing to clarify is that the plaquette and vertex measurements do not implement a measurement of the anyon types. They instead measure the flux and charge components separately. Nonetheless, we find that measuring the flux and charge separately is sufficient for two main reasons: \textbf{(1)} it does not change the anyon types, and \textbf{(2)} any dyon can be moved by successive flux and charge movement operators, thus allowing the removal of any excitation from the system.

To understand the first claim, we introduce the projector onto an anyon internal state at site $s$ labeled by a pair of a vertex $v$ and plaquette $p$:
\begin{eqs}
\label{anyon projector for internal state}
    \Pi^{C,R}_{c,ij} = \frac{|R|}{|Z(r)|}B_p^c \sum_{n \in Z(r)} \overline{R}(n)^*_{ij} A_v^{q_c n \overline{q}_c},
\end{eqs}
where $C$ is a conjugacy class with a pre-selected element $r$ (different choices of $r$ give isomorphic irreps, so the choice is arbitrary), $c\in C$, $R\in \mathrm{Irr}(Z(r))$ and $ij$ are the indices introduced in Eq.~\eqref{eq: transition operator}.
Here, $Z(r)$ is the centralizer of $r$ and $q_c \in G$ is an element satisfying $q_c r \bar{q}_c=c$ (for more details, see \cite{cui2018topological}). On the other hand, the charge and flux projectors were introduced in Appendix~\ref{app: charge and flux} as:\footnote{Note that we refer to $\Pi_{ij}^{R}$ as a projector, since it projects onto states with a charge $R$. However, the operator does not act as the identity in this subspace, instead it changes the internal state of the charge.}
\begin{eqs}
\label{charge and flux projector that we measure}
    \Pi_{ij}^{R} = \frac{|R|}{|G|} \sum_{g \in G} R(g)^*_{ij} A_v^g, \quad \Pi^C_c = B_p^c.
\end{eqs}

It is important to note that the charge and flux projectors above are not the same as the projectors obtained from restricting the anyon projector in Eq.~\eqref{anyon projector for internal state}. Explicitly, the restrictions are:
\begin{eqs}
    \Pi^{\{1\},R}_{1,(i,j)} = \frac{|R|}{|G|}B_p^1 \sum_{g \in G} \overline{R}(g)^*_{ij} A_v^{g} &\neq \Pi_{ij}^{R} \\
    \Pi^{C,\mathbf{1}_{Z(c)}}_{c,11} = \frac{1}{Z(C)}B_p^c \sum_{n \in Z(r)} A_v^{q_c n \overline{q}_c} &\neq \Pi^C_c.
\end{eqs}
The difference is that the restricted anyon projector not only projects onto the charge (flux) internal state, but also projects onto the trivial flux (charge) state. The vertex and plaquette measurements do not have this property. Equivalently, we can then write:
\begin{eqs}
    \Pi^{\{1\},R}_{1,(i,j)} = \Pi^{\{1\}}_1 \Pi_{ij}^{R}, \quad \Pi^{C,\mathbf{1}_{Z(c)}}_{c,11} = \Pi^C_c \Pi^{\mathbf{1}}_{11}.
\end{eqs}
Therefore, the anyon projector is not generally equal to the product of a plaquette projector and a vertex projector:
\begin{eqs}
    \Pi^{C,R}_{c,ij} \neq \Pi^C_c \Pi^R_{ij}.
\end{eqs}
Moreover, the above equation is not well defined when the conjugacy class is nontrivial, since in that case, the irrep on the left-hand side should be an irrep of $Z(r)$, while on the right-hand side, it is a $G$ irrep.

Despite not being projectors onto anyon types, we argue that the vertex and plaquette measurements commute with the anyon projectors. As a result, they do not change the anyon type. Since we are interested in the anyon type specifically, and not the internal state, we work with the projectors onto the anyon types (hence the lack of subscripts for internal state indices), where $\chi_R(g) = \mathrm{Tr}(R(g))$ is the $R$ character of $g$:
\begin{eqs}
    \Pi^{C,R} = \frac{|R|}{|Z(r)|} \sum_{c\in C} B_p^c \sum_{n \in Z(r)} \chi_R(\bar{n}) A_v^{q_c n \overline{q}_c}.
\end{eqs}
The difference from the internal state projector is that we are averaging over all internal states, by summing over all elements in the conjugacy class $C$, and taking the trace of each matrix element for the irrep $R$. It is well known that the $\Pi^{C,R}$ projectors are central in the quantum double algebra, i.e., they commute with all elements of the algebra \cite{gould1993quantum}. 
This means that, at a given site $s=(v,p)$, we have the commutation relations: 
\begin{eqs}
   \left [ \Pi^{C,R}_s, A^g_v \right ] = \left [ \Pi^{C,R}_s, B_p^h \right ] = 0, \quad \forall \, g,h \in G,
\end{eqs}
which intuitively says that no local operator generated by $A_v^g$ and $B_p^h$ can change the anyon type at a site. It then follows that $\Pi_{ij}^R$ and $\Pi^C_c$ also leave anyon types unchanged, confirming claim \textbf{(1)}: charge and flux measurements do not change the anyon types.

We now move on to the claim \textbf{(2)}: the charge and flux measurements, along with movement operators, can be used to infer the anyon types. This point is more subtle due to the following: the charge content of an anyon is given by an irrep of $Z(r)$ rather than an irrep of $G$, which is measured with $\Pi_{ij}^R$ for $R \in \mathrm{Irr}(G)$. 

Instead, assume that site $s=(v,p)$ hosts an anyon of type $(C,S)$, where $S\in \mathrm{Irr}(Z(r))$ for some $r\in C$. Upon measuring the charge content at vertex $v$, the outcome is some $G$ irrep $R$. More specifically, $R \in \mathrm{Ind}_{Z(r)}^G(S)$.\footnote{For an introduction to induction and restriction, see Appendix~\ref{app: ResInd}.} This implies that we cannot exactly determine the charge content of an anyon. Instead, assuming that we know the flux type at plaquette $p$, we can infer the subset of possible anyon charges from the vertex measurement. 

More concretely, given a flux measurement outcome $c$ and a charge measurement $R \in \mathrm{Irr}(G)$ (here ignoring the internal indices as they do not affect the argument), we can infer that the $Z(r)$ irrep $S$ must be such that:
\begin{eqs}
    \mathrm{Ind}_{Z(r)}^G(S) \supset R.
\end{eqs}
In general, this significantly narrows down the possible irreps $S$ at the site. One important example is that of a pure flux $(C,\mathbf{1}_{Z(r)})$, for which there is an explicit form for the induction~\cite{lu2026self}: 
\begin{eqs}
    \mathrm{Ind}_{Z(r)}^G(\mathbf{1}) \cong \bigoplus_{R\in \mathrm{Irr}(G)} m_R R, 
\end{eqs}
with $m_R$ defined as:
\begin{eqs}
     m_R = \frac{1}{|Z(r)|}\sum_{h\in Z(r)}\chi_R(h).
\end{eqs}
The above simplifies when $Z(r)$ is a normal subgroup, giving:
\begin{eqs}
    \mathrm{Ind}_{Z(r)}^C(\mathbf{1}) = \rho_{G/Z(r)} \circ \pi,
\end{eqs}
where $\rho_{G/H}$ is the regular representation on the quotient group $G/H$ (when the centralizer $Z(r)$ is normal, this is well defined) and $\pi$ is the projection map from the group to the quotient group $\pi: G \rightarrow G/Z(c)$.\footnote{The composition $\rho_{G/N} \circ \pi$ for a subgroup $N$ is called the inflation of the representation $\rho_{G/N}$ from the quotient group $G/N$ to the group $G$, and is denoted $\mathrm{Inf}_{G/N}^G(\rho_{G/N})$} As a reminder, the regular representation $\rho$ of $G$ has the direct sum decomposition:
\begin{eqs}
    \rho \cong \bigoplus_{R \in \mathrm{Irr}(G)} R^{\oplus d_R},
\end{eqs}
where $d_R$ is the dimension of $R$. 

We can now tackle the following question: is a measurement schedule composed of a round of plaquette measurements followed by a round of vertex measurements sufficient to distinguish anyon types $(C,S)$? The answer is generally no, due to the following fact: even if we assume that there is a protocol that, starting from the vertex measurements, allows us to exactly reconstruct the $G$ irrep $\mathrm{Ind}_{Z(c)}^G(S)$ for the $(C,S)$ anyon, knowing the induction of $S$ is in general not enough to deduce $S$. In fact, for a general subgroup $H\subset G$, and two irreps $S,R \in \mathrm{Irr}(H)$, the following condition holds:
\begin{eqs}
    \mathrm{Ind}_{H}^G(S) \cong \mathrm{Ind}_{H}^G(R) \iff \, \exists \, g\in G:\, R \cong S^g,
\end{eqs}
where $S^g$ is called the conjugate representation and is defined as:
\begin{eqs}
    S^g(n) = S(\bar{g}ng).
\end{eqs}

The above suggests that we cannot always distinguish anyon types. Nonetheless, what we actually need is to be able to move the anyons, either to fuse them  with a dual in the bulk, or to condense charges (fluxes) on the rough (smooth) boundaries. Indeed, since charges and fluxes can be separately moved deterministically (for charges we need the target vertex to not host another charge), it would be sufficient to show that the separate movement operators can be used to ``split'' any anyon $(C,S)$ into a pure charge and a pure flux, which can be moved independently. 

There are two important properties of the flux movement operator:
\begin{enumerate}[label=(\roman*)]
    \item The flux at the initial plaquette disappears deterministically (although the internal state at the final plaquette might be scrambled);
    \item Moving a flux commutes with all vertex stabilizers, so that no charges are created along the way.
\end{enumerate}
Within the flux detector cell of Sec.~\ref{app: charge and flux}, the presence of a charge in the vertex $v$ paired to plaquette $p$ at the site $(v,p)$ can only add phases, but not change any group labels.

As a result, even if a site contains an anyon with nontrivial charge content rather than a pure flux, after the flux movement operator we are ensured by property (i) that the initial site will host a trivial flux. That being the case, any remaining excitation at the initial site must be a pure charge, which can be moved independently from the flux. Moreover, by property (ii), since no other charges are created, that charge must be contained in $\mathrm{Ind}_{Z(c)}^G(S)$, which we can measure with the vertex stabilizer on the following round.

Ultimately, even if there is an anyonic excitation that is neither a pure flux nor a pure charge, we can separate the flux and charge content using the flux movement operator, and afterwards move the two independently to remove them from the system.
The above is what we need to remove errors from the system and stay in the code space, confirming claim $\bf{(2)}$: any dyon can be moved by successive flux and charge movement operators, thus allowing the removal of any excitation from the system.

\section{Clifford hierarchy for $G = D_{2^n}$}
\label{d2n appendix}

In Example~\ref{ex: transversal gates D4}, we showed that the $D_4$ GSC admits transversal logical Clifford gates using the logical left and right multiplication operators $L_\mathsf{L}^g$ and $R_\mathsf{L}^g$. In this appendix, we consider a generalization to the dihedral groups $D_{2^n}$. We find that they admit a transversal implementation of logical gates in the $n$th  level of the Clifford hierarchy. We start with a description of the structure of the dihedral groups. \\

\noindent \textbf{Group structure:} For the $D_{2^n}$ family of dihedral groups, we have a convenient subgroup relation:
\begin{align}
    D_{2^n} \triangleleft D_{2^{n+1}}, \, \forall \, n. 
\end{align}
Moreover, each pair of adjacent groups is related by a semidirect product:
\begin{align}
    D_{2^n} \cong D_{2^{n-1}} \rtimes \mathbb{Z}_2.
\end{align}
At the level of generators, for $D_{2^{n-1}} = \langle r,s \rangle $, we have $D_{2^{n-1}} = \langle r^2, s \rangle \triangleleft D_{2^{n}}$, where $r$ is the rotation and $s$ is the reflection.

Let us make the inclusion relation more explicit, starting with $D_4$. The $D_4$ generating set is:
\begin{eqs}
    D_4 &= (\mathbb{Z}_2 \times \mathbb{Z}_2) \rtimes \mathbb{Z}_2 \\
    &= (\langle s \rangle \times \langle r^2 \rangle) \rtimes \langle sr \rangle \\ 
    &= (\langle a \rangle \times \langle b \rangle) \rtimes \langle c \rangle,
\end{eqs}
where we have used the $a,b,c$ notation from Example~\ref{ex: transversal gates D4}.
Since $D_4 \triangleleft D_8$, we can write: 
\begin{eqs}
D_8  &= ((\mathbb{Z}_2 \times \mathbb{Z}_2) \rtimes \mathbb{Z}_2) \rtimes \mathbb{Z}_2 \\
&= ((\langle s \rangle \times \langle r^4 \rangle) \rtimes \langle sr^2 \rangle ) \rtimes \langle sr \rangle \\
&= ((\langle a \rangle \times \langle b \rangle) \rtimes \langle c_2 \rangle) \rtimes \langle c_3 \rangle.
\end{eqs}
The above semidirect product is well defined since $sr$ is not in the $D_4$ normal subgroup. We can therefore get a recursive construction for $D_{2^n}$, since we can always find a dihedral normal subgroup $D_{2^{n-1}}$, and repeat until we get to $D_4$. Below we will show that $L^{c_k}$ is in the $k$th level of the Clifford hierarchy. 

 More precisely: 
\begin{eqs}
D_{2^n} &= (((\mathbb{Z}_2 \times \mathbb{Z}_2) \rtimes \mathbb{Z}_2 \dots ) \rtimes \mathbb{Z}_2) \rtimes \mathbb{Z}_2 \\
 &\cong ((( \langle s \rangle \times \langle r^{2^{n-1}} \rangle) \rtimes \langle sr^{2^{n-2}} \rangle \dots ) \rtimes \langle sr^2 \rangle) \rtimes \langle sr \rangle \\
 & = (((\langle a \rangle \times \langle b \rangle) \rtimes \langle c_{2} \rangle \dots ) \rtimes \langle c_{n-1} \rangle ) \rtimes \langle c_n \rangle,   
\end{eqs}
where we defined $c_j = sr^{2^{n-j}}$.
As a result, any group element can be expressed in terms of $n+1$ qubits:
\begin{eqs}
\label{iterated semidirect element ket}
    g &= \ket{a^\alpha \, b^\beta \, c_{2}^{\gamma_{2}} \dots c_{n-1}^{\gamma_{n-1}} \, c_n^{\gamma_n}} \\ 
    & = \ket{\alpha, \beta, \gamma_{2}, \dots \gamma_{n-1}, \gamma_n}.
\end{eqs}
For each $n \geq 3$, we have the following relations ($n=1$ and $n=2$ are special cases of $n=3$):
\begin{align}\label{eq:D2n-commutation}
     c_n a c_n &= c_{n-1}, \quad c_n c_{n-1} c_n = a, \\ 
     c_n c_{n-j} c_n &= a c_{n-j} c_{n-1}~.
\end{align}
\\

\noindent \textbf{Transversal logical gates:} Using Eq.~\eqref{iterated semidirect element ket}, we can generalize Appendix~\ref{app: representation on qubits} to recursively define closed-form expressions for the left multiplication operators $L^{c_j}$, obtaining their representation on $n+1$ qubits. This allows us to easily deduce the level of $L^{c_k}$ in the Clifford hierarchy, for each $k$. Note that 
\begin{eqs}
    c_{k} g &= c_{k} (a^{\alpha} b^{\beta} c_{2}^{\gamma_{2}} \dots c_k^{\gamma_k} c_{k+1}^{\gamma_{k+1}} \dots c_n^{\gamma_n}) \\
    &=  (c_k a^{\alpha}c_k^{-1}) (c_k b^{\beta}c_k^{-1}) (c_k c_{2}^{\gamma_{2}}c_k^{-1}) \dots \nonumber \\ 
    &\,\,\,\,\,\,\,\,\,\,c_k^{\gamma_k+1}(c_{k+1}^{\gamma_{k+1}} \dots c_n^{\gamma_n}).
\end{eqs}
This tells us that $L^{c_k}$ is supported on at most $k+1$ qubits, but acts as a Pauli $X$ on the qubit corresponding to $c_k$. Therefore, when restricted to the remaining $k$ qubits, its action must be in the $k$th level $\mathcal{C}_{k}$. We will now give an inductive argument showing that $L^{c_k}$ acts in $\mathcal{C}_{k}/\mathcal{C}_{k-1}$ for $k=2,3, \dots , n$.

We have $L^a = X_1$ and $L^b = X_2$. From Eq.~\eqref{eq:D2n-commutation}, $L^{c_{2}}L^a L^{c_{2}} = L^{ab} = X_1 X_2$. Therefore, $L^{c_{2}}$ acts as a \textit{Clifford} operator $\mathrm{CX}_{12}$ on the first two qubits. Its complete action is $L^{c_{2}} = \mathrm{CX}_{12} X_3$, which agrees with the example of $D_4$ in the main text and appendix~\ref{app: representation on qubits}.

We claim that the operator $L^{c_{k}}$ is in $\mathcal{C}_k/\mathcal{C}_{k-1}$, for $k = 2,3, \dots , n$. We can proceed by induction, with $k=2$ as our base case. For the inductive step, assume that $L^{c_{k}}$ is in $\mathcal{C}_k/\mathcal{C}_{k-1}$.  Then, the group law implies that
\begin{equation}
    c_{k+1} a c^{-1}_{k+1} = sr^{2^{n-k-1}} s sr^{2^{n-k-1}} = c_{k}~.
\end{equation}
In other words, conjugating $L^a = X_1$ by $L^{c_{k+1}}$ turns it into $L^{c_{k}}$, which by assumption was in $\mathcal{C}_k/\mathcal{C}_{k-1}$; therefore, $L^{c_{k+1}}$ cannot be in $\mathcal{C}_k$. However, it is in $C_{k+1}$ because its non-trivial action is supported on at most $k+1$ qubits. Therefore, $L^{c_{k+1}}$ is in $\mathcal{C}_{k+1}/\mathcal{C}_{k}$, which establishes our claim by induction. In particular, $c_n$ is in $\mathcal{C}_n/\mathcal{C}_{n-1}$.

For small values of $k$, we can write $L^{c_{k}}$ explicitly:
\begin{align}
    L^{c_{2}} &= \mathrm{CX}_{12} X_3 ~;\\
    L^{c_{3}} &= \mathrm{CCX}_{132} \mathrm{SWAP}_{13} X_4~.
\end{align}
A similar procedure can be used to argue that the operators $R^{c_k}$ are in the $\mathcal{C}_{n-k+1}/\mathcal{C}_{n-k}$, for $k=0, 1, \dots , n-2$. In this case, the operator $R^a$ is in $\mathcal{C}_n/\mathcal{C}_{n-1}$. 

Finally, outer automorphisms of $D_{2^n}$ exchange $a$ and $c_n$; they  correspond to operators in $\mathcal{C}_{n+1}/\mathcal{C}_n$. These can also be implemented transversally in $D_{2^n}$ using automorphism gates. \\

\noindent \textbf{Arbitrary left multiplication operators:} It is possible to find the explicit qubit operations for the action of any $L^g_\mathsf{L}$ for $g\in D_{2^n}$. To do so, it is helpful to first obtain the bijective map between the generating sets $\{ r,s \}$ and $\{ a,b, c_{n-2}, \dots , c_0\}$. The reason to do this is that group multiplication is easy to compute in the $\{r,s\}$ basis, but the qubit mapping is not obvious, whereas in the $\{ a,b, c_{n-2}, \dots , c_0\}$ basis the qubit mapping is trivial but the group multiplication is complicated. As a result, to multiply two group elements and obtain the corresponding multi-qubit unitary, we can start by the $\{ a,b, c_{n-2}, \dots , c_0\}$ presentation, switch to the $\{ r,s \}$ presentation to do the multiplication, and finally switch back.

As a result, we require a map:
\begin{align}
    \phi: \{p,q\} \leftrightarrow \{\alpha, \beta, \gamma_{2}, \dots, \gamma_n\} 
\end{align}
between elements of the form of Eq.~\eqref{iterated semidirect element ket} and $\ket{s^p, r^q}$, with $p \in \{0,1\}$ and $q \in \{0, \dots 2^n-1\}$ .
The forward $\phi$ direction is simple: $p$ counts the parity of $g$, which is $0$ for rotations and $1$ for reflections. To compute $q$, we note the following:
\begin{align}
    sr^a sr^b = r^{-a+b}, \quad sr^a sr^b sr^c = sr^{a-b+c}
\end{align}
with the obvious generalization to $n > 3$ terms of the form $sr^{m_i}$.
We can now explicitly express $p$ and $q$ as:
\begin{eqs}
    \phi(p) &= \frac{1-(-1)^{m+\alpha}}{2}, \\
    \phi(q) &= (-1)^m \left [\sum_{i=1}^m (-2)^i) \right ] + 2^{n-1}\beta,
\end{eqs}
where $m$ indicates the parity of $sr^{m_i}$ terms:
\begin{align}
    m = \bigoplus_{i=2}^{n}\gamma_i.
\end{align}
Before obtaining the inverse map, we note that the first term of $q$ is the known integer sequence A065620 of the Online Encyclopedia of Integer Sequences (OEIS) \cite{oeis}, called the ``reversing binary representation'':
\begin{align}
    A065620(x) \equiv f(x) = (-1)^m \left [\sum_{i=1}^m (-2)^i) \right ],
\end{align}
where the sum is over all nonzero elements in the binary representation of $x$. By \cite{knuth2014art}, every nonzero integer has a unique ``reversing binary representation'' with increasing exponents, and two with decreasing exponents. Indeed, there is an inverse function given by sequence A065621, defined as:
\begin{align}
    A065621(x) \equiv f^{-1}(x) = (x-1) \, \oplus \, (2x-1),
\end{align}
where $\oplus$ indicates bitwise XOR on the binary representation of $x$. With these definitions, we have $f(f^{-1}(x)) = f^{-1}(f(x)) = x$.
Using the above functions we can explicitly give, after some calculations, the map $\phi^{-1}$:
\begin{eqs}
    \phi^{-1}(\alpha) &= p \oplus \mu, \\
    \phi^{-1}(\beta) &= \left \lfloor \frac{q + 2^{n-2}-2}{2^{n-1}} \right \rfloor \\
    \phi^{-1}(\Vec{\gamma_i}) &= f^{-1}(1-2^{n-1}\phi^{-1}(\beta)),
\end{eqs}
where we defined an auxiliary variable $\mu$, which is a function of $n$ and $q$, to simplify the notation:
\begin{align}
     \mu &= \left \lfloor \frac{q+ 2^{n-2}-1-2^{n-1}\phi^{-1}(\beta)}{2^{n-2}} \right \rfloor.
\end{align}
With these explicit isomorphisms, any logical action can be computed by starting with the iterated semidirect encoding, converting each group element to the $\langle r,s \rangle$ presentation, carrying out the multiplication there and then converting back to the iterated semidirect encoding. 

Using the above formulas we can compute the left and right multiplication by the generators $r$ and $s$ for the group $D_{2^4} = D_{16}$, which would be more challenging to do in the qubit encoding (i.e., the iterated semidirect product presentation): 
\begin{eqs}
L^s &= X_1,\\
L^r &= X_5\,
       \mathrm{CX}_{\bar{3}1}\,
       \mathrm{CCX}_{143}\,
       \mathrm{SWAP}_{14}\,
       \mathrm{C^3X}_{\bar{1}3\bar{4}2}\,
       \mathrm{CX}_{31},\\
R^s &= \mathrm{CX}_{54}\,
       \mathrm{CCX}_{4\bar{5}3}\,
       \mathrm{C^3X}_{3\bar{4}\bar{5}2}\,
       \mathrm{CCX}_{\bar{4}\bar{5}1},\\
R^r &= X_5\,
       \mathrm{CX}_{\bar{5}4}\,
       \mathrm{CCX}_{453}\,
       \mathrm{C^3X}_{35\bar{4}2}\,
       \mathrm{CCX}_{5\bar{4}1},
\end{eqs}
where a bar denotes that the operation is controlled on the $\ket{0}$ state, as opposed to the $\ket{1}$ state.
As expected, we obtain operations in the fourth level of the Clifford hierarchy, namely the 3-qubit controlled operations $\mathrm{C^3X}$, for the cases of $L^r$ and $R^s$.

\section{Condensing charges at interfaces}\label{app: ResInd}

In this appendix, we consider switching between an $H$-GSC and a $G$-GSC, for which $H$ is a subgroup of $G$. We show that performing the vertex measurements for $G$ can only produce certain charges. Specifically, the charges corresponding to irreps in the induction of the trivial $H$ representation, defined below. Consequently, we show that all such charges have a vacuum condensation channel on the $H|G$ interface, so that they can always be condensed at the interface. Moreover, we argue that if a cluster of $H$ charges with trivial total charge is present in the $H$-GSC, its total charge will be condensible on the $H|G$ interface, so that cleaning up charges is possible after performing the $G$ vertex measurements. These results underlie our charge-cleaning protocols during the extension of $H$- and $K$-GSC patches to a $G = H \knit K$ GSC patch in Section~\ref{sec: extension and splitting spacetime}. 

\subsubsection{Setup}
Before performing the $G$ vertex measurements in an $H$-GSC patch, we have to add local degrees of freedom of dimension $|G|/|H|$ at each edge for a $|G|$-dimensional local Hilbert space. We can interpret these extra degrees of freedom as indicating a given left $H$ coset in $G$. By doing so, we can use the fact that any element in $G$ can be written as
\begin{align}\label{eq:g=th}
    g = th, \quad t \in T, \quad h \in H,
\end{align}
where $T$ is a left transversal of $H$, that is to say a set of representatives for each left $H$ coset. For knit products, which are the focus of the main text, we have the simplification that $T$ is also a subgroup of $G$.

\subsubsection{Which charges can appear}

Starting from an $H$-GSC and initializing each $T$ qudit in the $\ket{1}$ state,\footnote{The presence of the trivial group element in $T$ does not require that $T$ is a group. This is because the set of left $H$ cosets, $G/H$, forms a partition of the group $G$, and as a result one of the cosets must contain the trivial element. We can thus always include the trivial element in $T$.} the corresponding configuration remains flux free in the $G$-GSC. This is because we start from a flux free configuration in the $H$ GSC, and the added $\ket{1}$ qudits do not change the flux around any plaquette.

Although the resulting state is in the $H$ code space $\mathcal{H}_C^H$, it is not in the $G$ code space $\mathcal{H}_C^G$ since it is a sum over group basis states belonging to the same gauge equivalence class in $H$, but not in $G$. 
In fact, if we apply a $T$ gauge transformation to some arbitrary vertex $v$ we get:
\begin{align}
    A_v^t \ket{\psi_H} \neq \ket{\psi_H}, \quad t \neq 1\in T, \quad\ket{\psi_H} \in \mathcal{H}_C^H,
\end{align}
since this configuration will contain some element $k \notin H$ on all edges connected to vertex $v$. Therefore, the state is not gauge invariant under the $G$ gauge transformations, and as a result, simply adding $T$ qudits in the $\ket{1}$ state at each edge does not land us in the $G$ code space. 

To make this more precise, we need to consider the representation of $G$ at the vertex $v$. We find that the representation is induced from the $H$ representation at $v$ (this is made rigorous throughout the rest of this appendix). This $H$ representation is trivial before extension since, by assumption, $A_v^h\ket{\psi_H}=\ket{\psi_H}$ for all vertices $v$ and all $h\in H$. 

Explicitly, we write a basis state for the code space $\mathcal{H}_C^H$ as follows:
\begin{align}
    \ket{h}_\mathsf{L} = \prod_v \sum_{k_v \in H}A^{k_v}_v\ket{h_\ell},
\end{align}
where $\ket{h_\ell}$ denotes the configuration for the $\ket{h}_\mathsf{L}$ logical state in the left gauge. After adding the initialized $T$ qudits to each edge, we can act on the state by $A_w^t$ with $t\in T$ on some vertex $w$ to get:
\begin{eqs}
    A_w^t \ket{h}_\mathsf{L} &= \prod_v \sum_{k_v \in H}A_w^t A_v^{k_v} \ket{h_\ell}\\
    &= \left ( \prod_{v\neq w} \sum_{k_v\in H} A_v^{k_v} \right ) \sum_{k_w\in H} A_w^t A_w^{k_w} \ket{h_\ell} \\
    &= \left ( \prod_{v\neq w} \sum_{k_v\in H} A_v^{k_v} \right ) \sum_{j\in tH} A_w^j \ket{h_\ell} \neq \ket{h}_\mathsf{L},
\end{eqs}
where the last summation is over elements of the $tH$ coset, since $tH = \{th\, | \,h \in H\}$. The above calculation shows that $A_w^t$ acts nontrivially, indicating the presence of an excitation at the site containing the $w$ vertex. Since the no-flux condition holds, the excitation must be a pure charge, that is to say a representation of $G$. 

To determine which representation this is, note that it acts locally by changing the summation to be over different left $H$ cosets. Therefore, this gives a $|G/H|=|G|/|H|$-dimensional permutation representation  $R^I$ of $G$ acting on the set $G/H$, with action:
\begin{eqs}
\label{Permutation representation of G on the coset space}
    R^I(g) aH = gaH, \quad a,g \in G,
\end{eqs}
where $aH, gaH \in G/H$ are left $H$ cosets, and the action is trivial on $1H=H$ for any $h \in H$. 
Therefore, to determine which charges are created when switching from the $H$- to the $G$-GSC, we need to decompose this $G$ representation $R^I$ into $G$ irreps, which correspond to pure charge labels. 

To do so, we have to introduce the concepts of induction and restriction of a representation between a group $G$ and one of its subgroups $H$. Restriction is straightforward: for any $G$ representation $R^G$,  we can define an $H$ representation $\mathrm{Res_G^H(R^G)}$ as:
\begin{eqs}
    \mathrm{Res}_G^H(R^G)ƒ(h) = R^G(h).
\end{eqs}
It is useful to note that, even if $R_G$ is irreducible in $G$, it is not generally true that its restriction is irreducible in $H$.

The definition of induction is slightly more involved, but we give a direct construction; see Section 3.3 of Ref.~\cite{serre1977linear} for further details. Let $H$ be a subgroup of $G$, and let $R^H$ be a representation of $H$ on a vector space $W$. For each left coset $\sigma\in G/H$, introduce a copy $W_\sigma$ of $W$, and define
\begin{eqs}
V = \bigoplus_{\sigma\in G/H} W_\sigma.
\end{eqs}
The induced representation
\begin{eqs}
\mathrm{Ind}_H^G(R^H)
\end{eqs}
is a representation of $G$ on $V$. 

To define its action explicitly, choose a representative $t_\sigma\in G$ for each coset $\sigma\in G/H$. For $g\in G$ and $w\in W_\sigma$, there is a unique coset $\sigma'\in G/H$ and an element $h\in H$ such that
\begin{eqs}
g t_\sigma = t_{\sigma'} h.
\end{eqs}
The action of $g$ is then defined by
\begin{eqs}
g : w \in W_\sigma \longmapsto R^H(h)w \in W_{\sigma'}.
\end{eqs}
Thus, $G$ acts by permuting the copies of $W$, with the original $H$-representation determining the action within each copy. The resulting representation is denoted
\begin{eqs}
\mathrm{Ind}_H^G(R^H).
\end{eqs}
In particular, its dimension is
\begin{eqs}
\dim \mathrm{Ind}_H^G(R^H)
=
|G/H| \dim R^H 
\end{eqs}
If $V$ has such a decomposition, we say that $R$ is induced by $R^H$:
\begin{eqs}
    R = \mathrm{Ind}_H^G(R^H).
\end{eqs}

Having given this definition, we can now see that Eq.~\eqref{Permutation representation of G on the coset space} shows that the $A_v^t$ action on the $H$-GSC ground state space corresponds to that of the induction of the $H$ action on the same space. Since $H$ acts trivially on the ground state space, the resulting $G$ representation is (we suppress $H$ and $G$ subscripts and superscripts for the rest of the appendix):
\begin{eqs}
    R^G = \mathrm{Ind}(\mathbf{1}_H),
\end{eqs}
where $\mathbf{1}_H$ is the trivial representation of $H$. However, this does not by itself tell us about the charges that can appear after switching to the $G$-GSC. To understand this, we must decompose $R^G$ into a direct sum of irreps, which will give the possible charge types in the system after the switch.

For this computation, we make use of Schur orthogonality via the inner product of characters. Given two representations $R^1$ and $R^2$ with characters $\chi_1$ and $\chi_2$,\footnote{For a representation $R$, its character $\chi$ is defined as the trace of the corresponding matrices. For example, $\chi(g) = \mathrm{Tr}[R(g)]$, $\forall g\in G$.} there is an inner product defined as: 
\begin{eqs}
\label{inner product of characters}
    \langle{\chi_1},{\chi_2}\rangle = \frac{1}{|G|}\sum_g \chi_1(g) \chi_2(g)^*.
\end{eqs}
The characters of irreps are orthonormal under this inner product:
\begin{eqs}
    \langle{\chi_1},{\chi_2}\rangle = \delta_{R^1,R^2}.
\end{eqs}
Moreover, for a direct sum (tensor product) the characters add (multiply):
\begin{eqs}
    \chi_{1 \oplus 2} = \chi_{1} + \chi_{2}, \quad \chi_{1 \otimes 2} = \chi_{1} \chi_{2}.
\end{eqs}
By slight abuse of notation, we will indicate this inner product directly using the representation labels rather than their characters:
\begin{eqs}
    \langle R^1, R^2 \rangle \equiv \langle \chi_1 , \chi_2 \rangle.
\end{eqs}
Thus, we can in theory compute the multiplicity of any $G$ irrep $R_{(\mu,\nu)}$ in $\mathrm{Ind}(\mathbf{1}_H)$ as: 
\begin{eqs}
    n_{(\mu,\nu)} = \langle{R_{(\mu,\nu)}},{\mathrm{Ind}(\mathbf{1}_H)}\rangle.
\end{eqs}
This does not seem of much use without the explicit decomposition of $\mathrm{Ind}(\mathbf{1}_H)$, but it can be transformed to an explicit computation by using Frobenius reciprocity, which states:
\begin{eqs}
    \langle{\mathrm{Ind}(S^H)},{R^G}\rangle_G= \langle{S^H},{\mathrm{Res}(R^G)}\rangle_H,
\end{eqs}
where the subscripts of each inner product denote whether we sum over $G$ or $H$ elements in Eq.~\eqref{inner product of characters}, $S^H$ is an $H$ representation, and $R^G$ is a $G$ representation. For our case, this transforms the computation of the multiplicities into:
\begin{eqs}
    n_{(\mu,\nu)} = \langle{\mathrm{Res}(R_{(\mu,\nu)})},{\mathbf{1}_H}\rangle.
\end{eqs}
This computation can be easily done for all $G$ irreps by computing their restrictions to $H$, determining which $G$ irrep can be observed after the switch to the $G$-GSC, with the only assumption of being in a $H$ code state before the switch.

\subsubsection{Condensing the charges that appear}

Thus far, we have shown that the charges that appear when we measure the $G$ vertex stabilizers, must belong to $\mathrm{Ind}(\mathbf{1}_H)$ in the absence of errors. We now argue that these charges admit a condensation channel on the $H|G$ interface.\footnote{We show further that the only charges $R$ that can condense on the interface are those satisfying $R \subset \mathrm{Ind}(\mathbf{1}_H)$.} This means that, in the process of extension, the charges that appear from measuring the $G$ vertex stabilizers can be condensed, and thus removed, at the $H|G$ interface.

To show that a $G$ charge of type $R\subset \mathrm{Ind}(\mathbf{1}_H)$ has a condensation channel on the spatial $H|G$ interface we proceed by an explicit lattice computation. Let us consider an intermediate step in the extension of an $H$- and $K$-GSC into a $G = H \knit K$ GSC. The $H|G$ interface is given by a column of $H$ vertices adjacent to a column of $G$ vertices:
\begin{eqs}
\label{HG boundary columns}
    \vcenter{\hbox{\includegraphics[scale=.35]{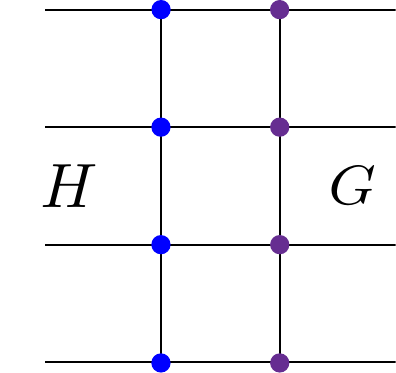}}}.  
\end{eqs}
The $G|K$ interface follows an analogous argument, so we focus on the $H|G$ interface.

We then consider a pair of adjacent vertices $v_H$ and $w_G$, belonging to the $H$- and $G$-GSCs, respectively.
To show that the $G$ charge can condense on the interface, we show that there exists a pair creation operator with endpoints $v_H$ and $w_G$ that commutes with $A_{v_H}^h$ for all $h\in H$. We consider, in particular, the pair creation operator $Z^{R^*_{ij}}$ supported on the edge between $v_H$ and $w_G$ and defined as:
\begin{eqs}
    Z^{R^*_{ij}} = \sum_{k\in G} R^*(k)_{ij} \ket{k}\bra{k}.
\end{eqs}

The commutativity would imply that $Z^{R^*_{ij}}$ creates a $G$ charge of type $R^*$ in the $G$-GSC but no $R$ charge on the $H$-GSC. In effect, the $R$ charge is condensed into the interface, while the $R^*$ charge created in the $G$-GSC can (probabilistically) fuse to the vacuum with an existing $R$ charge to remove it. 
The setup on the given pair of vertices is:
\begin{eqs}
\label{HG vertices with ZRij}
    \vcenter{\hbox{\includegraphics[scale=.60]{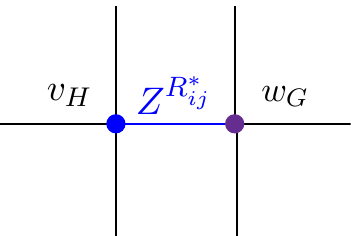}}}.  
\end{eqs}

To compute the commutation relations of $Z^{R^*_{ij}}$ and $A_{v_H}^h$, it is sufficient to consider their action on the edge between $v_H$ and $w_G$:
\begin{eqs}
    Z^{R^*_{ij}} A_{v_H}^h \ket{g} &=Z^{R^*_{ij}} \ket{h g} = R^*(hg)_{ij} \ket{hg}, \\
    A_{v_H}^h Z^{R^*_{ij}} \ket{g} &= A_{v_H}^h   R^*(g)_{ij}\ket{h g} = R^*(g)_{ij} \ket{hg}.
\end{eqs}
This shows that, for $Z^{R^*_{ij}}$ and $A_{v_H}^h$ to commute, we need the existence of matrix indices $i,j$ such that:
\begin{eqs}
\label{eq: initial condition on condensation}
    R^*(g)_{ij} = R^*(hg)_{ij}, \quad \forall\, h\in H, \, g\in G.
\end{eqs}
To find the conditions on $R^*$ to satisfy this, we can re-write the matrix elements as:
\begin{eqs}
    R^*(hg)_{ij} &= \bra{i} R^*(hg) \ket{j} = \bra{i} R^*(h) R^*(g) \ket{j},\\
    R^*(g)_{ij} &= \bra{i} \ket{R^*(g) | j},
\end{eqs}
so that the condition of Eq.~\eqref{eq: initial condition on condensation} reduces to the existence of an internal state $\ket{i}$ such that:
\begin{eqs} \label{eq: invariant internal state}
    R^*(h) \ket{i} &= \ket{i}, \quad \forall \, h\in H.
\end{eqs}

We argue now that there always exists an internal state $\ket{i}$ satisfying Eq~\eqref{eq: invariant internal state}. By assumption, $R$ satisfies:
\begin{eqs}
    \langle  \mathrm{Ind} (\mathbf{1}_H),R \rangle_G \geq 1.
\end{eqs}
Moreover, $\mathrm{Ind}(\mathbf{1}_H) \cong \mathrm{Ind}(\mathbf{1}_H)^*$ is self dual, which implies that:
\begin{eqs}
    \langle  \mathrm{Ind} \mathbf{1}_H,R^* \rangle_G \geq 1.
\end{eqs}
Frobenius reciprocity tells us that 
\begin{align}
    \langle \mathbf{1}_H, \mathrm{Res}(R^*) \rangle_H \geq 1,
\end{align}
which implies that $\mathbf{1}_H \subset \mathrm{Res}(R^*)$. Therefore, there is a one-dimensional $H$-invariant subspace when $R^*$ is restricted to $H$. We can always choose $\ket{i}$ to lie in this subspace, so that Eq.~\eqref{eq: invariant internal state} is satisfied.
With this choice of index, $Z^{R^*_{ij}}$ and $A_{v_H}^h$ commute, meaning that $Z^{R^*_{ij}}$ does not create any excitations on $v_H$.

We note that the argument above also shows that if $Z^{R^*_{ij}}$ and $A_{v_H}^h$ commute for some irrep $R^*$ of $G$, then $R^*$ must belong to $\mathrm{Ind}(\mathbf{1}_H)$. Explicitly, if the operators commute, then we have:
\begin{align}
    \langle \mathbf{1}_H, \mathrm{Res}(R^*) \rangle_H \geq 1.
\end{align}
Frobenius reciprocity then immediately implies that $R^* \subset \mathrm{Ind}(\mathbf{1}_H)$. Thus, the only charges that can condense on the interface are those that are in $\mathrm{Ind}(\mathbf{1}_H)$.

\subsubsection{$H$ charge cluster with trivial total charge.}

We have so far assumed that, when the $G$ vertex measurements are carried out, there are no $H$ charges in the system. We now treat the more general case, assuming that we measure a number of $H$ charges $R^H_a$ in a cluster, and that the cluster is isolated enough that it is composed by pairs of dual charges, so that the total charge is trivial:
\begin{align}
    \bigotimes_{a}R^H_a \supseteq \mathbf{1}_H.
\end{align}
If this assumption holds, the charges in this cluster deterministically fuse to vacuum.
We want to show that the fusion outcome of the cluster after switching to $G$ must be contained in $\mathrm{Ind}(\mathbf{1}_H)$, so that it can be also condensed on the $H|G$ wall. 

To show that the cluster can be condensed, it is sufficient to consider a pair of dual $H$ representations $R^{H}$ and $R^{H^*}$. Any cluster with trivial total charge must be composed of multiple such pairs, and the conclusion applies to each separately. 

We will need some mathematical identities to support our argument: first, by Frobenius reciprocity, for any $H$ representation $R^H$ and $G$ representation $R^G$ there is an isomorphism (sometimes called the projection formula):
\begin{eqs}
    \mathrm{Ind}(R^H) \otimes R^G \cong \mathrm{Ind}(R^H \otimes \mathrm{Res}(R^G)). 
\end{eqs}
For the $R^H, R^{H^*}$ pair present in the $H$ patch before the $G$ vertex measurement, this implies:
\begin{eqs}
\label{eq: Frobenius reciprocity projection formula}
    \mathrm{Ind}(R^H) \otimes \mathrm{Ind}(R^{H^*}) \cong 
    \mathrm{Ind}(R^H \otimes \mathrm{Res}(\mathrm{Ind}(R^{H^*}))).
\end{eqs}
We want to show that the right-hand side contains $\mathbf{1}_H$, which is equivalent to showing that, after switching to the $G$-GSC, the fusion of the charges coming from the induction of $R^H$ and ${R^H}^*$ can be condensed on the $H|G$ boundary. Again using Frobenius reciprocity, we have:
\begin{eqs}
    \langle{R^{H^*}},{\mathrm{Res}(\mathrm{Ind}(R^{H^*}))}\rangle  = \langle{\mathrm{Ind}(R^{H^*})} ,{\mathrm{Ind}(R^{H^*})}\rangle, 
\end{eqs}
which is clearly $\geq 1$, so that indeed:
\begin{eqs}
\label{eq: right hand side contains 1H}
    R^H \otimes \mathrm{Res}(\mathrm{Ind}(R^{H^*})) = & R^H \otimes (R^{H^*} \oplus \dots) \\
    = & \mathbf{1}_H \oplus \dots.
\end{eqs}
Plugging the result of Eq.~\eqref{eq: right hand side contains 1H} into the isomorphism of Eq.~\eqref{eq: Frobenius reciprocity projection formula} we finally have:
\begin{eqs}
    \mathrm{Ind}(R^H) \otimes \mathrm{Ind}(R^{H^*}) \cong \mathrm{Ind}(1_{H} \oplus \dots).
\end{eqs}
Since induction is additive, it then follows that:
\begin{eqs}
\label{eq: fused cluster after induction can be condensed}
    \langle{\mathrm{Ind}(R^H) \otimes \mathrm{Ind}(R^{H^*})},{\mathrm{Ind}(\mathbf{1}_H)}\rangle \geq 1.
\end{eqs}
This implies that there are irreps in $\mathrm{Ind}(R^H) \otimes \mathrm{Ind}(R^{H^*})$ which condense on the $H|G$ boundary.

There is one last step to make: $\mathrm{Ind}(R^H)$ is generally a reducible representation, whereas we measure irreps. As a result, we want to show that, whichever pair of $G$ irreps $S \subseteq \mathrm{Ind(R^H)}$ and $T^* \subseteq \mathrm{Ind(R^{H^*})}$ we measure, the fusion of this pair still has a condensation channel onto the $H|G$ boundary. Mathematically, we then want to show:
\begin{eqs}
\label{eq: induction of irreps in cluster is condensible}
    \langle{S \otimes T^*},{\mathrm{Ind}(\mathbf{1}_H)}\rangle &\geq 1 \, \\ \forall \, S \subseteq \mathrm{Ind}(R^H), \, T^* &\subseteq \mathrm{Ind}(R^{H^*}).
\end{eqs}
To argue this, we first note the following two properties:
\begin{eqs}
    \mathrm{Res(S\otimes T^*)} & \cong \mathrm{Res(S)} \otimes \mathrm{Res(T^*)}, \\
    \mathrm{Res(T^*)} &\cong \mathrm{Res(T)}^*.
\end{eqs}
Using these, we can rewrite the inner product of Eq.~\eqref{eq: induction of irreps in cluster is condensible} using Frobenius reciprocity:
\begin{eqs}
     \langle{S \otimes T^*},{\mathrm{Ind}(\mathbf{1}_H)}\rangle =& \langle{\mathrm{Res}(S \otimes T^*)},{\mathbf{1}_H}\rangle \\
     =& \langle{\mathrm{Res}(S) \otimes \mathrm{Res}(T)^*},{\mathbf{1}_H}\rangle \\
     =& \langle{\mathrm{Res}(S)},\mathrm{Res}(T)\rangle,
\end{eqs}
where in the last step we used the identity:
\begin{eqs}
    \langle A \otimes B, C\rangle = \langle A, C \otimes B^* \rangle.
\end{eqs}
To recap, Eq.~\eqref{eq: induction of irreps in cluster is condensible} is equivalent to:
\begin{eqs}
    \langle{\mathrm{Res}(S)},\mathrm{Res}(T)\rangle &\geq 1, \\ \forall \, S \subseteq \mathrm{Ind}(R^H), \, T^* &\subseteq \mathrm{Ind}(R^{H^*}).
\end{eqs}
By assumption, the restrictions of $S$ and $T$ contain $R$ for any $S$ and $T$, so that the above always holds, and we are done.

\end{appendix}

\bibliography{refs-LS}

\end{document}